\documentstyle[12pt,epsfig,rotating]{article}
\begin{document}
\renewcommand{\textfraction}{.1}
\renewcommand{\topfraction}{.9}
\renewcommand{\bottomfraction}{.9}
\renewcommand{\floatpagefraction}{.9}
%
%
%
\title{ Effect of Particle-Phonon Coupling on Pairing Correlations in Nuclei \\
      }
\author{
\small
J.~Terasaki \\[-1ex] 
\small
Department of Physics, University of Milan, Via Celoria 16,
I-20133 Milan, Italy; \\[-1ex]
\small
and INFN Sezione di Milano, Via Celoria 16, I-20133 Milan, Italy \\[-1ex]
\small
E-mail: terasaki@mi.infn.it \\
\small
F.~Barranco \\[-1ex]
\small
 Escuela Superior de Ingenieros Industriales, Universidad de Sevilla, \\[-1ex]
\small
Camino de los Descrubimientos, 41092 Sevilla, Spain \\[-1ex]
\small
E-mail: barranco@cica.es \\
\small
P.F.~Bortignon \\[-1ex]
\small
Department of Physics, University of Milan, Via Celoria 16,
I-20133 Milan, Italy \\[-1ex]
\small
E-mail: bortignon@mi.infn.it
 \\
\small
R.~A.~Broglia \\[-1ex]
\small
Department of Physics, University of Milan, Via Celoria 16,
\small
I-20133 Milan, Italy; \\[-1ex]
\small
INFN Sezione di Milano, Via Celoria 16, I-20133 Milan, Italy; \\[-1ex]
\small
and The Niels Bohr Institute, University of Copenhagen,
D-2100 Copenhagen, Denmark \\[-1ex]
\small
E-mail: broglia@mi.infn.it \\
\small
and \\
\small
E.~Vigezzi \\[-1ex]
\small
INFN Sezione di Milano, Via Celoria 16, I-20133 Milan, Italy \\[-1ex]
\small
E-mail: vigezzi@mi.infn.it \\
}
\date{
September 17, 2001
}
\maketitle
%
%
\begin{abstract}
The influence  of particle-phonon coupling on pairing correlations 
in nuclei is studied by solving the Dyson equation including 
the anomalous (pairing) Green function. 
We develop the formalism for solving the equation 
with the minimum of approximations. 
The solution of the Dyson equation is compared with
the diagonalization of 
particle-phonon coupled Hamiltonian in a small space. 
This comparison reveals that the effect of many-phonon states 
is incorporated  in the Dyson equation. 
We calculate the pairing gap of the neutron in $^{120}$Sn.
We compare analytically the present method 
with a simpler treatment based on Bloch-Horowitz perturbation theory.

\end{abstract}

\newpage
\section{Introduction}
The study of the nature of pairing correlations is one of the most fundamental
problems in nuclear physics. These correlations have been introduced in 
the study of nuclear structure in analogy with the case of metals \cite{Bo58}
and since then 
played  an important role in the understanding of many aspects of nuclear
behaviour, because they affect static and dynamic properties of 
atomic nuclei,
their structure as well as their reactions.
One can quote, for example,
the systematic staggering of the odd-even mass difference
(section 2-1 in \cite{Bo69}), 
the reduction of the moment of inertia of collective rotational bands
as compared to the rigid-body value (section 4-3 in \cite{Bo75}), 
the enhancement of the two-particle transfer reactions to specific 
final states \cite{Br73},
the appearance of signals in both one-particle pick-up and stripping reactions
for the same single-particle orbit near the Fermi level 
(section 5-3 in \cite{Bo75}),
the  lifetimes of $\alpha$ decay in the heavy mass region \cite{Ma62} 
and so on. 
Influence of the pairing gap can be also found in the spectra of 
deformed nuclei.$^1$
\footnote{
The influence can be recognized in average for low-lying levels.
Evidence of the correlations in each level is not always easy 
to find due to state dependence of the pairing gap.
}
Namely, the spacing of the low-lying non-collective excited states
(band heads) 
in odd nuclei is reduced in average due to the pairing gap \cite{Na65}. 
Two-quasiparticle states are of relatively high energy 
in heavy even-even nuclei, because breaking the pair costs energy.  
The feature analogous to the latter can be seen also in 
odd nuclei  
(sections 5-2 and 5-3 in \cite{Bo75}). 
The interplay between the pairing and the other degrees of freedom 
also plays an important role. 
For example, pairing correlations are suppressed by  
rotation \cite{MV60,So73}. 
%
%
%
%
%
%
%

Since pairing correlations play such an important role, it is 
crucial  to ask what is their origin.

Pairing correlations in finite nuclei have mostly been studied in mean field
theory through BCS and Hartree-Fock-Bogoliubov (HFB) calculations,
making use of phenomenological interactions,
%
the monopole pairing interaction having probably been the most often used 
(e.~g. \cite{Be69}). 
The list of the interactions may include 
the quadrupole pairing interaction 
(used for improving the monopole pairing interaction, see e.~g. 
\cite{WF78} for rotation and \cite{BB71} for the pairing vibration.),
the surface-delta interaction e.~g. \cite{KS63}, \cite{Pl66} and \cite{FP67},
density-dependent zero-range pairing forces \cite{BE91}, the 
Gogny force, \cite{Go75} and \cite{DG80}, and so on. 
%

Recently, several studies have been devoted to pairing 
in nuclei near the drip line, for which it is crucial that  
the pairing force   
treats correctly the interplay between the discrete and continuum
unperturbed single-particle levels correctly:
that is, 
summation of the unbound components of many-body wavefunction
caused by the interplay must be canceled out \cite{Doba},\cite{Te96}. 
%

The connections of pairing correlations in finite nuclei with 
the bare interactions, that is, on interactions which reproduce the 
properties of scattering of free nucleons, have been very scarce in
finite nuclei (for a few exceptions, cf.  \cite{Baldo},\cite{Barranco} and 
\cite{Go79}  ).
On the other hand, this represents an important line of research in 
nuclear matter  ( \cite{AO85}--\cite{Lombardo} and references therein ) 
and in its applications to nuclear astrophysics,
where for example pairing plays a very important role in  
the cooling of the neutron stars by neutrino emission 
(e.~g. \cite{Le00}).
 Also much studied have been the 
effects of medium polarization on pairing correlations. 
In terms of field theory  this effect is given by 
bubble diagrams of the particle-hole excitations  
\cite{BabuBrown}--\cite{Ba00}. 


In the limit of infinite number of the bubbles, one can think 
of an interaction between nucleons mediated by the exchange of 
phonons
(phonon-induced interaction). 
In this idea  the coupling, or vertex,
between particle$^2$
\footnote{
This terminology is used in different ways depending on the context. 
One is a single-particle sitting on a level above the Fermi level, 
and another is simply single-particle or nucleon. 
In this paper we use the terminology mainly in the latter sense.
}
%
 and phonon is crucially
important, and the influences of this coupling on many properties of 
phonons as well as of particles have been investigated intensively 
in finite nuclei, although not  its effect on pairing correlations.
In particular, the coupling with low-lying surface vibrations
renormalizes the single-particle motion in an essential way,
shifting the energy of the levels and changing the level density 
and the effective mass around the Fermi surface, and producing a significant
breaking of the single-particle strength for the levels lying far from
the Fermi surface \cite{Bernard},\cite{Ma85}.  
%
%
%

A formulation which can properly investigate the effect of the particle-phonon 
coupling on pairing correlations in nuclei 
was already proposed in 
1960's
by Belyaev et al.~\cite{Be61,Be66,Be65} and Migdal et al.~\cite{Mi67,Mi68} 
on the basis of field theory. 
The key point of the method is to introduce anomalous (pairing) Green functions 
in addition to normal Green functions. 
In condensed matter physics this formulation 
was introduced by Gor'kov \cite{Go58} and Nambu \cite{Na60}, 
and 
the original dynamical equation was rearranged
with approximations characteristic 
in the metal and is known as Eliashberg equation \cite{Sc64} or Gor'kov equation, 
which was shown to be powerful in the calculation of
the tunneling probability. 
A relatively recent review is given by \cite{Al82}. 
 The application of the particle-phonon coupling  
to pairing problems in finite nuclei 
has been, however, surprisingly rare. 
Recently 
Barranco et al.~solved BCS equation with the phonon-induced interaction 
obtained according to Bloch-Horowitz perturbation \cite{Ba99}. 
They have also shown the crucial importance of this coupling 
for the stability of the halo nucleus $^{11}Li$ \cite{Li11}.
Avdeenkov and Kamerdzhiev \cite{Av99} solve the Dyson equation 
in the linear approximation.  
( See also \cite{Ka87}. )
Keeping in mind  
the historical background described above, 
we investigate in this paper the effect of the phonon-induced 
interaction on nuclear pairing correlations.

 We solve the  Dyson equation consisting of the one-body 
as well as anomalous Green functions for particle-phonon coupled systems
and
calculate the pairing gaps.  
The phonon is treated as an independent degree of freedom.
In nuclei, this is a technique to treat complicated diagrams of interaction 
nonperturbatively. 
Introduction of the anomalous Green function can be understood also as 
this kind of technique. 
It is emphasized that in the field-theoretical approach
it is possible to include  effects beyond the mean field as well as 
the mean pair-field approximation, which 
manifest themself in the fragmentation of the single-particle 
strength. 
After making a choice of proper self-energy diagrams, 
we solve Dyson equation with the minimum of approximations, 
being careful not to suppress
the effects beyond BCS.  
Thus the most significant point of this study is that 
we investigate the pairing correlation without assuming a priori 
that the single-particle or quasiparticle picture is good. 
In addition a class of diagrams is included nonperturbatively 
in the calculation. 
We shall pay attention also to relation between this approach 
and diagonalization of particle-phonon coupled  Hamiltonian 
in the many-body plus phonon space.  
%

 Finally let us mention two more groups concerning the Dyson equation
 in nuclear-structure. 
One of them is M\"{u}ther and Dickhoff et al. 
They developed an approximate way to solve Dyson equation \cite{Mu88} 
( without the anomalous Green function )
and applied their method to doubly-magic nuclei
 \cite{Di92}--\cite{Am96} 
with a nucleon-nucleon interaction, 
emphasizing fragmentation of the single-particle strength. 
An application was made also to nuclear matter \cite{Vo91}. 
Another group which we note is Waroquier et al. 
They proposed another way to approximate a solution of Dyson equation 
and applied it to some nuclei \cite{Ne90,Ne91,Ne93}. 
They performed calculation also for an open-shell nucleus \cite{Sl93}
using a formulation which includes a concept of the pairing gap, however, 
the gap was not discussed in their paper. 
This group uses both nucleon-nucleon interaction and 
particle-phonon coupling. 
A comparison was made of the two methods of the groups in \cite{De97}.

 This paper is organized as follows:
In section 2 we discuss the formulation and describe our way to solve
the Dyson equation. 
We also consider the relation between diagrams with and without 
the anomalous Green functions. 
In the next section detailed comparisons are made 
between the solution of Dyson equation and the diagonalization 
of the particle-phonon coupled Hamiltonian. 
%
%
Section 4 shows results of the calculation of the neutrons 
in $^{120}$Sn using Dyson equation. 
We discuss the spectral functions, pairing gaps, 
the single-particle energy shifts and the pairing energy. 
BCS approximation is discussed briefly in section 5. 
The last section is devoted to  a summary. 
The equations used in this paper are derived thoroughly 
in appendices. 
Appendix A is for the derivation of the diagram rule. 
The self-energy used in the calculation is derived in Appendix B. 
The next appendix is about the general form of the particle 
Green functions, and some useful relations are derived. 
Appendix D treats equations for calculating the poles and 
the residues. 
The equations of the total energy and the pairing gaps are 
derived in Appendices E and F, respectively.

\section{Formulation}
The core of this study is the Dyson equation
\begin{equation}
 G^{-1}_\mu (\omega) = {G^0_\mu}^{-1} (\omega) - \Sigma_\mu(\omega) \ ,
 \label{Dy}
\end{equation}
where each term is a 2$\times$2 matrix, i.~e.,
\begin{eqnarray}
&&G_\mu(\omega) =
\left(
 \begin{array}{cc}
  G_\mu^{11}(\omega) & G_\mu^{12}(\omega) \vspace{1.2ex} \\
  G_\mu^{21}(\omega) & G_\mu^{22}(\omega) 
 \end{array}
\right) \ , \\
&&G_\mu^0(\omega) =
\left(
 \begin{array}{cc}
  {G_\mu^0}^{11}(\omega) & {G_\mu^0}^{12}(\omega) \vspace{1.2ex} \\
  {G_\mu^0}^{21}(\omega) & {G_\mu^0}^{22}(\omega) 
 \end{array}
\right) \ , \\
&&\Sigma_\mu(\omega) =
\left(
 \begin{array}{cc}
  \Sigma_\mu^{11}(\omega) & \Sigma_\mu^{12}(\omega) \vspace{1.2ex} \\
  \Sigma_\mu^{21}(\omega) & \Sigma_\mu^{22}(\omega) 
 \end{array}
\right) \ , 
\end{eqnarray}
with $\omega$ being an energy variable. 
$G_\mu(\omega)$ is a perturbed nucleon Green function obtained
by solving the equation.
$\mu$ denotes the spherical good quantum numbers $(nlj)_\mu$ 
of a single-particle orbit throughout this paper 
except for Appendix A which treats the most general case.
We assume that the system considered has spherical symmetry. 
The off-diagonal elements of $G_\mu(\omega)$ are anomalous Green functions.
$G_\mu^0(\omega)$ is an unperturbed nucleon Green function 
\begin{eqnarray}
&&\frac{1}{\hbar}{G_\mu^0}^{11}(\omega) =
\frac{e^{i\eta\omega}}
     { \omega - ( \varepsilon_\mu^0 - \varepsilon_F) + i\eta_\mu } \ , \\
&& \frac{1}{\hbar}{G_\mu^0}^{12}(\omega) =
   \frac{1}{\hbar}{G_\mu^0}^{21}(\omega) = 0 \ , \\
&& \frac{1}{\hbar}{G_\mu^0}^{22}(\omega) =
\frac{e^{-i\eta\omega}}
     { \omega + ( \varepsilon_\mu^0 - \varepsilon_F) - i\eta_\mu } \ , 
\end{eqnarray}
where $\varepsilon_\mu^0$ denotes an unperturbed single-particle energy, 
and 
$\varepsilon_F$ is Fermi level.
$\eta_\mu$ is a parameter as 
\begin{equation}
\eta_\mu = 
\left\{
 \begin{array}{cc}
  -\eta \ , & \varepsilon_\mu^0 < \varepsilon_F \\
   \eta \ , & \varepsilon_\mu^0 > \varepsilon_F
 \end{array}
\right.
\end{equation}
where $\eta$ is a real positive parameter.
In this study we consider a self-energy $\Sigma_\mu(\omega)$ 
given by the diagram in Fig.~1.

\begin{figure}
\begin{center}
\epsfig{file=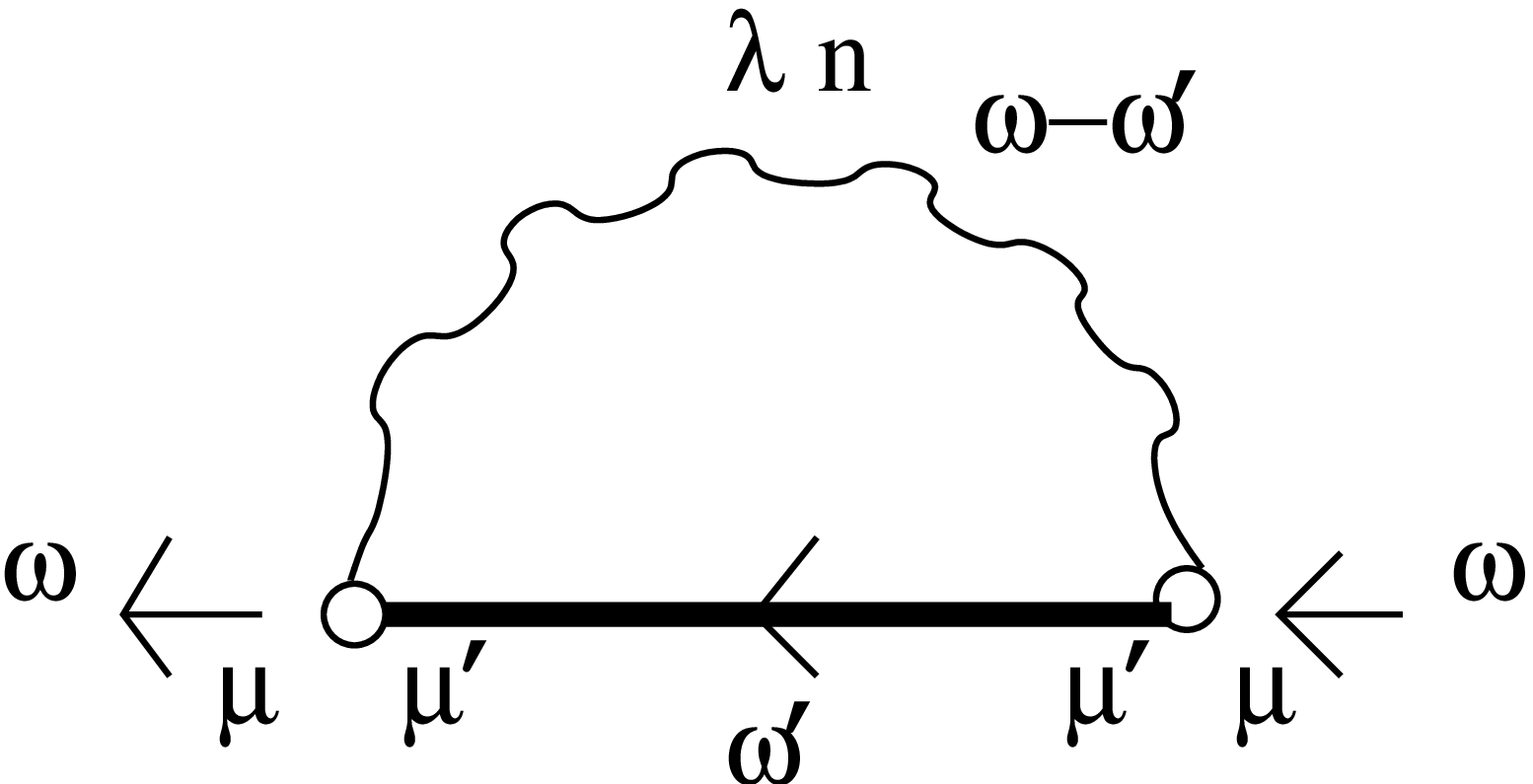,height=3cm} 

\vspace{2ex}

\parbox{13cm}{
      \baselineskip=1.1ex
      \small
Figure 1.
The self-energy. 
$\omega$ and $\omega^\prime$ denote the energy variable.
$\lambda$ stands for the multipolarity of phonon, and 
$\mu$ and $\mu^\prime$ are the single-particle indices.
 }

\end{center}
\end{figure}

The wavy line in Fig.~1 denotes an unperturbed phonon Green function 
\begin{equation}
 \frac{i}{\hbar} D_{\lambda n}^0 (\omega-\omega^\prime) = 
 \frac{i}{ \omega-\omega^\prime -\hbar\Omega_{\lambda n} + i\eta_D }
 + 
 \frac{i}{-\omega+\omega^\prime -\hbar\Omega_{\lambda n} + i\eta_D } \ ,
\end{equation}
where $\hbar\Omega_{\lambda n}$ is the phonon energy with the multipolarity 
$\lambda$ and is uniquely specified by another label $n$, 
if many modes have the same $\lambda$. 
$\eta_D$ stands for a real positive parameter. 
The thick  line indicates the perturbed nucleon Green function 
$\frac{1}{\hbar} G_{\mu^\prime}(\omega^\prime)$. 
The small open circle is an unperturbed vertex of the phonon 
and the nucleon: 
\begin{equation}
 (-)^{j_\mu + j_{\mu^\prime}}
 \sqrt{\frac{\hbar}{ 2\Omega_{\lambda n} B_{\lambda n}} }
 \langle \mu || R_0 \frac{dU}{dr} Y_\lambda ||\mu^\prime
 \rangle \ ,
\label{vertex}
\end{equation}
$B_{\lambda n}$ is the inertia parameter of the phonon. 
$R_0$ and $U(r)$ are a nuclear radius and a nuclear potential, 
respectively. 
From the diagram we obtain the equation of the self-energy
\begin{eqnarray}
\hbar\Sigma_\mu(\omega)
&=& 
 \int_{-\infty}^\infty \frac{d\omega^\prime}{2\pi} 
 \sum_{\mu^\prime}
 \tau_3
 \frac{1}{\hbar} G_{\mu^\prime}(\omega^\prime)
 \tau_3
 \sum_{\lambda n} 
 \frac{\hbar}{ 2\Omega_{\lambda n} B_{\lambda n} }
 \frac{1}{2j_\mu+1} \nonumber \\
&&
 \times \left|
 \langle \mu || R_0 \frac{dU}{dr} Y_\lambda ||\mu^\prime
 \rangle \right|^2
 \frac{i}{\hbar} D_{\lambda n}^0 (\omega-\omega^\prime) \ , \label{s-eint} 
\end{eqnarray}
where
$
\tau_3 = 
\left(
{\scriptsize
\begin{array}{rr}
1 & 0 \\
0 & -1 
\end{array}
}
\right). 
$
The derivation of the diagram rule is given in Appendix A, 
and the self-energy in a spherical nucleus is derived in Appendix B.

 For solving Eq.~(\ref{Dy}) we put
\begin{eqnarray}
&&
\frac{1}{\hbar}G_\mu^{11}(\omega) =
\sum_a
 \left(
  \frac{R_{\mu a}^{11}( \omega_{G+}^{\mu a})}{\omega - \omega_{G+}^{\mu a} }
 +\frac{R_{\mu a}^{11}(-\omega_{G+}^{\mu a})}{\omega + \omega_{G+}^{\mu a} }
 \right) e^{i\omega\eta} \ ,
 \label{G11} \\
&&
\frac{1}{\hbar}G_\mu^{12}(\omega) =
\sum_a
\left(
  \frac{R_{\mu a}^{12}( \omega_{G+}^{\mu a})}{\omega - \omega_{G+}^{\mu a} }
 +\frac{R_{\mu a}^{12}(-\omega_{G+}^{\mu a})}{\omega + \omega_{G+}^{\mu a} }
\right)
\ . \ 
 \label{G12}
\end{eqnarray}
Assuming the time-reversal invariance of the ground state, we can put 
\begin{eqnarray}
&&
\frac{1}{\hbar}G_\mu^{22}(\omega) =
\sum_a
 \left(
  \frac{R_{\mu a}^{11}(-\omega_{G+}^{\mu a})}{\omega - \omega_{G+}^{\mu a} }
 +\frac{R_{\mu a}^{11}( \omega_{G+}^{\mu a})}{\omega + \omega_{G+}^{\mu a} }
 \right) e^{-i\omega\eta} \ ,
 \label{G22} \\
&&
\frac{1}{\hbar}G_\mu^{21}(\omega) =
\sum_a
\left(
  \frac{{R_{\mu a}^{12}}^\ast( \omega_{G+}^{\mu a})}{\omega - \omega_{G+}^{\mu a} }
 +\frac{{R_{\mu a}^{12}}^\ast(-\omega_{G+}^{\mu a})}{\omega + \omega_{G+}^{\mu a} }
\right)
\ , \ 
 \label{G21}
\end{eqnarray}
where $\omega_{G+}^{\mu a}$ denotes a pole of the nucleon Green function 
( Re\,$\omega_{G+}^{\mu a}$ $>$ 0 ). 
$a$ is a label to distinguish poles associated with $\mu$.
$R^{11}_{\mu a}(\pm\omega_{G+}^{\mu a})$ are residues of
$G^{11}_\mu(\omega)/\hbar$ 
at the poles $\pm\omega_{G+}^{\mu a}$. 
Eqs.\ (\ref{G11})--(\ref{G21}) are explained in detail in Appendix C.
Inserting the above equations to Eq.~(\ref{s-eint})
we obtain
\begin{eqnarray}
\hbar\Sigma_\mu^{11}(\omega) &=&
 \sum_{\mu^\prime a} \sum_{\lambda n} 
 \frac{\hbar}{ 2\Omega_{\lambda n} B_{\lambda n} } 
 \left|
 \langle \mu ||
 R_0 \frac{dU}{dr} Y_\lambda
 || \mu^\prime \rangle \right|^2 
 \frac{1}{2j_\mu+1} \:
 \nonumber \\
&&
 \times
 \left\{
  \frac{R^{11}_{\mu^\prime a}(-\omega_{G+}^{\mu^\prime a})}
   {\omega + \omega_{G+}^{\mu^\prime a} + \hbar\Omega_{\lambda n} - i \eta_D}
  +
  \frac{R^{11}_{\mu^\prime a}( \omega_{G+}^{\mu^\prime a})}
   {\omega - \omega_{G+}^{\mu^\prime a} - \hbar\Omega_{\lambda n}+ i \eta_D}
 \right\} \ , \label{slf-e-f1}\\
\hbar\Sigma_\mu^{12}(\omega) &=&
 \sum_{\mu^\prime a} \sum_{\lambda n} 
 \frac{\hbar}{ 2\Omega_{\lambda n} B_{\lambda n} } 
 \left|
 \langle \mu ||
 R_0 \frac{dU}{dr} Y_\lambda
 || \mu^\prime \rangle\right|^2 
 \frac{-1}{2j_\mu+1} \:
 \nonumber \\
&&
 \times
 \left\{
  \frac{R^{12}_{\mu^\prime a}(-\omega_{G+}^{\mu^\prime a})}
   {\omega + \omega_{G+}^{\mu^\prime a} + \hbar\Omega_{\lambda n} - i \eta_D}
  +
  \frac{R^{12}_{\mu^\prime a}( \omega_{G+}^{\mu^\prime a})}
   {\omega - \omega_{G+}^{\mu^\prime a} - \hbar\Omega_{\lambda n} + i \eta_D}
 \right\} \ . \label{slf-e-f2}
\end{eqnarray}
$\Sigma_\mu^{22}(\omega)$ and $\Sigma_\mu^{21}(\omega)$ can be obtained 
in the same way. 
( If $R^{12}_{\mu a}(\pm\omega_{G+}^{\mu a})$'s are real, then 
$\Sigma_\mu^{12}(\omega)$ =
$\Sigma_\mu^{12}(-\omega)$. )

 Keeping the idea that stationary states are treated, 
we calculate $\omega_{G+}^{\mu a}$ and 
$R^{11}_{\mu a}(\pm \omega_{G+}^{\mu a})$ etc. in the following way. 
First $\pm {\rm Re}\,\omega_{G+}^{\mu a}$ are determined by 
searching for the points, numerically on the real $\omega$ axis, satisfying 
\begin{equation}
\det \overline{ G}_\mu^{-1} (\pm {\rm Re}\,\omega_{G+}^{\mu a}) = 0 \ ,
\label{eqdet}
\end{equation}
where 
\begin{equation}
\hbar\overline{G}_\mu^{-1} (\omega) = 
\left(
 \begin{array}{cc}
  \omega -\tilde{\varepsilon}_\mu^0 - {\rm Re}\,\hbar\Sigma_\mu^{11}(\omega) &
  -\hbar\Sigma^{12}_\mu(\omega) \vspace{1.2ex} \\
  -\hbar\Sigma^{21}_\mu(\omega) &
  \omega +\tilde{\varepsilon}_\mu^0 - {\rm Re}\,\hbar\Sigma_\mu^{22}(\omega)
 \end{array}
\right) \ ,
\label{Gbarinv}
\end{equation}
with $\tilde{\varepsilon}_\mu^0$ = $\varepsilon_\mu^0$ $-$ $\varepsilon_F$. 
The above equation is the same as $\hbar G_\mu^{-1}(\omega)$ except for the 
non-hermitian component of the diagonal elements eliminated. 
We identify the residue of $\overline{G}_\mu^{11}(\omega)/\hbar$ with 
that of $G_\mu^{11}(\omega)/\hbar$: 
\begin{equation}
R^{11}_{\mu a}(\pm \omega_{G+}^{\mu a}) = 
\left.
\frac
{  \omega + \tilde{\varepsilon}_\mu^0
 - \hbar\,{\rm Re}\,\Sigma_\mu^{22}(\omega) }
{ \frac{d}{d\omega}\, \hbar^2 \det \overline{G}_\mu^{-1}(\omega) }
\right|_{\omega=\pm{\rm Re}\,\omega_{G+}^{\mu a}}
\ ,  \label{Z11}
\end{equation}
where the derivative is taken on the real axis. 
For the pairing part the residue is calculated
\begin{equation}
R^{12}_{\mu a}(\omega_{G+}^{\mu a}) = 
 \left.
 \frac{   \Sigma_\mu^{12}(\omega) }
  { \hbar\frac{d}{d\omega} {\rm det}\,{\overline{G}}^{-1}_\mu(\omega) }
 \right|_{ \omega={\rm Re}\,\omega_{G+}^{\mu a}}
             \ .  \label{Z12}
\end{equation}
The imaginary part of the pole may be approximated by
\begin{equation}
{\rm Im} \,\omega_{G+}^{\mu a} \simeq
{\rm Im}\, \frac{\hbar}{G_\mu^{11}({\rm Re}\,\omega_{G+}^{\mu a})}
R^{11}_{\mu a}(\omega_{G+}^{\mu a}) \ , \label{Impole}
\end{equation}
or one can set Im$\;\omega_{G+}^{\mu a}$ equal to  a small constant. 
In section 4 the latter method is used. 
Eqs.~(\ref{Z11})--(\ref{Impole}) are an extension of the argument in the sections 
 3.4 and 4.3.6 in \cite{Ma85}, and are 
discussed in Appendix D.

 Eq.~(\ref{Dy}) can be solved by iteration using 
the BCS Green function ( section 7-2 in \cite{Sc64} ) 
as an initial guess of a solution: 
\begin{eqnarray}
\frac{1}{\hbar} G^{\rm B}_\mu{}^{11}(\omega) &=& 
\left(
\frac{v_\mu^2}{\omega + {\cal E}_\mu - i\eta} 
+
\frac{u_\mu^2}{\omega - {\cal E}_\mu + i\eta}
\right) e^{i\eta\omega}
\ , \label{BCS-G11} \\ 
\frac{1}{\hbar} G^{\rm B}_\mu{}^{22}(\omega) &=& 
\left(
\frac{u_\mu^2}{\omega + {\cal E}_\mu - i\eta} 
+
\frac{v_\mu^2}{\omega - {\cal E}_\mu + i\eta}
\right) e^{-i\eta\omega}
\ , \label{BCS-G22} \\ 
\frac{1}{\hbar} G^{\rm B}_\mu{}^{12}(\omega) &=& 
\frac{-u_\mu v_\mu}{\omega + {\cal E}_\mu - i\eta} 
+
\frac{ u_\mu v_\mu}{\omega - {\cal E}_\mu + i\eta} \ , \label{BCS-G12}
\end{eqnarray}
where ${\cal E}_\mu$ =
$\sqrt{ (\varepsilon_\mu^0 - \varepsilon_F)^2 + \Delta^2 } $, 
$u_\mu$ and $v_\mu$ being the occupation factors of BCS theory. 
$\Delta$ is a parameter. 
The BCS Green function is inserted to
Eqs.~(\ref{slf-e-f1}) and (\ref{slf-e-f2}), 
i.e. 
\begin{eqnarray}
&& {\cal E}_\mu \rightarrow {\rm Re}\,\omega_G^\mu{}_+\ , \label{Ew} \\
&& -\eta \rightarrow {\rm Im}\, \omega_G^\mu{}_+ \ , \\
&& v_\mu^2 \rightarrow R^{11}_\mu(-\omega_G^\mu{}_+) \ , \\
&& u_\mu^2 \rightarrow R^{11}_\mu( \omega_G^\mu{}_+) \ , \\
&& u_\mu v_\mu \rightarrow R^{12}_\mu(\omega_G^\mu{}_+) \label{uvZ12}\ .
\end{eqnarray}
Note that Eqs.~(\ref{slf-e-f1}) and (\ref{slf-e-f2}) do not have 
the summation $\sum_a$ at this stage. 
Then the new $G_\mu(\omega)$ can be calculated by the matrix inversion of  the 
right-hand side of Eq.~(\ref{Dy}). 
Finding the real points satisfying Eq.~(\ref{eqdet}) and
calculating Eqs.~(\ref{Z11}) and (\ref{Z12}),
one can get new poles $\omega_{G+}^{\mu a}$ and
$R^{ij}_{\mu a}(\pm \omega_{G+}^{\mu a})$, 
($i,j=1,2$). 
These are substituted to Eqs.~(\ref{slf-e-f1}) and (\ref{slf-e-f2}) again,
and the procedure is repeated.
The converged $G_\mu(\omega)$ and $\Sigma_\mu(\omega)$ give a 
solution of Eq.~(\ref{Dy}). 
$\varepsilon_F$ is determined so as to obtain a right expectation 
value of the nucleon number $\langle \hat{N} \rangle$ for 
the nucleus under consideration. 

 It is worthy to discuss what would happen if the energy denominators 
in Eqs.~(\ref{slf-e-f1}) and (\ref{slf-e-f2}) become very small around 
the poles of the particle Green function.   
This can happen when the unperturbed energy levels are very dense. 
Let us ignore pairing correlations for simplicity, and 
consider the pole $-\omega_{G+}^{\mu a}$. 
Then we have 
\begin{equation}
R_{\mu a}^{11}(-\omega_{G+}^{\mu a}) = 
 \frac{1}
  { 1-\left.\frac{d}{d\omega} \hbar \Sigma_\mu^{11}(\omega)\right|
   _{ \omega=-\omega_{G+}^{\mu a} }
  }
\ .
\end{equation}
Thus if the energy denominator of $\Sigma_\mu^{11}(\omega)$ is very small 
around the pole, 
then so is $R_{\mu a}^{11}(-\omega_{G+}^{\mu a})$. 
On the other hand the sum rule of the single-particle strength must be 
satisfied: 
\begin{equation}
\sum_a (\,
R_{\mu a}^{11}(-\omega_{G+}^{\mu a})
+
R_{\mu a}^{11}( \omega_{G+}^{\mu a})
\,)
= 1 \ . \label{sumrule}
\end{equation}
Therefore if the small-energy denominators are encountered so often, 
then the single-particle strength  
becomes distributed among   many poles, leading to a strong  
fragmentation \cite{Di92}. 
%
%
This mechanism means a technical advantage of Dyson method because  
the small-energy denominators do not cause severe problems as can happen in 
Rayleigh-Schr\"{o}dinger perturbation theory. 

%
If the $\omega$-dependence of the self-energy $\Sigma_\mu(\omega)$ is 
ignored, then det$\,\overline{G}_\mu^{-1}(\omega)$, see Eq.~(\ref{Gbarinv}), is 
a parabola, and there are only two roots in Eq.~(\ref{eqdet}) --- that is BCS. 
With the $\omega$-dependence of $\Sigma_\mu(\omega)$ switched on, 
det$\,\overline{G}_\mu^{-1}(\omega)$ behaves in the way schematically
shown in 
Fig.~2a. 

\begin{figure}
\begin{center}
\epsfig{file=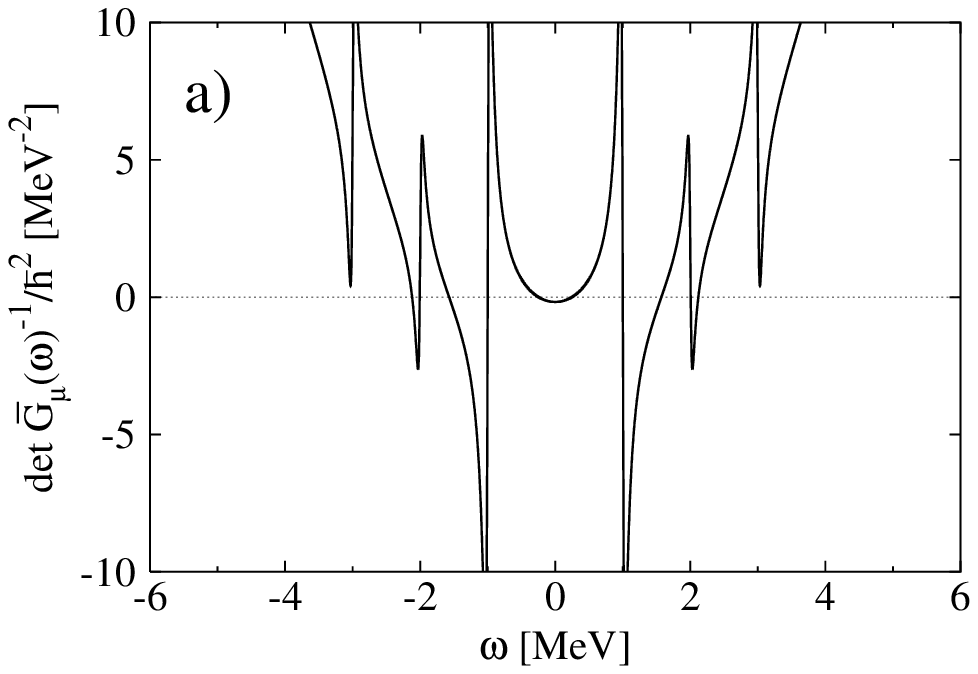,width=6.0cm} 
\epsfig{file=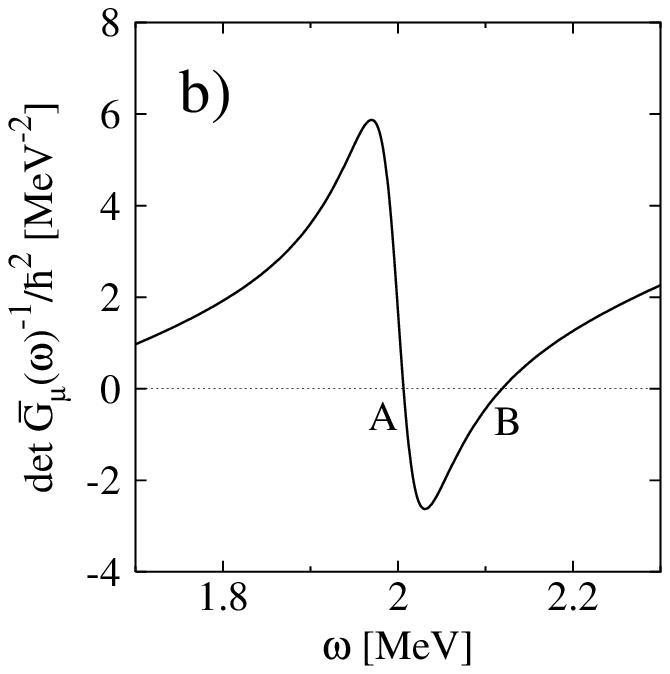,width=6.0cm} 

\vspace{2ex}

\parbox{13cm}{
      \baselineskip=1.1ex
      \small
Figure 2.
a) Schematic behavior of det$\,\overline{G}_\mu^{-1}(\omega)$ 
on the real axis of $\omega$. 
b) Magnification in a neighborhood of a singularity. 
 }

\end{center}
\end{figure}

It is seen that the singularities of $\Sigma_\mu(\omega)$  create many 
poles of $G_\mu(\omega)$. 
In Fig.~2b we illustrate a magnification in a neighborhood of a singularity. 
When the averaging parameter $\eta$ and $\eta_D$ are finite, 
the determinant is a continuous function on the real $\omega$ axis. 
It is noted that the zero-point A in Fig.~2b is a spurious pole, 
because the curve becomes discontinuous at the point A with  
$\eta, \eta_D \rightarrow 0$, and thus the pole disappears. 
In our experience the residue $R_{\mu a}^{11}(\pm\omega_{G+}^{\mu a})$  for 
the point A, calculated using Eq.~(\ref{Z11}), is usually negative.

 We note that effect of the off-diagonal Green functions 
$G_{(nlj)(n^\prime\ne n lj)}$ are assumed small and neglected in this paper. 
( Cf. \cite{Sl93} and \cite{Ec90} )

%
Now let us consider a field theoretical question related to the 
anomalous Green function. 
 Is it possible to express diagrams including the anomalous 
Green functions in terms of the normal Green functions?
Let us assume an equation 
\begin{equation}
\raisebox{-3.0ex}{ \epsfig{file=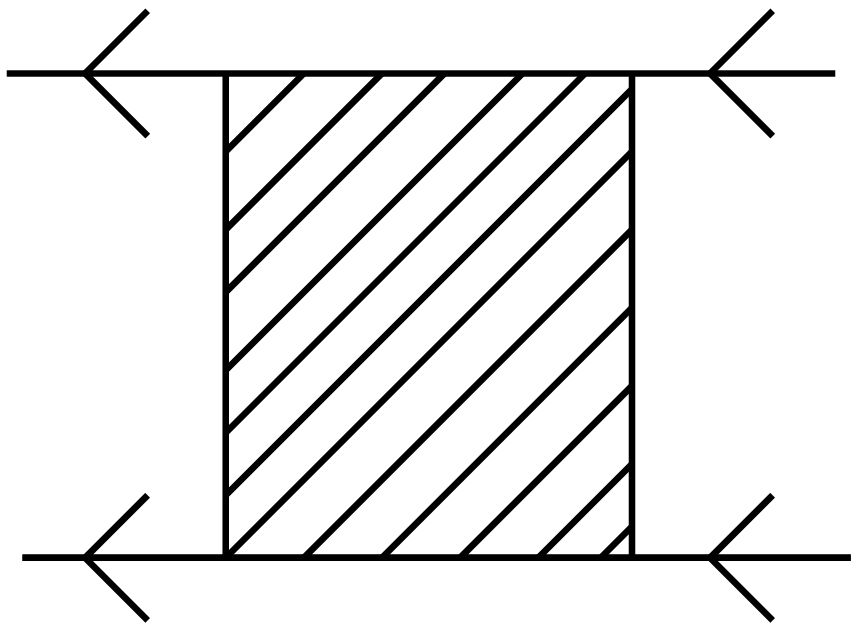,height=1.5cm} }
\raisebox{0.5ex}{\Large $\simeq$}
\raisebox{-2.5ex}{ \epsfig{file=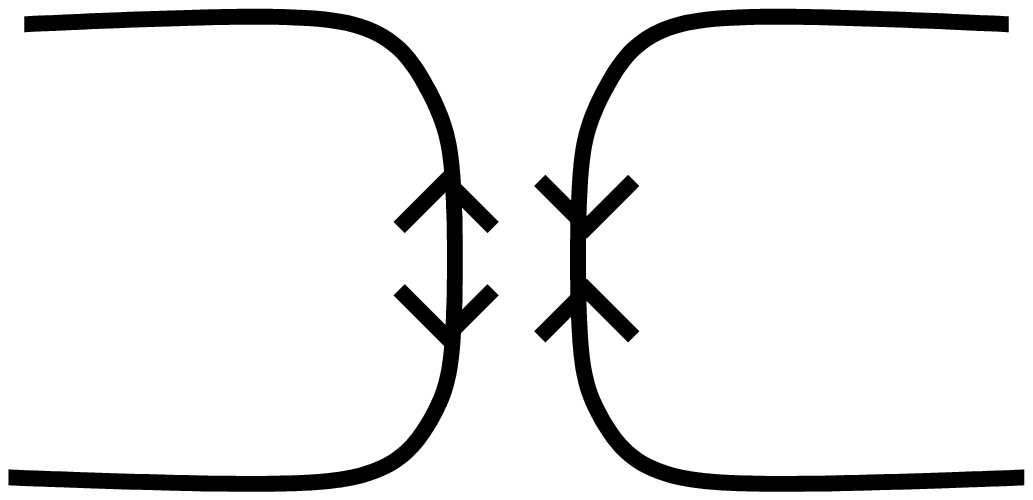,height=1.3cm} }
\end{equation}
The left diagram is a two-body Green function carrying 
the pairing correlations expressed without the anomalous 
Green functions (see section 34, chapter 7 in \cite{AGD75}).
The ladder diagram of the phonons may be included in the complicated 
diagram. 
The right-hand side indicates the anomalous Green functions
in the notation of Migdal. 
( Cf section 1.3.4 in \cite{Mi67}. ) 
In this discussion, a line represents not a 2 by 2 matrix but 
either a normal or a anomalous nucleon Green function. 
One of components of the correlation energy associated with
the solution of Dyson equation can be written, up to a factor, as 
\begin{eqnarray}
\raisebox{1.0ex}{
\epsfig{file=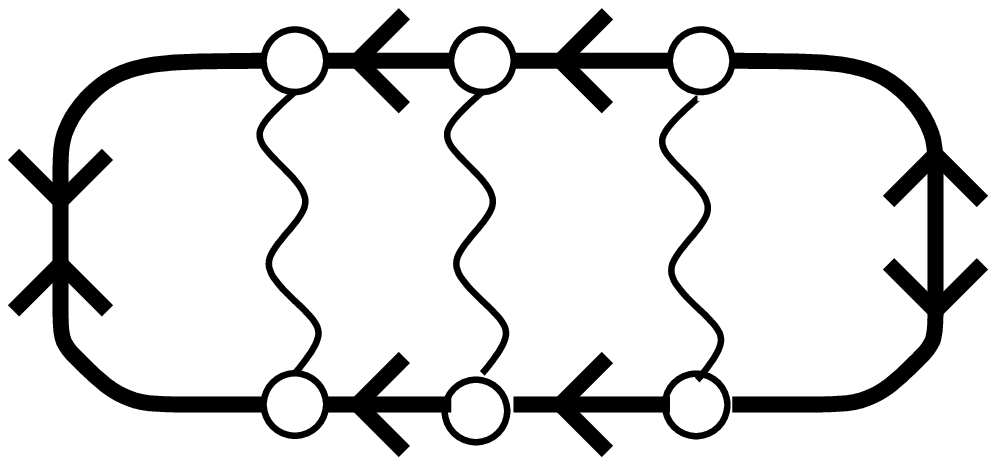,height=1.5cm} 
               }
&\raisebox{4.0ex}{\Large $=$}&
\epsfig{file=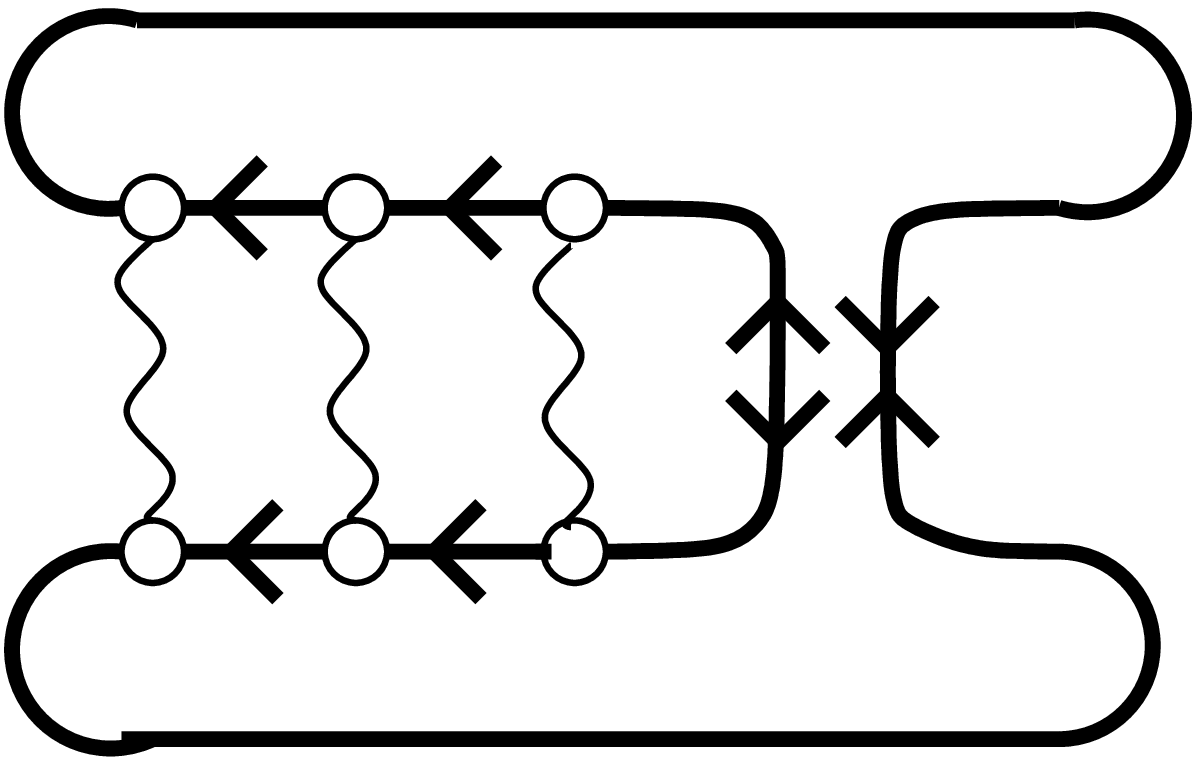,height=2cm} 
\raisebox{4.0ex}{\Large $\;\;\simeq\;\;$}
\epsfig{file=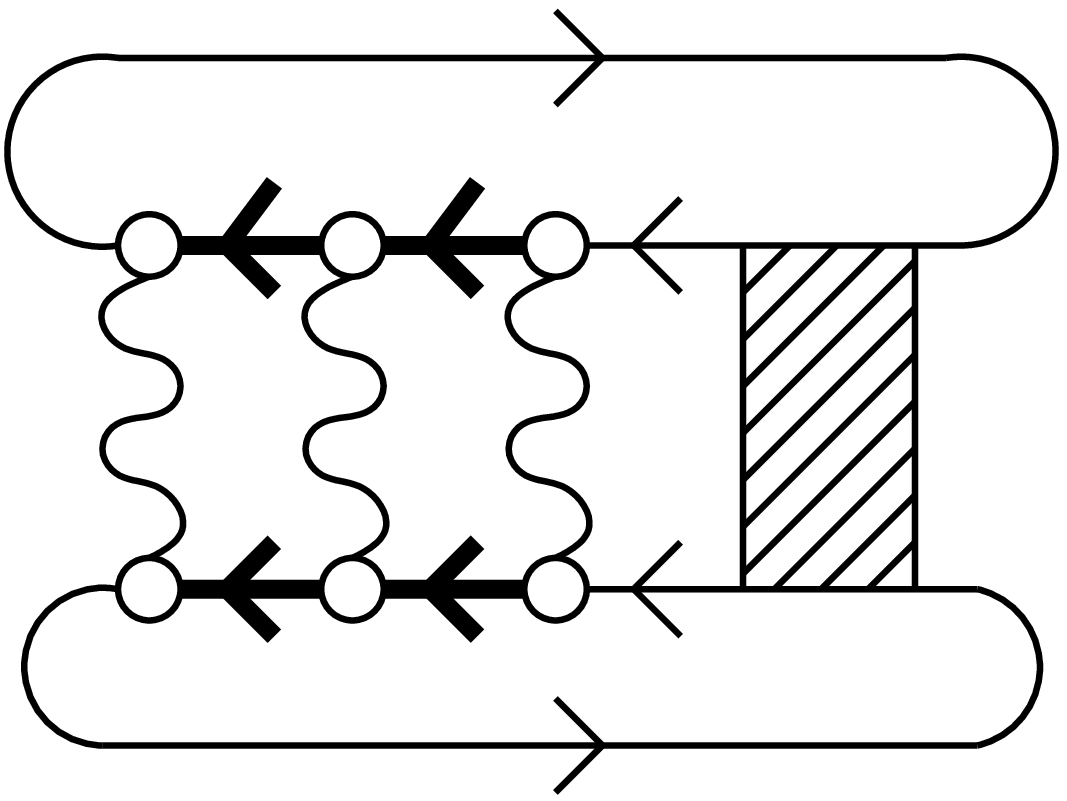,height=2cm} 
\nonumber\\
&\raisebox{3.0ex}{\Large $=$}&
\raisebox{-1.5ex}{
\epsfig{file=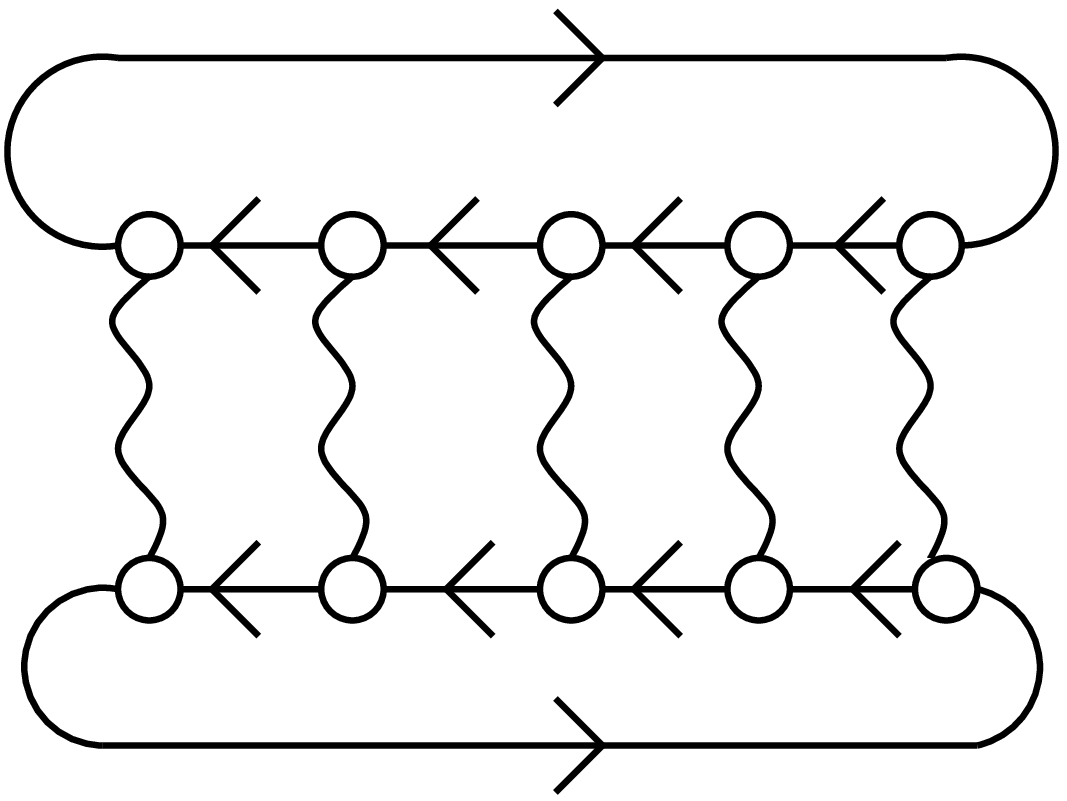,height=2cm}
 }
\raisebox{3.0ex}{\Large $+ \cdots$}
%
\end{eqnarray}
where the thick lines are perturbed Green functions, and 
the thin lines are unperturbed ones. 
Therefore it is seen that the diagrams of the following type are 
included in the proper self-energy in the Dyson equation: 
\begin{equation}
\raisebox{-2.0ex}{ \epsfig{file=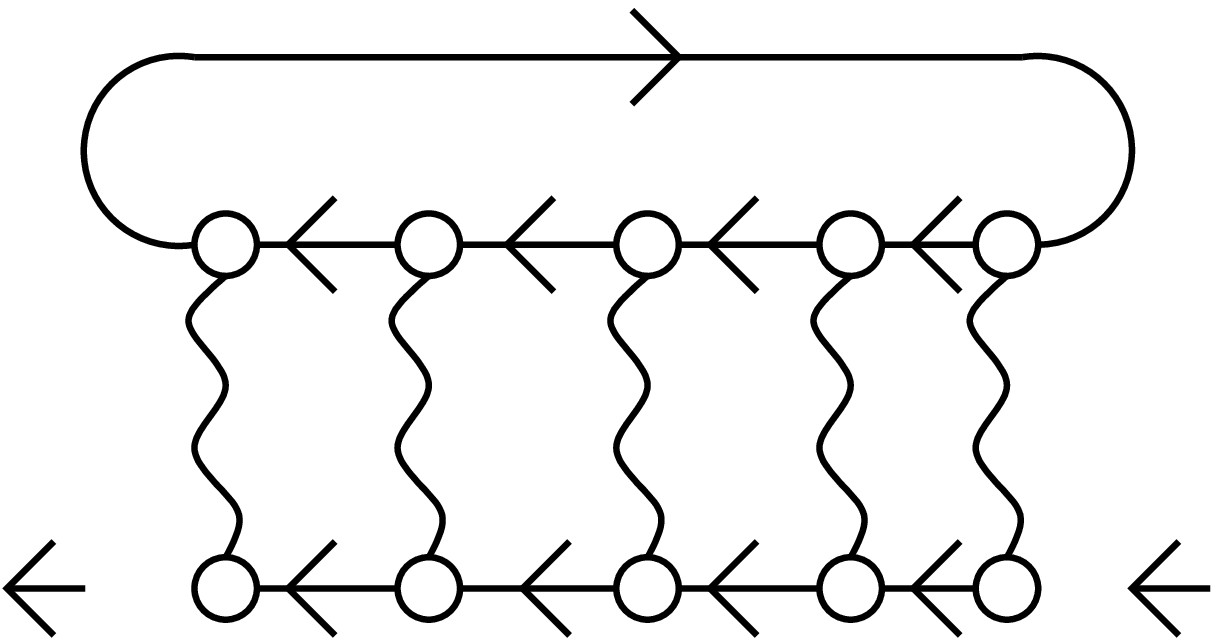,height=2.0cm} }
\raisebox{1.5ex}{\Large $+ \cdots$}
\label{s-e_lad}
\end{equation}
( See also chap.~2 in \cite{Mi67}. )

\section{Comparison with a diagonalization of
 the particle-phonon coupled  Hamiltonian}
 In order to understand this field-theoretical calculation from a viewpoint of 
a method more familiar to the nuclear-structure physics,
we made a comparison with a diagonalization of the Hamiltonian 
\begin{equation}
 H = H_{\rm par} + H_{\rm coup} + H_{\rm pho} \ , 
\end{equation}
where $H_{\rm par}$ and $H_{\rm pho}$ are unperturbed particle and 
phonon Hamiltonians, respectively. 
The particle-phonon coupling $H_{\rm coup}$  was taken from 
$H^\prime$ in section 6-5a in \cite{Bo75}. 
We diagonalized $H$ in a space of 
two particle $\otimes$ 0--2 phonons coupled to 
the angular momentum $J$ = 0.
For the particle space we took
\{ g$_{9/2}$, d$_{5/2}$, j$_{15/2}$ \} above the 
$N$ = 126 shell gap calculated with a  Woods-Saxon potential \cite{Ba99}. 
For phonons we used the lowest 2$^+$ and 3$^-$ solutions of 
a QRPA calculation performed using the multipole-multipole 
force with strengths adjusted so as to reproduce
the observed transition probabilities in $^{208}$Pb.
The coupling strength can also  be obtained from the QRPA calculation  
(cf. sections 6-2c and 6-5a in \cite{Bo75}). 

%
%

The input energies and vertex are shown in Table I and II, respectively.

\begin{table}
\parbox{13cm}{
      \small
      \baselineskip=1.1ex
\noindent
 Table I. The input energies used in the calculation.    
$\varepsilon_\mu^0$ is the unperturbed single-particle energy, and 
$\hbar\Omega_\lambda$ is the phonon energy.
}

\vspace{2.0ex}

\begin{center}
\begin{tabular}{llll}
 \hline
 \noalign{\vspace{1ex}}
$\mu$ & $\varepsilon_\mu^0$ & $\lambda^\pi$ & $\hbar\Omega_\lambda$  \\
      &  [MeV]              &               & [MeV]                 \\
 \noalign{\vspace{1ex}}
 \hline
 \noalign{\vspace{1ex}}
g$_{9/2}$  & $-$4.314 & $2^+$ & 4.10 \\
j$_{15/2}$ & $-$2.490 & $3^-$ & 2.10 \\
d$_{5/2}$  & $-$1.992 &       &      \\
 \noalign{\vspace{1ex}}
 \hline
\end{tabular}
\end{center}

\vspace{1.5ex}


\parbox{13cm}{
      \small
      \baselineskip=1.1ex
\noindent
 Table II. The matrix elements of the vertex. 
The vertex matrix is real and Hermite. 
}

\vspace{2.0ex}

\begin{center}
\begin{tabular}{lllc}
 \hline
 \noalign{\vspace{1.0ex}}
$\mu$ &
$\mu^\prime$ &
$\lambda^\pi$ &
$ \left|
\sqrt{\frac{\hbar}{2\Omega_\lambda B_\lambda}}
\langle \mu || R_0 \frac{dU}{dr} Y_\lambda || \mu^\prime \rangle
\right| $ \\
 \noalign{\vspace{1ex}}
      &                     &               & [MeV]                 \\
 \noalign{\vspace{1ex}}
 \hline
 \noalign{\vspace{1ex}}
g$_{9/2}$  & g$_{9/2}$  & $2^+$ & 1.36 \\
d$_{5/2}$  & d$_{5/2}$  & $2^+$ & 0.87 \\
j$_{15/2}$ & j$_{15/2}$ & $2^+$ & 2.24 \\
g$_{9/2}$  & d$_{5/2}$  & $2^+$ & 1.26 \\
j$_{15/2}$ & g$_{9/2}$  & $3^-$ & 4.02 \\
 \noalign{\vspace{1ex}}
 \hline
\end{tabular}
\end{center}

\vspace{1.4ex}

\end{table}

%
We compare two kinds of quantities. 
One is the occupation probability of the orbits
\begin{eqnarray}
v_\mu^2 ({\rm Dyson}) &=& 
\sum_{m_\mu}
\langle c_{\mu m_\mu}^\dagger c_{\mu m_\mu} \rangle / (2j_\mu + 1) \nonumber\\
&=&
\sum_a R_{\mu a}^{11}(-\omega_{G+}^{\mu a}) \ ,\\
v_\mu^2({\rm Diag}) &=& 
\sum_X (a_{\mu^2\,X})^2 \frac{2}{2j_\mu+1}
+
\sum_Y (a_{\mu\,Y})^2 \frac{1}{2j_\mu+1} \ ,
\end{eqnarray}
where $a_{\mu^2\,X}$ and $a_{\mu\,Y}$ are the amplitudes 
of the components of the ground state $|{\rm g\,s\,}\rangle$
in the diagonalization method: 
\begin{eqnarray}
&&|{\rm g\,s\,}\rangle = \sum_X a_{\mu^2\,X} |X \rangle
                     + \sum_Y a_{\mu\,Y} |Y \rangle \ ,\\
&& |X\rangle = [\,|\mu^2\rangle \otimes
  | 0\,,1\,{\rm or}\, 2\,{\rm phonon} \rangle ]_{J\pi=0+} \ ,\\
&& |Y\rangle = [\,|\mu\rangle
  \otimes |\mu^\prime\ne\mu\rangle
  \otimes | 0\,,1\,{\rm or}\, 2\,{\rm phonon} \rangle ]_{J\pi=0+} \ .
\end{eqnarray}
Another quantity to compare is the correlation energy 
\begin{equation}
E_{\rm cor} = E_0 - E_{\rm unp}^0 \ , 
\end{equation}
where $E_{\rm unp}^0$ is the unperturbed ground-state energy, 
which is equal to 2$\varepsilon_{g9/2}^0$ in the present model. 
The total energy $E_0$  is given in the Dyson method by          
\begin{equation}
E_0 = \sum_{\mu a} R_{\mu a}^{11}(-\omega_{G+}^{\mu a})
      ( 2j_\mu + 1 ) (-\omega_{G+}^{\mu a})
      + \varepsilon_F \langle \hat{N} \rangle \ .
\end{equation}
This equation is derived in Appendix E.
The counterpart in the diagonalization method is 
\begin{equation}
E_0 = {\rm the}\; {\rm lowest}\; {\rm eigenvalue}\; {\rm of}\; H
  - \langle H_{\rm pho} \rangle \ .
\end{equation}
The result is shown in Table III. 

\begin{table}
\parbox{13cm}{
      \small
      \baselineskip=1.1ex
\noindent
 Table III. The occupation probabilities of the single-particle orbits
$v_\mu^2$ and the correlation energies $E_{\rm cor}$ of the two methods.
$\eta$ = $\eta_{\rm D}$ = 1 keV and $\varepsilon_F$ = $-$4.7908 MeV
were used in the Dyson calculation. 
}

\vspace{2.0ex}

\begin{center}
\begin{tabular}{lllll}
 \hline
 \noalign{\vspace{1.0ex}}
    &           & $v_\mu^2$  &           &\hspace{1.4ex} $E_{\rm cor}$ \\
    & g$_{9/2}$ & j$_{15/2}$ & d$_{5/2}$ &\hspace{1.4ex}  [MeV] \\
 \noalign{\vspace{1ex}}
 \hline
 \noalign{\vspace{1ex}}
 Dyson           & 0.178 & 0.013 & 0.001 & \hspace{1.4ex} $-$1.62 \\
 Diagonalization & 0.179 & 0.013 & 0.001 & \hspace{1.4ex} $-$1.53 \\
 \noalign{\vspace{1ex}}
 \hline
\end{tabular}
\end{center}

\vspace{1.5ex}

\end{table}

It is seen that the two calculations give a nearly-identical result. 
In order to see difference between the two methods, 
we used another set of the unperturbed energies artificially changed
( see Table IV )
 keeping 
the matrix elements of the particle-phonon coupling unchanged. 

\begin{table}
\parbox{13cm}{
      \small
      \baselineskip=1.1ex
\noindent
 Table IV. Another set of the input energies. 
}

\vspace{2.0ex}

\begin{center}
\begin{tabular}{llll}
 \hline
 \noalign{\vspace{1ex}}
$\mu$ & $\varepsilon_\mu^0$ & $\lambda^\pi$ & $\hbar\Omega_\lambda$ \\
      &  [MeV]              &               & [MeV]                 \\
 \noalign{\vspace{1ex}}
 \hline
 \noalign{\vspace{1ex}}
g$_{9/2}$  & $-$3.00  & $2^+$ & 0.50 \\
j$_{15/2}$ & $-$2.50  & $3^-$ & 0.50 \\
d$_{5/2}$  & $-$2.00  &       &      \\
 \noalign{\vspace{1ex}}
 \hline
\end{tabular}

\vspace{1.5ex}

\end{center}
\end{table}

This choice was made in accordance with the discussion on the 
fragmentation in section 2. 
The physical difference between the two models is clear in the 
distribution of the single-particle strength in Fig.~3. 

\begin{figure}
\begin{center}

\epsfig{file=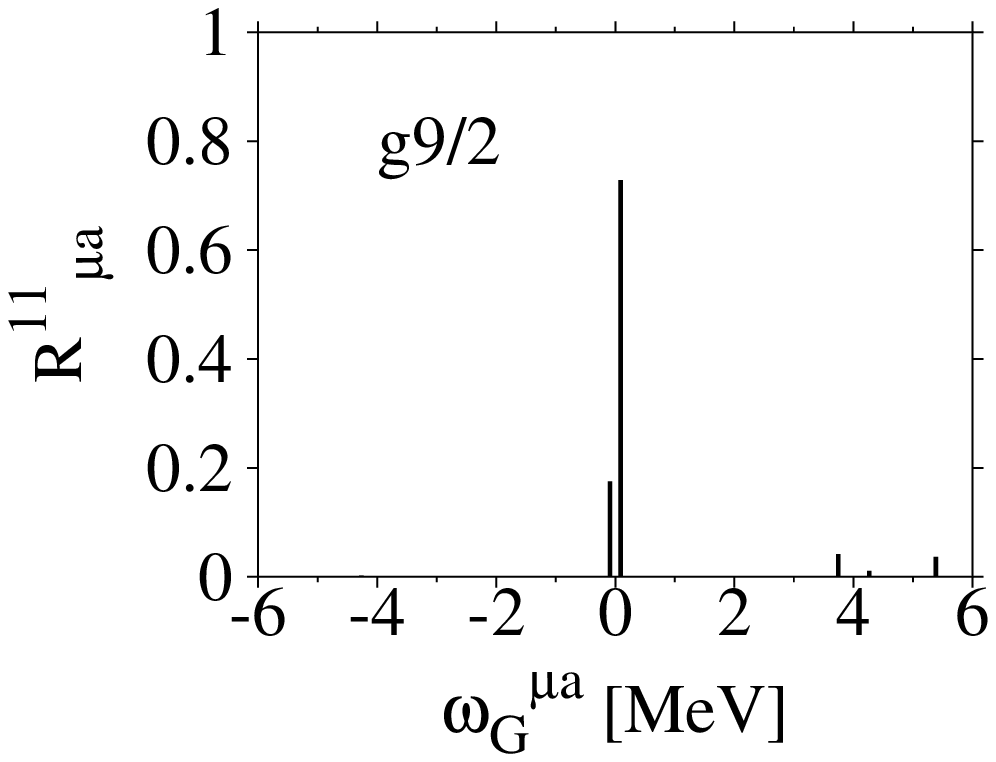,width=6.0cm} 
\epsfig{file=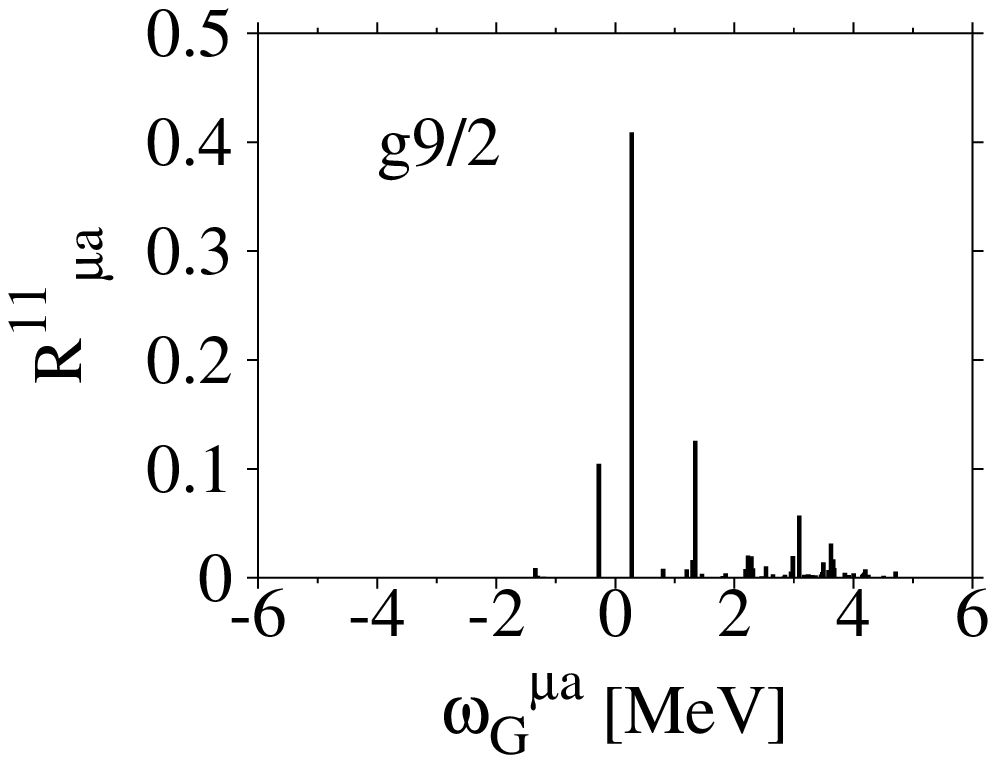,width=6.0cm} 

\vspace{1.5ex}

\epsfig{file=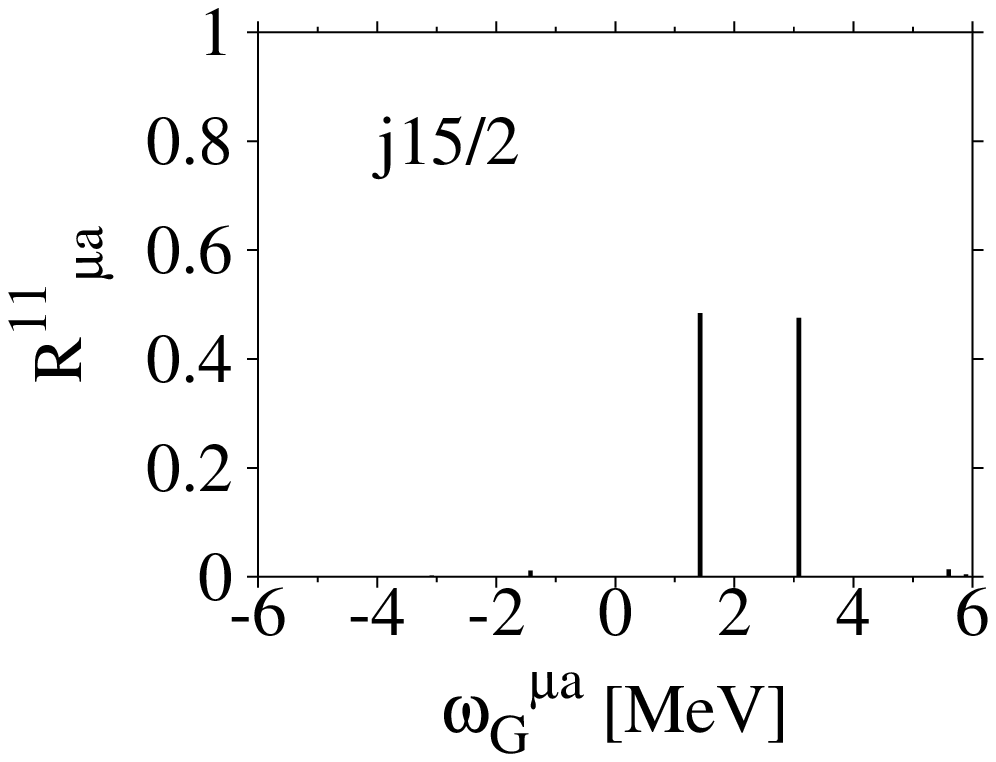,width=6.0cm} 
\epsfig{file=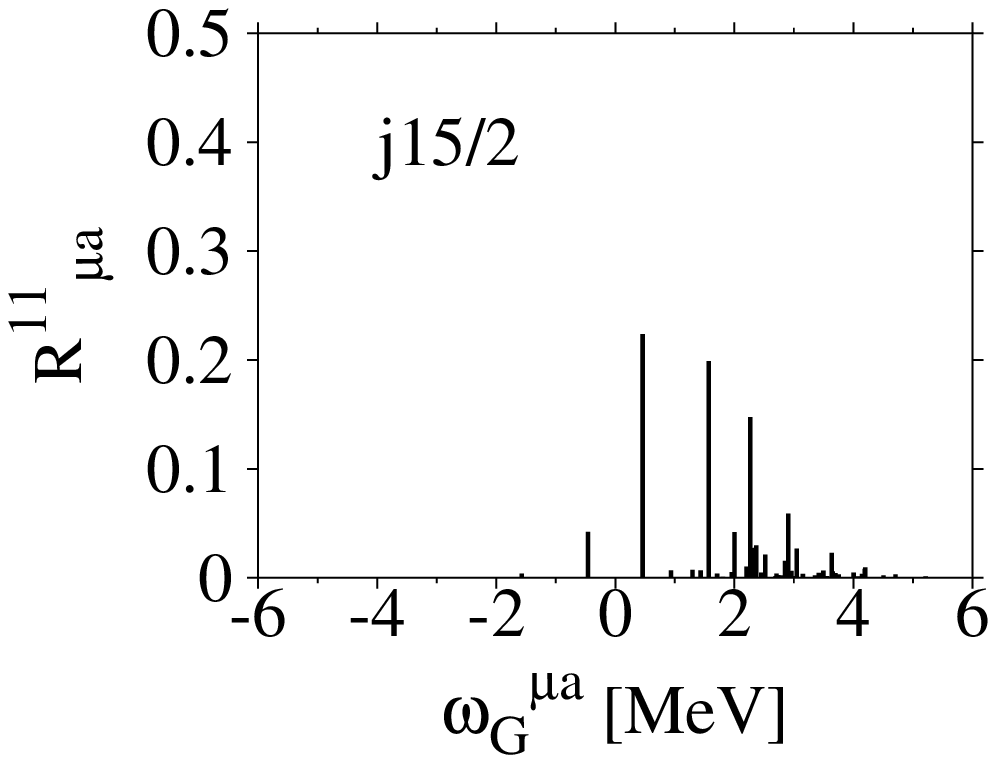,width=6.0cm} 

\vspace{1.5ex}

\epsfig{file=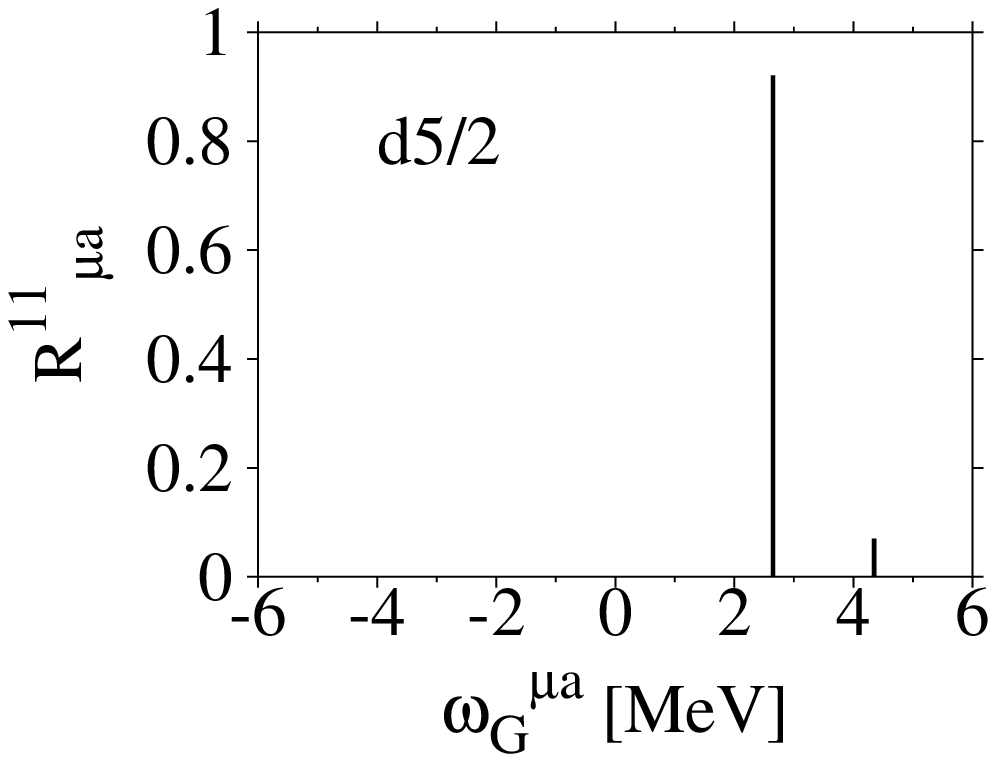,width=6.0cm} 
\epsfig{file=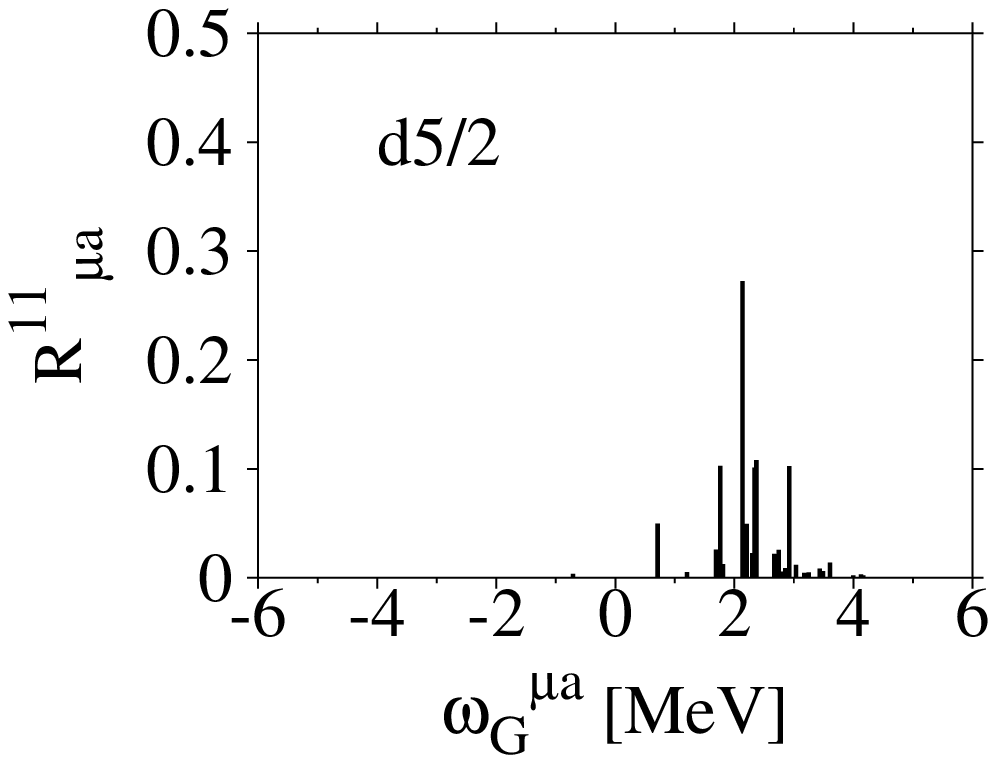,width=6.0cm} 

\vspace{2ex}

\parbox{13cm}{
      \baselineskip=1.1ex
      \small
Figure 3.
The single-particle strength distribution of the three orbits. 
$\omega_G^{\mu a}$ means $\pm\omega_{G+}^{\mu a}$. 
The left three panels are results of the first model (Table I),
and the right three are of the second model (Table IV). 
Note the difference in the vertical scale.

 }

\end{center}
\end{figure}

Given that there are only three particle levels,
the latter model (Table IV) is an extremely fragmentation-enhanced model. 
 The correlation energies of the second model is plotted in Fig.~4 (p.13). 

\begin{figure}
\begin{center}
 \epsfig{file=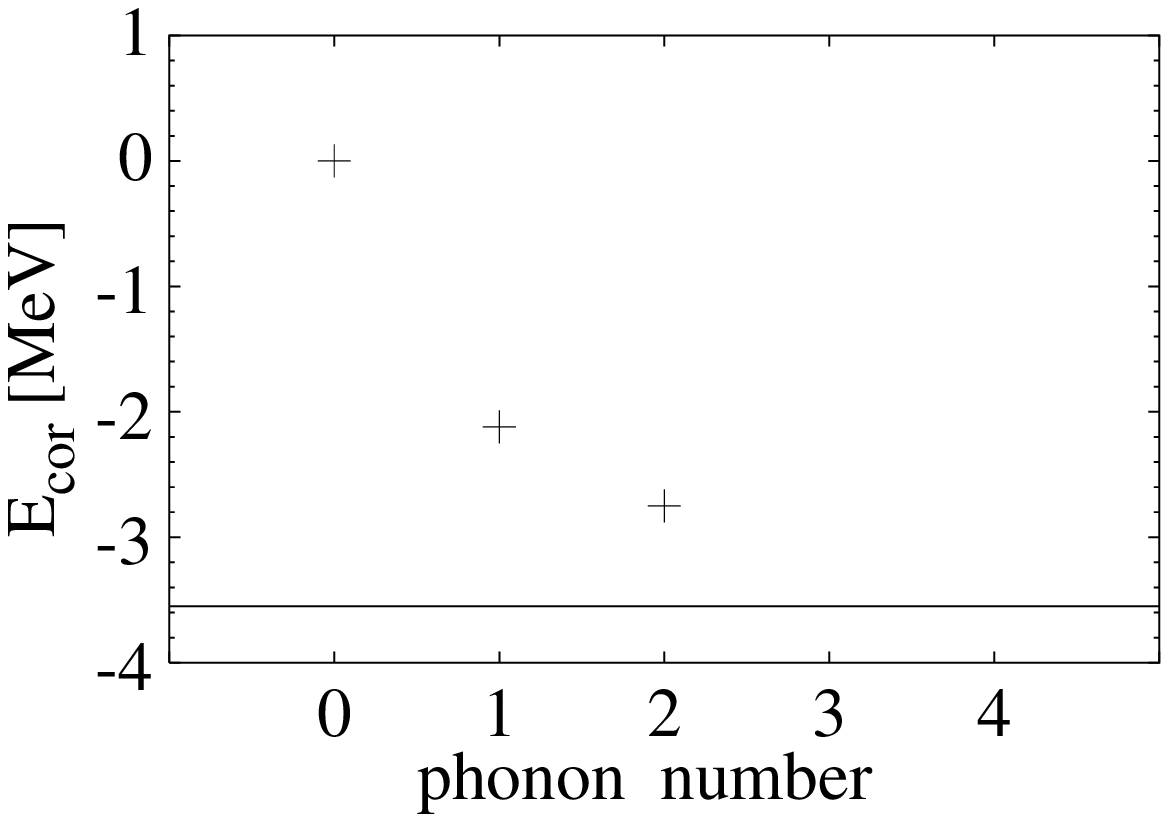,height=5cm} 

\vspace{2ex}

\parbox{13cm}{
      \baselineskip=1.1ex
      \small
Figure 4.
The correlation energies of the Dyson equation 
( $E_{\rm cor}$ = $-$3.55 $\pm$ 0.01 MeV, the mean value is shown by the line.
For the error, see text. ) and
the diagonalization method (the crosses) obtained for the input in Table IV. 
The horizontal axis is the maximum phonon number in the basis states used 
in the diagonalization, 
e.~g., phonon number = 1 means that 
the basis state is 
$
[\, |\mu\mu^\prime \rangle \otimes | 0\, {\rm or}\, 1 {\rm phonon} \rangle
\, ]_{J\pi=0+}
$. 
$E_{\rm cor}$ of the Dyson method corresponds to a large phonon number 
(see the text). 
$\varepsilon_F$ = $-$4.280 MeV was used. 
$\eta$ and $\eta_{\rm D}$ are unchanged. 

 }

\end{center}
\end{figure}

We emphasize that Dyson equation is solved nonperturbatively including
many poles, 
so that the many-phonon diagrams are included in the solution. 
Figure 4 clearly shows this many-phonon effect. 
It is also noted that the present solution of Dyson equation does not have 
the vertex correction, while the diagonalization method includes the effect 
within the order of 2-phonon. 
Thus the truncation scheme in the actual calculation is in fact different 
between the two methods.
Given this difference, the comparisons show that the two methods are
consistent.

 In our solution, the number of the poles of the particle Green functions can vary 
from one step to another in the iteration process, and due to this reason small 
fluctuations are  unavoidable. 
In most calculations presented in this paper, including those of the next section,
the errors in the pairing gaps and perturbed single-particle energies of
the quasiparticle poles in the valence shell are negligible.

\section{Pairing gaps in $^{120}$Sn}
 In the wake of the above calculation, we have performed a more 
realistic study, calculating the neutron pairing gap of $^{120}$Sn 
using the Dyson equation. 
(A part of this calculation has been published in a compact way in 
\cite{Te01}. )
In this calculation the single-particle basis 
covers all of the bound energy region starting from the orbital 1s$_{1/2}$. 
We used the unperturbed single-particle spectra
\begin{equation}
 \varepsilon_\mu^0 = 
 \frac{m}{m_{\rm k}} ( \varepsilon_\mu^{\rm WS} - \varepsilon_F )
 + \varepsilon_F \ ,
\end{equation}
with $m_{\rm k}$ = 0.7$m$ and 
$\varepsilon_\mu^{\rm WS}$ being Woods-Saxon spectra, 
for avoiding double counting of the particle-phonon coupling effect. 
( Cf. section 4.6.3 in \cite{Ma85}. )

 The computation time needed for our calculations depends strongly on 
the number of phonon modes $\lambda n$ included.
The full QRPA response for the multipolarities
$\lambda^{\pi} =2^+,3^-,4^+,5^-$ in the energy interval 
0--20 MeV used in ref.\ \cite{Ba99} consists of about two hundreds
phonon modes of energies $\hbar\Omega_{\lambda n }$ and zero-point amplitudes 
$\beta_{\lambda n}$ for each multipolarity. 
We include the four lowest phonons, one for each multipolarity, 
which give the largest contributions to the induced phonon interaction. 
We account for the effects of the other roots including only a few 
effective phonons of energy $\hbar\Omega^{\rm eff}_{\lambda n}$, distributed 
in the interval 0--20 MeV, choosing their effective strength 
so that when they are used in the Bloch-Horowitz calculation
of ref.\ \cite{Ba99} 
they reproduce the state-dependent gap obtained there.
This is obtained, considering that the sum 
of the (asymmetrized) matrix elements of the induced interaction between 
two pairs $(j_{\nu})^2_{J=0},(j_{\nu'})^2_{J=0} $ due to the phonons
lying in an energy interval $[\Omega_a,\Omega_b]$, 
calculated according to the Bloch-Horowitz formalism, is given by 
\begin{equation}
v_{\nu\nu^\prime} = 
\frac { |\langle \nu'|| R_0 
{\frac{\partial U}{\partial r}} || \nu \rangle|^2 }
{(2j_\nu +1)(2j_{\nu'}+1)(2\lambda+1)}
\sum_{\Omega_{\lambda n}= \Omega_a}^{\Omega_b} 
{\frac {4\beta^2_{\lambda n}}
{E_{\rm cor} - (e_{\nu} + e_{\nu'} + \hbar \Omega_{\lambda n})}}\ ,
\label{veff1}
\end{equation}
where $E_{\rm cor}$ and $e_{\nu }$ are the correlation energy of
the ground state
and the absolute value of the single-particle energy 
with respect to the Fermi level, respectively. 
The effective strength of the phonon  
representing this interval is then chosen so as to satisfy the equations 
\renewcommand{\theequation}{\arabic{equation}a}
\begin{equation}
{\frac{(\beta^{\rm eff}_{\lambda n})^2}
{E_{\rm cor} - (e_{\nu} + e_{\nu'} + \hbar \Omega^{\rm eff}_{\lambda n})}}= 
\sum_{\Omega_{\lambda n}= \Omega_a}^{\Omega_b} 
{\frac{\beta^2_{\lambda n}}
{E_{\rm cor} - (e_{\nu} + e_{\nu'} + \hbar \Omega_{\lambda n})}}\ ,
\label{veff2}
\end{equation}
\begin{equation}
\addtocounter{equation}{-1}
\renewcommand{\theequation}{\arabic{equation}b}
\hbar\Omega^{\rm eff}_{\lambda n} = \hbar\Omega_b\ , 
\label{veff2b}
\end{equation}
\renewcommand{\theequation}{\arabic{equation}}
for the pairs $(j_{\nu})^2_{J=0},(j_{\nu'})^2_{J=0}$ giving the 
largest contribution to the
pairing gap for the multipolarity $\lambda$. 
The energies and zero-point amplitudes of the effective 
phonons are listed in Table V, divided by 
$\sqrt{2\lambda+1}$. 

\begin{table}
\begin{center}
\parbox{12cm}{\small
\baselineskip=1.5ex
Table V. 
The energies of the phonon modes $\hbar\Omega^{\rm eff}_{\lambda n}$
and coupling strength 
$\beta^{\rm eff}_{\lambda n}/\sqrt{2\lambda+1}$. 
Tables a), b), c) and d) are for $\lambda^\pi = 2^+$, 3$^-$, 4$^+$ and 
5$^-$, respectively. The lowest-energy modes ($n = 1$) were taken from a QRPA
calculation directly. 
The coupling strengths as well as the energies of the other modes 
were determined by the procedure shown in
Eqs.\ (\ref{veff1}) and (47). 
}
\end{center}

\begin{center}
\small
\begin{tabular}{lll}
\multicolumn{3}{l}{a)}\\
\noalign{\vspace{1ex}}
\hline
\noalign{\vspace{1ex}}
\multicolumn{3}{c}{$\lambda^\pi=2^+$}\\
\noalign{\vspace{1ex}}
$n$ & $\hbar\Omega^{\rm eff}_{\lambda n}$ &
$\beta^{\rm eff}_\lambda/\sqrt{2\lambda+1}$ \\
    & [MeV] & \\
\noalign{\vspace{1ex}}
\hline
\noalign{\vspace{1ex}}
1   & \hspace{1ex}1.173 & \hspace{1ex}0.0554 \\
2   & \hspace{1ex}5.2   & \hspace{1ex}0.0134 \\
3   &            12.5   & \hspace{1ex}0.0447 \\
\noalign{\vspace{1ex}}
\hline
\noalign{\vspace{1.2em}}
\end{tabular}
\hspace{2em}
\begin{tabular}{lll}
\multicolumn{3}{l}{b)}\\
\noalign{\vspace{1ex}}
\hline
\noalign{\vspace{1ex}}
\multicolumn{3}{c}{$\lambda^\pi=3^-$}\\
\noalign{\vspace{1ex}}
$n$ & $\hbar\Omega^{\rm eff}_{\lambda n}$ &
$\beta^{\rm eff}_\lambda/\sqrt{2\lambda+1}$ \\
    & [MeV] & \\
\noalign{\vspace{1ex}}
\hline
\noalign{\vspace{1ex}}
1   & \hspace{1ex}2.423 & \hspace{1ex}0.0591 \\
2   & \hspace{1ex}5.57  & \hspace{1ex}0.0317 \\
3   &            10.0   & \hspace{1ex}0.0238 \\
4   &            21.0   & \hspace{1ex}0.0291 \\
\noalign{\vspace{1ex}}
\hline
\end{tabular}
\par
\mbox{}\par
\mbox{}\par
\begin{tabular}{lll}
\multicolumn{3}{l}{c)}\\
\noalign{\vspace{1ex}}
\hline
\noalign{\vspace{1ex}}
\multicolumn{3}{c}{$\lambda^\pi=4^+$}\\
\noalign{\vspace{1ex}}
$n$ & $\hbar\Omega^{\rm eff}_{\lambda n}$ &
$\beta^{\rm eff}_\lambda/\sqrt{2\lambda+1}$ \\
    & [MeV] & \\
\noalign{\vspace{1ex}}
\hline
\noalign{\vspace{1ex}}
1   & \hspace{1ex}2.470 & \hspace{1ex}0.0248  \\
2   & \hspace{1ex}8.0   & \hspace{1ex}0.0300  \\
3   &            12.0   & \hspace{1ex}0.0300  \\
4   &            15.0   & \hspace{1ex}0.0270  \\
\noalign{\vspace{1ex}}
\hline
\end{tabular}
\hspace{2em}
\begin{tabular}{lll}
\multicolumn{3}{l}{d)}\\
\noalign{\vspace{1ex}}
\hline
\noalign{\vspace{1ex}}
\multicolumn{3}{c}{$\lambda^\pi=5^-$}\\
\noalign{\vspace{1ex}}
$n$ & $\hbar\Omega^{\rm eff}_{\lambda n}$ &
$\beta^{\rm eff}_\lambda/\sqrt{2\lambda+1}$ \\
    & [MeV] & \\
\noalign{\vspace{1ex}}
\hline
\noalign{\vspace{1ex}}
1   & \hspace{1ex}2.402 & \hspace{1ex}0.0250  \\
2   & \hspace{1ex}8.0   & \hspace{1ex}0.0365  \\
3   &            13.0   & \hspace{1ex}0.0166  \\
4   &            21.0   & \hspace{1ex}0.0232  \\
\noalign{\vspace{1ex}}
\hline
\end{tabular}

\vspace{1.5ex}

\end{center}
\end{table}

Then the coupling strength
$$
\sqrt{ \frac{\hbar}{2\Omega^{\rm eff}_{\lambda n}B^{\rm eff}_{\lambda n}} }
= 
\frac{\beta^{\rm eff}_{\lambda n}}{ \sqrt{2\lambda+1} }\ ,
$$
is used for Eq.\ (\ref{vertex}).  
The BCS+Bloch-Horowitz calculation performed with this restricted 
ensemble of phonons reproduces the state-dependent pairing 
gaps of ref.\ \cite{Ba99} within a few per cent.   

Below, results obtained  with the fixed imaginary parameter
$\eta=\eta_0=-{\rm Im\,}\omega_{G+}^{\mu a}=$ 1 keV are shown. 
In the present calculation 
the maximum number of poles with respect to $\mu$ is $\sim$200. 
The sum rule of the single-particle strength, 
Eq.~(\ref{sumrule}), is satisfied by more than 90 \% in average 
in the orbits in the valence shell. The computations have been performed
on a parallel computer. In fact the Green function method is well suited
for parallel computation, because the pole search of each orbit can be done 
separately. 

 A few self-checks are possible on the accuracy of 
the solution of Dyson equation.
The self-energy is a functional of the particle Green function, 
and if this Green function, represented by Eqs.~(\ref{G11})--(\ref{G21}),
is identical to 
$(\,{G_\mu^0}^{-1}(\omega) - \Sigma_\mu(\omega)\,)^{-1}$, 
then the Green function is a solution. 
Such a direct check is shown in Fig.~5 for the orbits 
in the valence shell around the Fermi level. 

\begin{figure}
\begin{center}

\epsfig{file=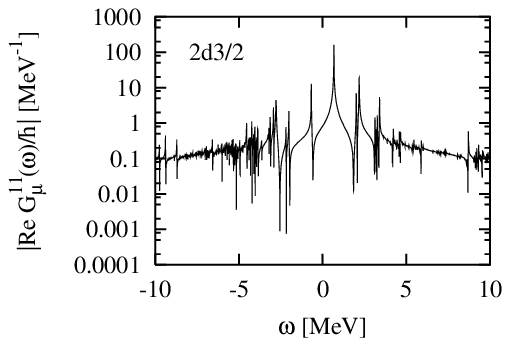,height=4.0cm} 
\hspace{-4ex}
\epsfig{file=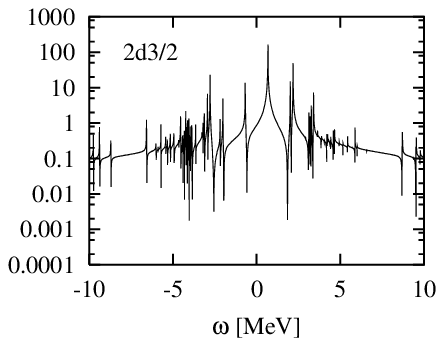,height=4.0cm} 

\epsfig{file=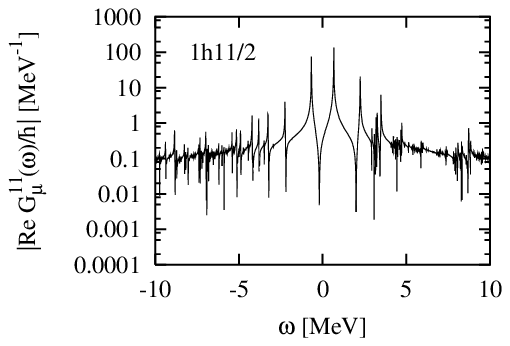,height=4.0cm} 
\hspace{-4ex}
\epsfig{file=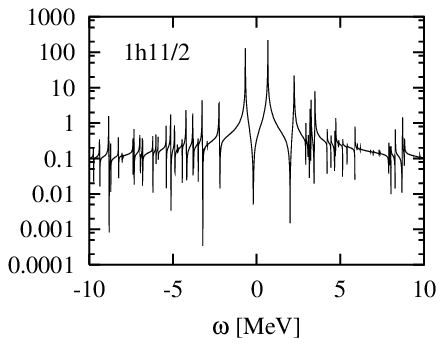,height=4.0cm} 

\epsfig{file=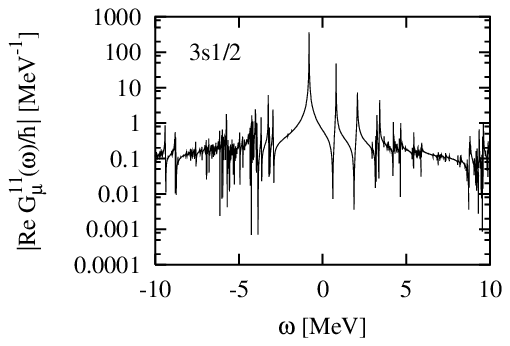,height=4.0cm} 
\hspace{-4ex}
\epsfig{file=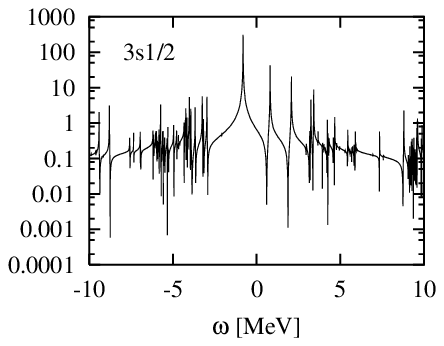,height=4.0cm} 

\epsfig{file=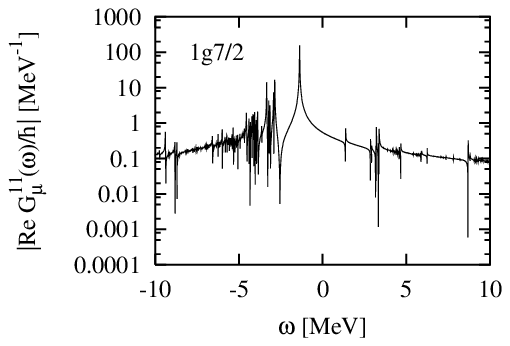,height=4.0cm} 
\hspace{-4ex}
\epsfig{file=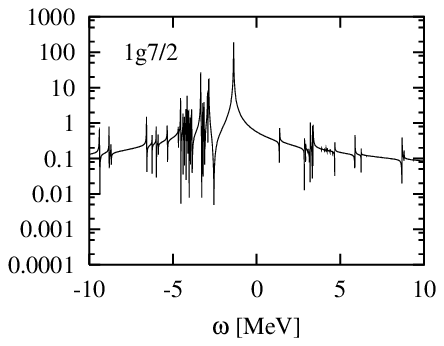,height=4.0cm} 

\epsfig{file=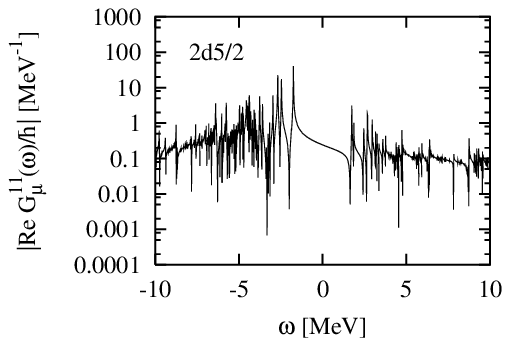,height=4.0cm} 
\hspace{-4ex}
\epsfig{file=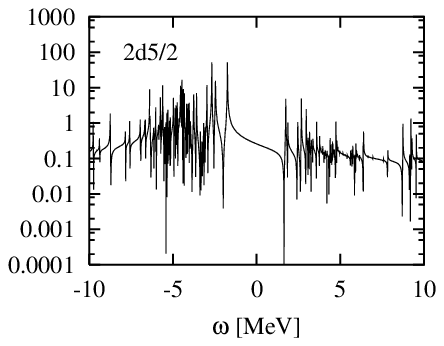,height=4.0cm} 

\vspace{2ex}

\parbox{13cm}{
      \baselineskip=1.1ex
      \small
Figure 5.
Absolute value of the real part of $G_\mu^{11}(\omega)/\hbar$ 
in log scale. 
The left column shows 
$\left|\left[(\,\hbar{G_\mu^0}^{-1}(\omega) -
\hbar\Sigma_\mu(\omega)\,)^{-1} 
\right]_{11}\right|$, 
and the right column illustrates  
those given by Eq.~(\ref{G11}). 
 }

\end{center}
\end{figure}

It is seen that the accuracy is good.
%
%
%
Another check is Eq.~(\ref{Z11Z22Z12}) in Appendix C. 
This equation is satisfied with good accuracy. 

 We show the calculated pairing gaps in Fig.~6 (p.17). 

\begin{figure}
\begin{center}

  \epsfig{file=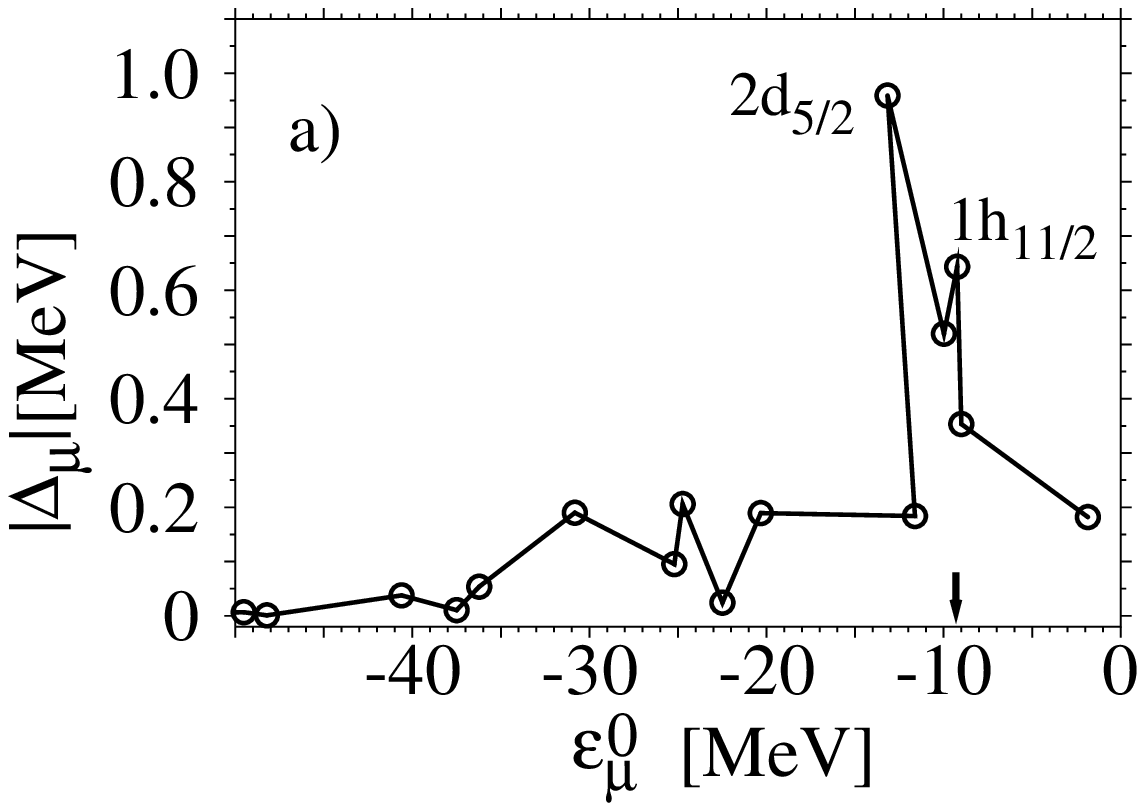,height=5cm} 
  \epsfig{file=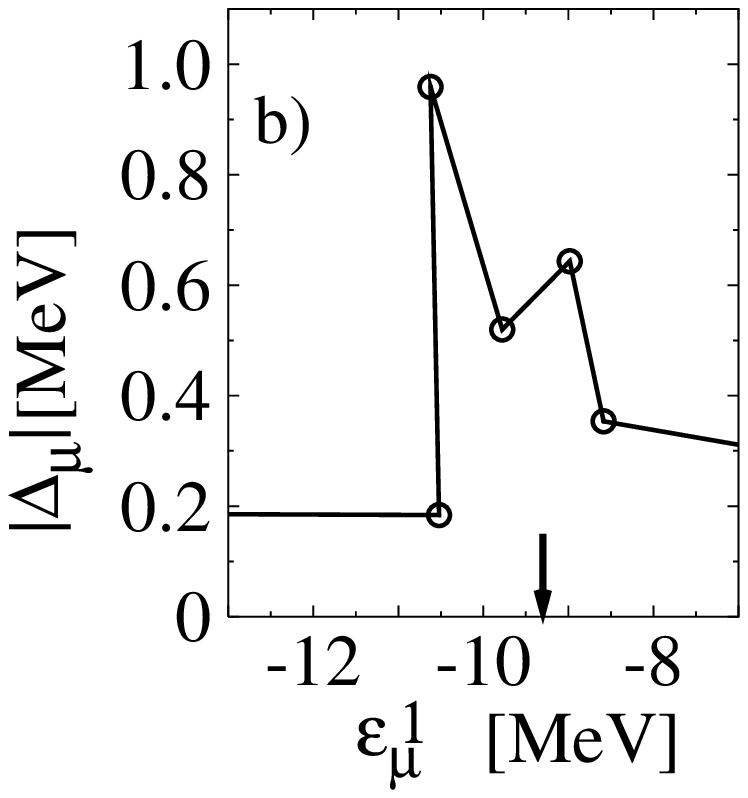,height=5cm} 

\vspace{2ex}

\parbox{13cm}{
      \baselineskip=1.1ex
      \small

Figure 6.
a) Absolute values of the pairing gaps of the neutron 
of $^{120}$Sn in the calculation of the Dyson method 
shown as a function of $\varepsilon_\mu^0 $. 
The gap is obtained according to Eq.~(\ref{gap}) in Appendix F. 
The arrow indicates the location of the Fermi level. 
b) $\left| \Delta_\mu\right|$ as a function of the perturbed 
single-particle energy $\varepsilon_\mu^1$ 
(see Eq.~(\ref{per-spe}) in Appendix F) 
in the vicinity of the Fermi level. 
 }

\end{center}
\end{figure}

The value of the gap of 2d$_{5/2}$ is mostly determined by
the off-diagonal element of 
Eq.~(\ref{vertex}) with 
$\mu^\prime$ = 1h$_{11/2}$ and $\lambda^\pi$ = 3$^-$. 
Therefore 
$ | R_{\rm 1h11/2}^{12} (\omega_{G+}^{{\rm 1h11/2}\,a_0}) | $ = 0.35  
of the quasiparticle pole$^3$ 
\footnote{ 
 A pole, or a pair of poles if the pairing gap is not zero, 
 which carries the largest, or major,  single-particle strength  
$ R_{\mu a}^{11} ( \omega_{G+}^{\mu a}) + 
  R_{\mu a}^{11} (-\omega_{G+}^{\mu a}) $. 
} 
%
$a_0$ plays an important role not only for 
$\Delta_{\rm 1h11/2}$ but also for $\Delta_{\rm 2d5/2}$. 
On the other hand 
$ | R_{\rm 2d5/2}^{12} (\omega_{G+}^{{\rm 2d5/2}\,a_0}) | $ is 0.04, 
thus the influence of 2d$_{5/2}$ on $\Delta_{\rm 1h11/2}$ is small. 

 The average of the pairing gaps in the valence shell is 0.54 MeV in 
the present calculation. 
This may be compared with the value of 1.39 MeV deduced from the observed 
odd-even mass difference  as well as
the average gap of 0.82 MeV obtained in ref.~\cite{Ba99}.

The poles $\pm \omega_{G+}^{\mu a}$ of the Green function $G^{11}$  are 
distributed symmetrically with respect to the Fermi surface ($\omega$ = 0). 
The spectral function of the orbits in the valence shell are illustrated 
in Fig.~7 (p.18).
The single-particle picture is rather good for most orbits,
except for the $d_{5/2}$ whose sinlge-particle strength is split into several
peaks. Most orbits have a pronounced hole- or particle-character, except for
for the $h_{11/2}$ which displays two large peaks on either side of the 
Fermi energy. It is to be noted that for this orbit the single-particle 
strength for the two quasi-particle poles is only 0.73.
The third not negligible peak close to $\omega= $2 MeV may be associated with the 
quasiparticle plus a collective state.

 It is seen from Eq.~(\ref{G11w}) in Appendix C that 
the peaks in $\omega$ $<$ 0 part of the spectral function are associated with
a pick-up reaction and 
those in the  $\omega$ $>$ 0 part with a stripping reaction. 
Thus the spectral functions of Fig.~7, can be rearranged as shown into 
Fig.~8 (p.19) for the case of $h_{11/2}$. 

\begin{figure}
\begin{center}

\epsfig{file=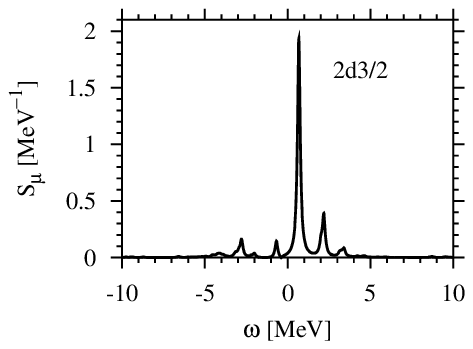, height=5.0cm}
\epsfig{file=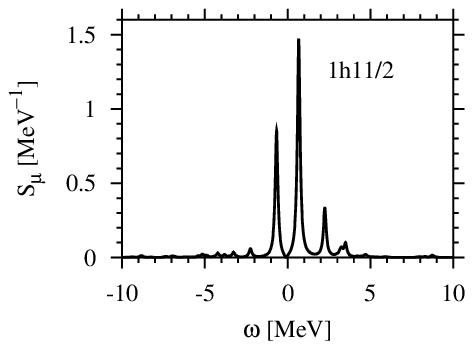, height=5.0cm}

\vspace{1ex}

\epsfig{file=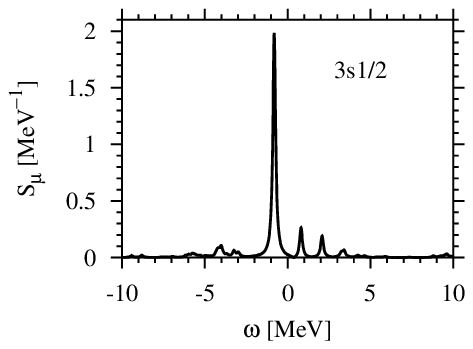, height=5.0cm}
\epsfig{file=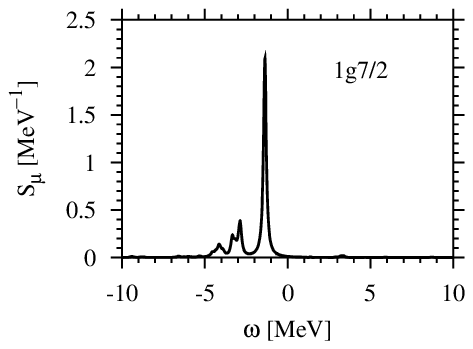, height=5.0cm}

\vspace{1ex}

\epsfig{file=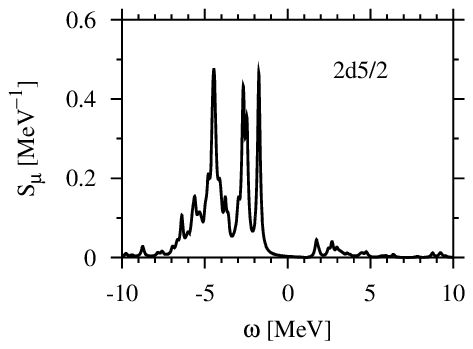, height=5.0cm}

\vspace{2ex}

\parbox{13cm}{
      \baselineskip=1.1ex
      \small
Figure 7.
The spectral function 
$S_\mu(\omega)$ = $\left|{\rm Im}\;G_\mu^{11}(\omega)/\hbar\right|/\pi$
for the orbits in the valence shell. 
The poles and residues of the solution were 
inserted into Eq.~(\ref{G11}) with  Im$\,\omega_{G+}^{\mu a}$ = 
$-$100 keV. 
$\omega$ = 0 corresponds to the Fermi level. 
 }

\end{center}
\end{figure}

\begin{figure}
\begin{center}

\epsfig{file=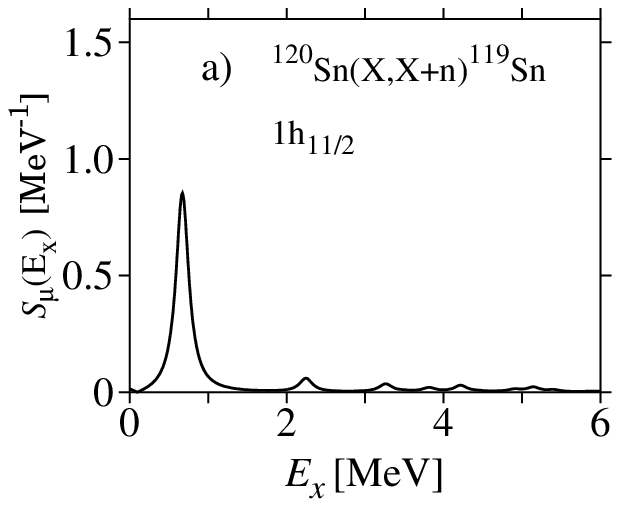,height=4.0cm} 
\epsfig{file=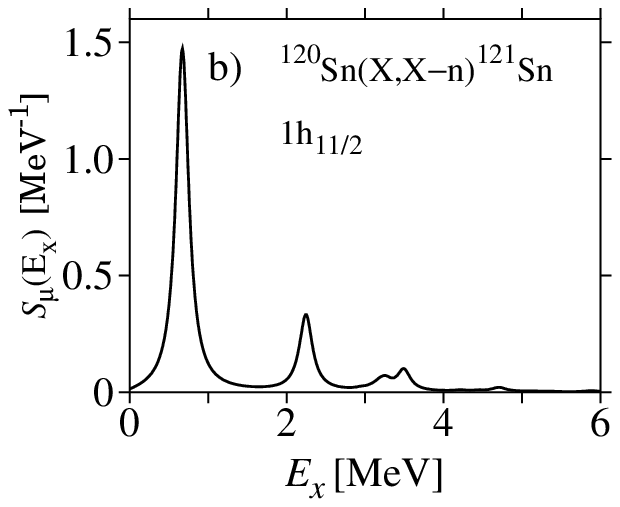,height=4.0cm} 

\vspace{2ex}

\parbox{13cm}{
      \baselineskip=1.1ex
      \small
Figure 8. 
The spectral functions of the neutron 1h$_{11/2}$ as functions of 
$E_x$ = $\left| \omega \right|$. 
a) hole type b) particle type. 
 }

\end{center}
\end{figure}

If each excited state can be resolved,
the definition of $E_x$ is $E_0$ $-$ $E_i$,
where $E_0$ is the ground-state total energy of the neutrons in $^{120}$Sn
relative to $\varepsilon_F\langle \hat{N}\rangle$, and
$E_i$ is defined in the same way as $E_0$ but for various states of 
$^{119}$Sn and $^{121}$Sn. 
If strong signals are observed
in both pick-up and stripping reactions for the same orbit 
near the Fermi level 
with the same $E_x$, this would be a clear indication of pairing correlations. 

 The particle-phonon coupling influences of course also 
the single-particle spectrum. 
We show the perturbed as well as unperturbed spectra in Fig.~9 (p.19). 

\begin{figure}
\begin{center}
\epsfig{file=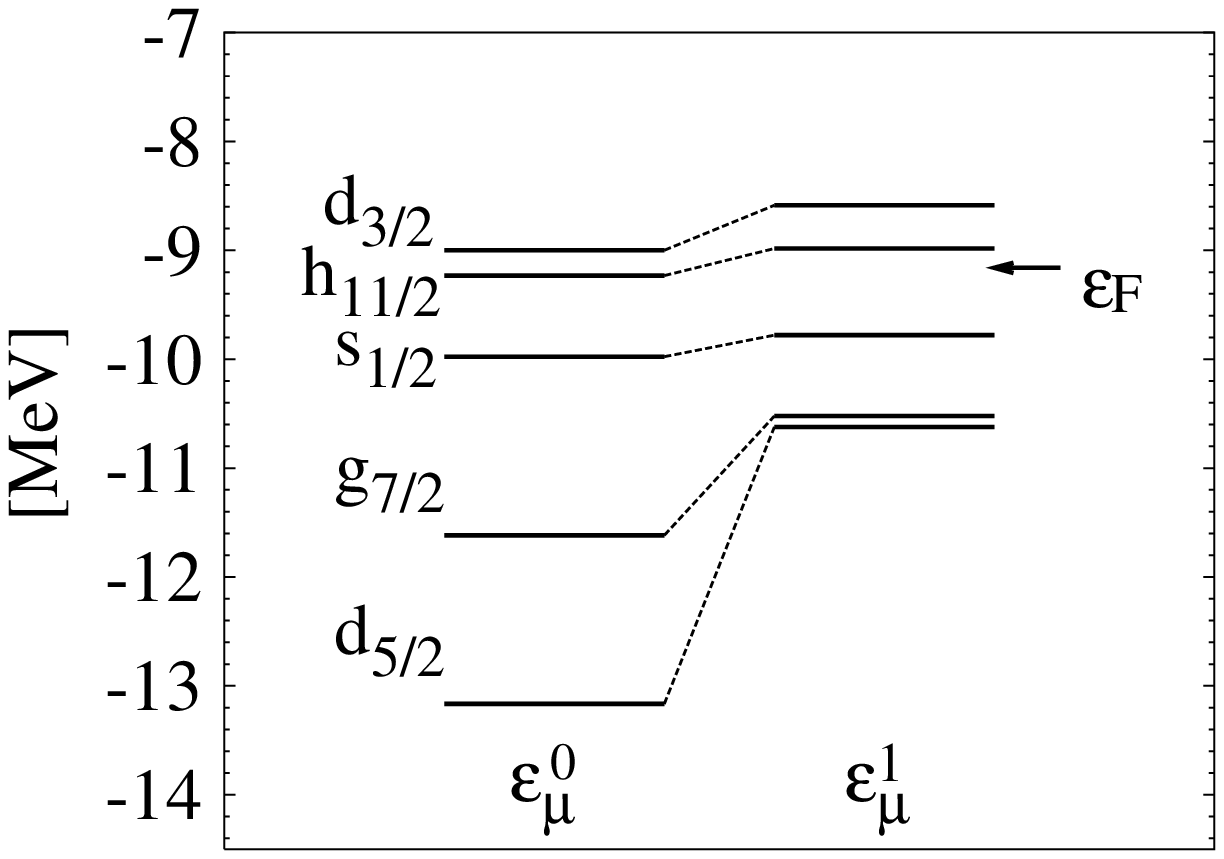,height=5.0cm} 

\vspace{2ex}

\parbox{13cm}{
      \baselineskip=1.1ex
      \small

Figure 9.
The unperturbed ($\varepsilon_\mu^0$) and 
perturbed ($\varepsilon_\mu^1$) single-particle spectra 
associated with the quasiparticle poles. 
$\varepsilon_\mu^1$ is defined in Eq.~(\ref{per-spe}) in Appendix F. 
 }

\end{center}
\end{figure}

The increase in the level density in the vicinity of the Fermi level 
is a known effect of the coupling (section 4 in \cite{Ma85}). 

 We calculated also the pairing energy (see Appendix E):
\begin{equation}
E_{\rm pair} = 
-i \sum_\mu (2j_\mu + 1)\frac{1}{2} \int_{-\infty}^\infty \frac{d\omega}{2\pi}
\left(
 \hbar \Sigma_\mu^{12}(\omega) \frac{1}{\hbar} G_\mu^{21}(\omega) e^{ i\eta\omega}
+\hbar \Sigma_\mu^{21}(\omega) \frac{1}{\hbar} G_\mu^{12}(\omega) e^{-i\eta\omega}
\right) \ ,\label{Epair}
\end{equation}
and obtained $E_{\rm pair}$ = $-$3.9 MeV.

\section{BCS approximation}
This approximation was introduced in section 2 as the initial 
guess for solving Dyson equation. 
From the viewpoint of Green function method, 
BCS approximation is a normalized-quasiparticle approximation. 
Thus, generally speaking,  if the single-particle strength is well concentrated on 
the quasiparticle pole, then BCS may be a good approximation. 
As was discussed in section 2, if the self-energy has no $\omega$-dependence, 
then the single-particle strength is distributed to only two poles at most. 

 Next we compare analytically the present Dyson calculation and 
the BCS + Bloch-Horowitz calculation in \cite{Ba99}. 
One of the differences is, as was mentioned, that
$R_{\mu a_0}^{12}(\omega_{G+}^{\mu a_0})$ 
($a_0$ denotes the quasiparticle pole)
differs from $u_\mu v_\mu$ in BCS due to the fragmentation 
of the single-particle strength. 
The second difference is the deviation of the function 
$ Z_\mu(\omega_{G+}^{\mu a_0})$, Eq.~(\ref{calC}) in Appendix F, from 1. 
A third difference  is the energy denominator in the gap equation. 
If one puts for simplicity
$\hbar\Sigma_\mu^{12}(\omega_{G+}^{\mu a_0}) \simeq \Delta_\mu$, 
$R_{\mu a_0}^{12}(\omega_{G+}^{\mu a_0}) \simeq u_\mu v_\mu$ and 
$\omega_{G+}^{\mu a_0} \simeq {\cal E}_\mu$, 
the pairing gap in Dyson method can be written as 
\begin{eqnarray}
\Delta_\mu
&\simeq&
-\sum_{\lambda n}\sum_{\mu^\prime}
\frac{\hbar}{ 2\Omega_{\lambda n} B_{\lambda n} }
\frac{1}{ 2j_\mu + 1 }
\left| 
\langle \mu || R_0 \frac{dU}{dr} Y_\lambda ||
\mu^\prime\rangle\right|^2 
u_{\mu^\prime} v_{\mu^\prime} \nonumber\\
&&
\times
\left\{
\frac{1}{-{\cal E}_\mu - {\cal E}_{\mu^\prime} - \hbar\Omega_{\lambda n} }
+
\frac{1}{ {\cal E}_\mu - {\cal E}_{\mu^\prime} - \hbar\Omega_{\lambda n} }
\right\} \ . \label{gapDycmp}
\end{eqnarray}
On the other hand, the gap of BCS + Bloch-Horowitz calculation 
leads to 
\begin{eqnarray}
\Delta_\mu({\rm BCS+BH})
&=&
-\sum_{\lambda n}\sum_{\mu^\prime}
\frac{\hbar}{ 2\Omega_{\lambda n} B_{\lambda n} }
\frac{1}{ 2j_\mu + 1 }
\left| 
\langle \mu
|| R_0 \frac{dU}{dr} Y_\lambda ||
\mu^\prime\rangle\right|^2 
u_{\mu^\prime} v_{\mu^\prime} \nonumber\\
&&
\times
\frac{2}
{ E_{\rm cor} -
 ( |\tilde{\varepsilon}_\mu^0| + |\tilde{\varepsilon}_{\mu^\prime}^0|
  + \hbar\Omega_{\lambda n} ) }
\ , \label{gapBCSBHcmp}
\end{eqnarray}
where $E_{\rm cor}$ is the correlation energy of the perturbed ground state. 
Thus, the essential difference, in the no correlation limit, is 
the sign of ${\cal E}_\mu$ in the second energy denominator in
Eq.~(\ref{gapDycmp}). 

Eq.~(\ref{Epair}) may be compared to the pairing energy in the BCS 
approximation:
\begin{equation}
E_{\rm pair}({\rm BCS}) = 
 \sum_{\mu\mu^\prime} 
 \frac{(2j_\mu + 1)(2j_{\mu^\prime} + 1)}{4}
 u_\mu v_\mu
 u_{\mu^\prime} v_{\mu^\prime}
 V_{\mu\mu^\prime} \ ,
\label{EpairBCS}
\end{equation}
\begin{equation}
V_{\mu\mu^\prime} = 
 \langle (\mu^\prime m_{\mu^\prime})\overline{(\mu^\prime m_{\mu^\prime})} |
 V
 | (\mu m_\mu) \overline{(\mu m_\mu)} \rangle_{\rm antisymmetrized} \ . 
\end{equation}
We obtained $E_{\rm pair}$(BCS) = $-$4.6 MeV
in the BCS + Bloch-Horowitz calculation.

\section{Summary}
In this paper we have investigated the effect of particle-phonon 
coupling on nuclear pairing correlations in detail. 
The dynamical equation treated is Dyson equation including 
the anomalous Green function.
We showed the formulation from the viewpoint to clarify 
how to solve the equation in such a way that 
the effect beyond BCS is not suppressed. 
The comparison was made between the solutions of 
Dyson equation and diagonalization of the particle-phonon 
coupled Hamiltonian, and we clarified both difference and similarity. 
The difference is in the truncation scheme,
and
nonperturbative effect can be taken into account 
for a class of diagrams in Dyson method by solving 
the equation iteratively. 
The similarity of the solutions is clear
when the model does not have strong the fragmentation effect 
of the single-particle strength. 
We have solved Dyson equation for the neutrons in $^{120}$Sn and 
calculated the pairing gaps. 
The average pairing gap near the Fermi level arising 
from the phonon-induced interaction is around 40 \% of
the observed gap from the odd-even mass difference. 
According to the spectral functions the quasiparticle picture is
not perfect but acceptable 
for many of the single-particle orbits near the Fermi level 
except for 2d$_{5/2}$. 
This level is more than 2 MeV below the Fermi level, however, 
the orbit has the appreciable magnitude of the pairing gap. 
This gap arises mainly from the $3^-$ phonon coupling, of which 
the vertex matrix has only off-diagonal elements. 
For this reason the pairing gap 
as a function of the unperturbed single-particle energy
does not have a completely-smooth shape peaked at the Fermi level. 
As for the overall shape, 
the pairing gap arising from the phonon-induced interaction is
of the surface type.

We have discussed also the BCS + Bloch-Horowitz calculation. 
The mechanism to avoid divergence possibly arising from small 
energy denominator is different 
between Dyson calculation and BCS + Bloch-Horowitz method. 
The latter method has always negative, i.~e. non-zero, 
energy denominators in the pairing gap. 

Several further steps are needed, in order to obtain a complete quantitative 
assessment of the contribution of the 
particle-phonon coupling to the nuclear pairing correlations. 
%
%
First, the vertex correction of the self-energy should be investigated. 
In fact, our present calculation is partially nonperturbative and 
partially perturbative. 
Needless to say, it is a nonperturbative effect that 
the pairing gaps were obtained using the single-particle basis. 
On the other hand, the vertex was treated in the lowest perturbation.

 It will be necessary also to combine the phonon-induced interaction with
a bare NN interaction into 
the self-energy for understanding the nuclear pairing correlations 
fully microscopically, and 
to look for an adequate Hartree-Fock basis consistent with 
the effective mass arising from the particle-phonon coupling. 
One may be also concerned with  breaking of the particle-number conservation. 
Since we calculate the pairing gap in a finite-body system, 
the particle number is not conserved.

%


\vspace{2em}
%
%
%
\noindent
{\large \bf Appendix A -- diagram rule --}

The original Nambu-Gor'kov formulation assumes that the spin up or down is a good quantum 
number  \cite{Sc64}  . 
In this appendix we show in detail a derivation of the diagram rule 
without this condition. 
We derive the rule by  making an expansion of a perturbed nucleon Green function 
using products of unperturbed nucleon Green functions \cite{FW71}. 
It is noted that the nonperturbative effect of Dyson equation is 
essential for getting the anomalous Green function if unpaired states 
are used as the basis. 

We put
\begin{equation}
\Psi_\nu =
 \left(
 \begin{array}{l}
  c_\nu \\
  c_{\overline{\nu}}^\dagger
 \end{array}
 \right) \ . 
\end{equation}
In this appendix $\nu$ is a single-particle index of a complete basis,
and
$\overline{\nu}$ is a time-reversed state of $\nu$.
Note that 
$\{\nu_i\}_{i=\cdots}$ = $\{\overline\nu_i\}_{i=\cdots}$,
and we do not impose any explicit good quantum number 
on the basis as long as a general diagram rule is discussed.  
Let us consider the particle-phonon coupling Hamiltonian 
( see sections 6.2b and 6.3a in \cite{Bo75} ): 
\begin{equation}
 \hat{H}_{\rm int} = \sum_{lm} \kappa_l \hat{\alpha}_{lm} 
 \sum_{\mu\nu} 
 \langle \mu | F_{lm} ^\dagger | \nu \rangle
 \frac{1}{2} \Psi_\mu^\dagger \tau_3 \Psi_\nu \ ,
\label{h_int}
\end{equation}
where 
\begin{equation}
F_{lm} = -R_0 \frac{1}{\kappa_l}
    \frac{d U(r)}{dr}
    Y_{lm}(\theta,\varphi) \ , 
\end{equation}
\begin{equation}
\hat{\alpha}_{lm} = (\alpha_l)_0\, ( c_{lm}^\dagger + c_{\overline{lm}} ) \ ,
\end{equation}
\begin{equation}
(\alpha_l)_0 = \langle lm | F_{lm} | 0 \rangle = 
\sqrt{ \frac{\hbar}{2B_l \Omega_l} } \ , 
\end{equation}
\begin{equation}
\kappa_l = \int dr \: r^2
R_0 \frac{d \rho_0(r)}{dr} R_0 \frac{d U(r)}{dr}  \ . 
\end{equation}
$R_0$ is a mean nuclear radius, and $U(r)$ is a nuclear potential. 
$c_{lm}^\dagger$ is a creation operator of a phonon specified by 
the multipolarity $l$ and its $z$-component $m$.
We define $c_{\overline{lm}}$  = $(-)^m c_{l\,-m}$.  
$(\alpha_l)_0$  is a zero-point amplitude of the phonon, 
which is related to the inertia parameter $B_l$ and 
the phonon energy $\hbar\Omega_l$. 
$|lm\rangle$ and $|0\rangle$  are a one-phonon and zero-phonon state, respectively. 
$\rho_0(r)$ denotes a nuclear density in the equilibrium. 

With a relation 
\begin{eqnarray}
\langle \mu | F_{lm}^\dagger | \nu \rangle
&=&
\langle \overline{\nu} | F_{lm}^\dagger | \overline{\mu} \rangle \ , 
\end{eqnarray}
it follows that 
\begin{eqnarray}
\sum_{\mu\nu} \langle \mu | F_{lm}^\dagger | \nu \rangle 
\frac{1}{2}\Psi_\mu^\dagger \tau_3 \Psi_\nu
&=&
\sum_{\mu\nu} \langle \mu | F_{lm}^\dagger | \nu \rangle 
\frac{1}{2}( c_\mu^\dagger c_\nu - c_{\overline{\mu}} c_{\overline{\nu}}^\dagger ) \\
&=&
\sum_{\mu\nu} \langle \mu | F_{lm}^\dagger | \nu \rangle 
c_\mu^\dagger c_\nu + {\rm const} \ .
\end{eqnarray}
That is, since the particle-phonon coupling considered here does not 
depend on the spin, Nambu-Gor'kov formalism can be introduced without 
much extension. 

We introduce a time-dependent coupling in the interaction picture: 
\begin{equation}
 \hat{H}_{\rm int}(t) = \sum_{lm} \kappa_l \hat{\alpha}_{lm}(t) 
 \sum_{\mu\nu} 
 \langle \mu | F_{lm} ^\dagger | \nu \rangle
 \frac{1}{2} \Psi_\mu^\dagger(t) \tau_3 \Psi_\nu(t) \ ,
 \label{Hint(t)}
\end{equation}
where the time-dependent operators are defined as 
\begin{eqnarray}
&& \hat{\alpha}_{lm}(t) = 
e^{i\hat{H}_0^{\rm ph}t/\hbar}
\hat{\alpha}_{lm}
e^{-i\hat{H}_0^{\rm ph}t/\hbar} \ , \\
&& \Psi_\mu^\dagger (t) = 
e^{i\hat{H}_0t/\hbar}
\Psi_\mu^\dagger
e^{-i\hat{H}_0t/\hbar} \ . 
\end{eqnarray}
$ \hat{H}_0^{\rm ph} $ and $ \hat{H}_0 $ are unperturbed Hamiltonians of 
the phonon and single-particle relative to the Fermi level, respectively. 
The basic formula from which we start is
\begin{eqnarray}
iG(t) &=& 
\left(
\begin{array}{ll}
iG_{\nu\mu}^{11}(t)            & iG_{\nu\overline{\mu}}^{12}(t)\vspace{1ex}\\
iG_{\overline{\nu}\mu}^{21}(t) & iG_{\overline{\nu}\overline{\mu}}^{22}(t) \\
\end{array}
\right)
\nonumber \\
&=&
\sum_{m=0}^\infty \left( \frac{-i}{\hbar}\right)^m
\frac{1}{m!}
\int_{-\infty}^\infty dt_1 \cdots \int_{-\infty}^\infty dt_m \nonumber \\
&& \times
\langle \Phi_0 | 
{\rm T} \left[
\hat{H}_{\rm int}(t_1) \hat{H}_{\rm int}(t_2) \cdots \hat{H}_{\rm int}(t_m)
\Psi_\nu(t) \Psi_\mu^\dagger 
 \right]
| \Phi_0 \rangle_{\rm connected} \ . \label{basic}
\end{eqnarray}
T$[\cdots]$ is the time-ordered product, and the meaning of 
``connected" shall be clearest in the diagram later. 
$|\Phi_0\rangle$ denotes an unperturbed ground state
of the system of the single-particles and phonons. 
It is assumed 
$c_\mu |\Phi_0\rangle$ = 
$c_{lm} |\Phi_0\rangle$ = 0.  
The $m$ = 0 term is 
$\langle \Phi_0 | {\rm T} [ \Psi_\nu(t) \Psi_\mu^\dagger ] | \Phi_0 \rangle $ 
= $iG_0(t)$. 
Since the unperturbed Hamiltonian does not have the coupling, 
we have 
\begin{equation}
|\Phi_0\rangle = 
|\Phi_0^{\rm n}\rangle |\Phi_0^{\rm ph}\rangle \ , 
\end{equation}
where $|\Phi_0^{\rm n}\rangle$ and $|\Phi_0^{\rm ph}\rangle$ are 
the ground states of the single-particles and phonons, respectively. 
The derivation of the formula (\ref{basic}) is not affected by the inclusion of 
the anomalous Green function. 
(See, e.~g., \cite{FW71} for the derivation.)

 By inserting Eq.~(\ref{Hint(t)}) to Eq.~(\ref{basic}) and using 
Wick theorem, it follows that
\begin{eqnarray}
\lefteqn{
\left(
\begin{array}{ll}
iG_{\nu\mu}^{11}(t)            & iG_{\nu\overline{\mu}}^{12}(t)\vspace{1ex}\\
iG_{\overline{\nu}\mu}^{21}(t) & iG_{\overline{\nu}\,\overline{\mu}}^{22}(t) \\
\end{array}
\right)
         } \nonumber\\
= &&
\sum_{m=0}^\infty \left( \frac{-i}{\hbar}\right)^m \frac{1}{m!}
\int_{-\infty}^\infty dt_1 \cdots
\int_{-\infty}^\infty dt_m
\frac{1}{2^m} \nonumber \\
&& \times
\sum_{l_1 m_1} \kappa_{l_1}
\sum_{\nu_1\mu_1} \langle \mu_1 | F_{l_1m_1}^\dagger | \nu_1 \rangle 
\cdots
\sum_{l_m m_m} \kappa_{l_m}
\sum_{\nu_m\mu_m} \langle \mu_m | F_{l_m m_m}^\dagger | \nu_m \rangle 
\nonumber \\
&& \times
\langle \Phi_0^{\rm ph} | {\rm T}
\left[
 \hat{\alpha}_{l_1 m_1}(t_1)
 \hat{\alpha}_{l_2 m_2}(t_2)
 \cdots
 \hat{\alpha}_{l_m m_m}(t_m)
\right]
| \Phi_0^{\rm ph} \rangle 
\nonumber \\
&& \times
\langle \Phi_0^{\rm n} | {\rm T}
\left[
 \Psi_{\mu_1}^\dagger (t_1) \tau_3 \Psi_{\nu_1}(t_1)
 \Psi_{\mu_2}^\dagger (t_2) \tau_3 \Psi_{\nu_2}(t_2)
 \cdots
 \Psi_{\mu_m}^\dagger (t_m) \tau_3 \Psi_{\nu_m}(t_m)
 \Psi_\nu(t) \Psi_\mu^\dagger
\right]
| \Phi_0^{\rm n} \rangle_{\rm conn.}
\nonumber \\
= &&
\sum_{m=0}^\infty \left( \frac{-i}{\hbar}\right)^m \frac{1}{m!}
\int_{-\infty}^\infty dt_1 \cdots
\int_{-\infty}^\infty dt_m
\frac{1}{2^m}
\nonumber \\
&& \times
\sum_{l_1 m_1} \kappa_{l_1}
\sum_{\nu_1\mu_1} \langle \mu_1 | F_{l_1m_1}^\dagger | \nu_1 \rangle 
\cdots
\sum_{l_m m_m} \kappa_{l_m}
\sum_{\nu_m\mu_m} \langle \mu_m | F_{l_m m_m}^\dagger | \nu_m \rangle 
\nonumber \\
&& \times
\langle \Phi_0^{\rm ph} | 
\hat{\alpha}_{l_1m_1}(t_1)
\hspace{-2em}
               {
        \rule[-2.5ex]{0.1ex}{1.5ex}
        \hspace{-0.1ex}
        \rule[-2.5ex]{7em}{0.1ex}
        \hspace{-0.1ex}
        \rule[-2.5ex]{0.1ex}{1.5ex} 
               }
\hspace{-5em}
\hat{\alpha}_{l_2m_2}(t_2)
\hspace{-2em}
     { \rule[-1.6ex]{0.1ex}{0.7ex}
        \hspace{-0.1ex}
        \rule[-1.6ex]{6em}{0.1ex}
        \hspace{-0.1ex}
        \rule[-1.6ex]{0.1ex}{0.7ex}
      }
\hspace{-3em}
\cdots
\hspace{2em}
\hat{\alpha}_{l_m m_m}(t_m)
\hspace{-6em}
     { \rule[-2.5ex]{0.1ex}{1.5ex}
        \hspace{-0.1ex}
        \rule[-2.5ex]{4em}{0.1ex}
        \hspace{-0.1ex}
        \rule[-2.5ex]{0.1ex}{1.5ex}
      }
\nonumber \\
&& \mbox{\hspace{2ex}}+
\mbox{all other possible full contractions} | \Phi_0^{\rm ph} \rangle 
\nonumber \\
&& \times
\langle \Phi_0^{\rm n} | 
\Psi_{\mu_1}^\dagger(t_1)
\hspace{-2.5em}
     { \rule[-2.5ex]{0.1ex}{1.5ex}
        \hspace{-0.1ex}
        \rule[-2.5ex]{7.2em}{0.1ex}
        \hspace{-0.1ex}
        \rule[-2.5ex]{0.1ex}{1.5ex}
      }
\hspace{-4.5em}
\tau_3 \Psi_{\nu_1}(t_1)
\hspace{-2.5em}
     { \rule[-1.5ex]{0.1ex}{0.7ex}
        \hspace{-0.1ex}
        \rule[-1.5ex]{4em}{0.1ex}
        \hspace{-0.1ex}
        \rule[-1.5ex]{0.1ex}{0.7ex}
      }
\hspace{-0.7em}
\cdots
\hspace{1.5em}
\Psi_{\mu_m}^\dagger(t_m)
\hspace{-5em}
     { \rule[-2.0ex]{0.1ex}{0.7ex}
        \hspace{-0.1ex}
        \rule[-2.0ex]{3em}{0.1ex}
        \hspace{-0.1ex}
        \rule[-2.0ex]{0.1ex}{0.7ex}
      }
\hspace{2.3em}
\tau_3 \Psi_{\nu_m}(t_m)
\hspace{-10em}
     { \rule[-2.5ex]{0.1ex}{1.5ex}
        \hspace{-0.1ex}
        \rule[-2.5ex]{8em}{0.1ex}
        \hspace{-0.1ex}
        \rule[-2.5ex]{0.1ex}{1.5ex}
      }
\hspace{2.0em}
\Psi_\nu(t)
\hspace{-10em}
     { \rule[-3.5ex]{8.3em}{0.1ex}
       \hspace{-0.1ex}
       \rule[-3.5ex]{0.1ex}{2.5ex}
      }
\hspace{1.7em}
\Psi_\mu^\dagger
\hspace{-10em}
     { \rule[-4.0ex]{9.4em}{0.1ex}
       \hspace{-0.1ex}
       \rule[-4.0ex]{0.1ex}{2.8ex}
      }
\nonumber \\
&& \mbox{\hspace{1em}} + 
\mbox{all other possible full contractions} | \Phi_0^{\rm n} \rangle_{\rm conn}
\ .
\label{16.1}
\end{eqnarray}
Let us show an example how the contraction of the type
$\Psi_\nu
\hspace{-0.5em}
     { \rule[-1.5ex]{0.1ex}{1.0ex}
        \hspace{-0.1ex}
        \rule[-1.5ex]{3.3em}{0.1ex}
        \hspace{-0.1ex}
        \rule[-1.5ex]{0.1ex}{1.0ex}
      }
\hspace{-2.6em}
(\cdots) \,\Psi_{\nu^\prime}$
can be treated: 
\begin{eqnarray}
\lefteqn{
 \langle \Phi_0^{\rm n} | 
 \Psi_{\mu_1}^\dagger (t_1) \tau_3 \Psi_{\nu_1}(t_1) (\cdots)
 \Psi_{\mu_i}^\dagger (t_i) \tau_3 \Psi_{\nu_i}(t_i) \cdots
 | \Phi_0^{\rm n} \rangle
\hspace{-18.0em}
      { \rule[-3.0ex]{0.1ex}{2.0ex}
        \hspace{-0.1ex}
        \rule[-3.0ex]{4em}{0.1ex}
      }
 \hspace{-0.3em}
      { \rule[-2.0ex]{0.1ex}{1.0ex}
        \hspace{-0.1ex}
        \rule[-2.0ex]{9em}{0.1ex}
        \hspace{-0.1ex}
        \rule[-2.0ex]{0.1ex}{1.0ex}
      }
 \hspace{-4.0em}
      { \rule[-3.3ex]{0.1ex}{2.0ex}
        \hspace{-0.1ex}
        \rule[-3.3ex]{7em}{0.1ex}
        \hspace{-0.1ex}
        \rule[-3.3ex]{0.1ex}{2.0ex}
       }
         }
\nonumber \\
&& = 
 \langle \Phi_0^{\rm n} | 
 \Psi_{\mu_1}^\dagger (t_1) \tau_3 
\left(
\begin{array}{l}
  c_{\nu_1}(t_1) (\cdots)
  c_{\mu_i}^\dagger
  c_{\nu_i}(t_i)
- c_{\nu_1}(t_1) (\cdots)
  c_{\overline{\mu}_i}(t_i)
  c_{\overline{\nu}_i}^\dagger(t_i)
\hspace{-19em}
      { \rule[-1.7ex]{0.1ex}{0.8ex}
        \hspace{-0.1ex}
        \rule[-1.7ex]{6.0em}{0.1ex}
        \hspace{-0.1ex}
        \rule[-1.7ex]{0.1ex}{0.8ex}
      }
\hspace{-1.5em}
      { \rule[-2.3ex]{0.1ex}{1.3ex}
        \hspace{-0.1ex}
        \rule[-2.3ex]{3em}{0.1ex}
      }
\hspace{2.5em}
      { \rule[-1.7ex]{0.1ex}{0.8ex}
        \hspace{-0.1ex}
        \rule[-1.7ex]{7.0em}{0.1ex}
        \hspace{-0.1ex}
        \rule[-1.7ex]{0.1ex}{0.8ex}
      }
\hspace{-2.5em}
      { \rule[-2.3ex]{0.1ex}{1.3ex}
        \hspace{-0.1ex}
        \rule[-2.3ex]{3em}{0.1ex}
      }
\vspace{1em}
\\

  c_{\overline{\nu}_1}^\dagger(t_1) (\cdots)
  c_{\mu_i}^\dagger(t_i)
  c_{\nu_i}(t_i)
- c_{\overline{\nu}_1}^\dagger(t_1) (\cdots)
  c_{\overline{\mu}_i}(t_i)
  c_{\overline{\nu}_i}^\dagger(t_i)
\hspace{-20.5em}
      { \rule[-1.7ex]{0.1ex}{0.8ex}
        \hspace{-0.1ex}
        \rule[-1.7ex]{7.5em}{0.1ex}
        \hspace{-0.1ex}
        \rule[-1.7ex]{0.1ex}{0.8ex}
      }
\hspace{-2.5em}
      { \rule[-2.3ex]{0.1ex}{1.3ex}
        \hspace{-0.1ex}
        \rule[-2.3ex]{3em}{0.1ex}
      }
\hspace{3.3em}
      { \rule[-2.0ex]{0.1ex}{0.8ex}
        \hspace{-0.1ex}
        \rule[-2.0ex]{7.5em}{0.1ex}
        \hspace{-0.1ex}
        \rule[-2.0ex]{0.1ex}{0.8ex}
      }
\hspace{-3.0em}
      { \rule[-2.5ex]{0.1ex}{1.3ex}
        \hspace{-0.1ex}
        \rule[-2.5ex]{3em}{0.1ex}
      }
\hspace{1em}
\end{array}
\right)
\cdots
| \Phi_0^{\rm n} \rangle
\hspace{-29em}
      { \rule[-3em]{0.1ex}{5ex}
        \hspace{-0.1ex}
        \rule[-3em]{10em}{0.1ex}
      }
\hspace{10.0em}
\nonumber \\
&& = 
\langle \Phi_0^{\rm n} |
\Psi_{\mu_1}^\dagger (t_1)
\hspace{-2em}
      { \rule[-2.7em]{0.1ex}{5ex}
        \hspace{-0.1ex}
        \rule[-2.7em]{6em}{0.1ex}
      }
\hspace{-3.8em}
\tau_3
\left(
\begin{array}{ll}
c_{\nu_1}(t_1)(\cdots) c_{\overline{\nu}_i}^\dagger(t_i)
\hspace{-6.8em}
      { \rule[-1.8ex]{0.1ex}{0.8ex}
        \hspace{-0.1ex}
        \rule[-1.8ex]{5.0em}{0.1ex}
        \hspace{-0.1ex}
        \rule[-1.8ex]{0.1ex}{0.8ex}
      }
\hspace{2em}
& 
c_{\nu_1}(t_1)(\cdots) c_{\nu_i}(t_i)
\hspace{-6.8em}
      { \rule[-1.8ex]{0.1ex}{0.8ex}
        \hspace{-0.1ex}
        \rule[-1.8ex]{5.0em}{0.1ex}
        \hspace{-0.1ex}
        \rule[-1.8ex]{0.1ex}{0.8ex}
      }
\hspace{2em}
\vspace{1em}
\\
c_{\overline{\nu}_1}^\dagger(t_1)(\cdots) c_{\overline{\nu}_i}^\dagger(t_i)
\hspace{-6.8em}
      { \rule[-1.8ex]{0.1ex}{0.8ex}
        \hspace{-0.1ex}
        \rule[-1.8ex]{5.0em}{0.1ex}
        \hspace{-0.1ex}
        \rule[-1.8ex]{0.1ex}{0.8ex}
      }
\hspace{2em}
& 
c_{\overline{\nu}_1}^\dagger(t_1)(\cdots) c_{\nu_i}(t_i) 
\hspace{-6.8em}
      { \rule[-1.8ex]{0.1ex}{0.8ex}
        \hspace{-0.1ex}
        \rule[-1.8ex]{5.0em}{0.1ex}
        \hspace{-0.1ex}
        \rule[-1.8ex]{0.1ex}{0.8ex}
      }
\hspace{2em}
\end{array}
\right)
\left(
\begin{array}{l}
  c_{\overline{\mu}_i}(t_i)
\vspace{3ex}
\\
 -c_{\mu_i}^\dagger(t_i)
\end{array}
\right)
\hspace{-3em}
      { \rule[-2.5em]{0.1ex}{1.3ex}
        \hspace{-0.1ex}
        \rule[-2.5em]{3em}{0.1ex}
      }
\nonumber \\
&& \hspace{2em}
\cdots | \Phi_0^{\rm n} \rangle
\nonumber \\
&& = 
\langle \Phi_0^{\rm n} |
\Psi_{\mu_1}^\dagger(t_1)
\hspace{-2.5em}
      { \rule[-1.3em]{0.1ex}{1.5ex}
        \hspace{-0.1ex}
        \rule[-1.3em]{3em}{0.1ex}
      }
\hspace{-0.3em}
\tau_3
\Psi_{\nu_1}(t_1) (\cdots) \Psi_{\overline{\nu}_i}^\dagger(t_i)
\hspace{-7em}
      { \rule[-1em]{0.1ex}{1ex}
        \hspace{-0.1ex}
        \rule[-1em]{5em}{0.1ex}
        \hspace{-0.1ex}
        \rule[-1em]{0.1ex}{1ex}
      }
\hspace{2em}
\tau_3
\Psi_{\overline{\mu}_i}^\dagger(t_i)
\hspace{-2.0em}
      { \rule[-1.0em]{0.1ex}{0.8ex}
        \hspace{-0.1ex}
        \rule[-1.0em]{3em}{0.1ex}
      }
\hspace{-0.8em}
\cdots | \Phi_0^{\rm n} \rangle \ .
\end{eqnarray}
Here we used 
$c_{\overline{\overline{\nu}}}$ = $-c_\nu$. 
The summation indices $(\mu_i,\nu_i)$ can be replaced by 
$(\overline{\nu}_i,\overline{\mu}_i)$. 
Thus Eq.~(\ref{16.1}) reads
\begin{eqnarray}
\lefteqn{
\left(
\begin{array}{ll}
iG_{\nu\mu}^{11}(t)            & iG_{\nu\overline{\mu}}^{12}(t)\vspace{1ex}\\
iG_{\overline{\nu}\mu}^{21}(t) & iG_{\overline{\nu}\,\overline{\mu}}^{22}(t) \\
\end{array}
\right)
         } \nonumber \\
= &&
\sum_{m=0}^\infty \left( \frac{-i}{\hbar}\right)^m \frac{1}{m!}
\int_{-\infty}^\infty dt_1 \cdots
\int_{-\infty}^\infty dt_m
\frac{1}{2^m} \nonumber \\
&& \times
\sum_{l_1 m_1} \kappa_{l_1} \sum_{\nu_1\mu_1} 
( \langle \mu_1 | F_{l_1 m_1}^\dagger | \nu_1 \rangle  + 
  \langle \overline{\nu}_1 | F_{l_1 m_1}^\dagger | \overline{\mu}_1 \rangle  )
\nonumber \\
&& \times \cdots
\nonumber \\
&& \times
\sum_{l_m m_m} \kappa_{l_m} \sum_{\nu_m\mu_m} 
( \langle \mu_m | F_{l_m m_m}^\dagger | \nu_m \rangle  +
  \langle \overline{\nu}_m | F_{l_m m_m}^\dagger | \overline{\mu}_m \rangle  )
\nonumber \\
&& \times
\langle \Phi_0^{\rm ph} | [
 \hat{\alpha}_{l_1 m_1}(t_1)
\hspace{-2.5em}
      { \rule[-2.5ex]{0.1ex}{1.5ex}
        \hspace{-0.1ex}
        \rule[-2.5ex]{8.0em}{0.1ex}
        \hspace{-0.1ex}
        \rule[-2.5ex]{0.1ex}{1.5ex}
      }
\hspace{-5.5em}
 \hat{\alpha}_{l_2 m_2}(t_2)
\hspace{-3em}
      { \rule[-2.0ex]{0.1ex}{1.0ex}
        \hspace{-0.1ex}
        \rule[-2.0ex]{4em}{0.1ex}
        \hspace{-0.1ex}
        \rule[-2.0ex]{0.1ex}{1.0ex}
      }
\hspace{-0.3em}
\cdots
\hspace{1em}
 \hat{\alpha}_{l_m m_m}(t_m)
\hspace{-5.5em}
      { \rule[-2.0ex]{0.1ex}{1.0ex}
        \hspace{-0.1ex}
        \rule[-2.0ex]{2.5em}{0.1ex}
        \hspace{-0.1ex}
        \rule[-2.0ex]{0.1ex}{1.0ex}
      }
\hspace{1em}
\nonumber \\
&& \hspace{1.1ex}
+ \mbox{all other possible full contractions }
 ] | \Phi_0^{\rm ph} \rangle
\nonumber \\
&& \times
\langle \Phi_0^{\rm n} | [
 \Psi_{\mu_1}^\dagger(t_1) 
\hspace{-2.5em}
      { \rule[-2.5ex]{0.1ex}{1.5ex}
        \hspace{-0.1ex}
        \rule[-2.5ex]{8em}{0.1ex}
        \hspace{-0.1ex}
        \rule[-2.5ex]{0.1ex}{1.5ex}
      }
\hspace{-5.4em}
\tau_3
 \Psi_{\nu_1}(t_1)
\hspace{-2em}
      { \rule[-2.0ex]{0.1ex}{1.0ex}
        \hspace{-0.1ex}
        \rule[-2.0ex]{3em}{0.1ex}
        \hspace{-0.1ex}
        \rule[-2.0ex]{0.1ex}{1.0ex}
      }
\hspace{0.3em}
\cdots
\hspace{1.2em}
 \Psi_{\mu_m}^\dagger(t_m)
\hspace{-4.5em}
      { \rule[-2.0ex]{0.1ex}{1.0ex}
        \hspace{-0.1ex}
        \rule[-2.0ex]{2em}{0.1ex}
        \hspace{-0.1ex}
        \rule[-2.0ex]{0.1ex}{1.0ex}
      }
\hspace{3em}
\tau_3
\Psi_{\nu_m}(t_m)
\hspace{-3.7em}
      { \rule[-1.7ex]{2em}{0.1ex}
        \hspace{-0.1ex}
        \rule[-1.7ex]{0.1ex}{0.7ex}
      }
\hspace{1.7em}
\Psi_{\nu}(t)
\hspace{-5em}
      { \rule[-2.3ex]{4em}{0.1ex}
        \hspace{-0.1ex}
        \rule[-2.3ex]{0.1ex}{1.5ex}
      }
\hspace{1.2em}
\Psi_{\mu}^\dagger
\hspace{-3.5em}
      { \rule[-3.0ex]{3em}{0.1ex}
        \hspace{-0.1ex}
        \rule[-3.0ex]{0.1ex}{2.0ex}
      }
\nonumber \\
&& \hspace{1em} + 
\mbox{ all other possible full contractions between $\Psi$ and $\Psi^\dagger$}
 ] | \Phi_0^{\rm n} \rangle_{\rm conn}
\ .
\label{18.1}
\end{eqnarray}

 Now let us introduce a permutation $\cal P$ of the integers (1,2, $\cdots$ m,0) 
$\rightarrow$ ($p(1)$,$p(2)$, $\cdots$, $p(m)$, $p(0)$)
in order to express the contractions explicitly: 
\begin{eqnarray}
{\cal P} &=&
\left(
\begin{array}{ccccc}
 1   &  2   & \cdots &  m  &  0 \\
p(1) & p(2) & \cdots & p(m)& p(0) 
\end{array}
\right)
\nonumber \\
&=&
\left(
\begin{array}{cccc}
 X(1)  &  X(2)  & \cdots &  X(m_1) \\
 X(2)  &  X(3)  & \cdots &  X(1) 
\end{array}
\right) \nonumber \\
&& \times
\left(
\begin{array}{cccc}
 X(m_1+1)  &  X(m_1+2)  & \cdots &  X(m_2) \\
 X(m_1+2)  &  X(m_1+3)  & \cdots &  X(m_1+1) 
\end{array}
\right)
\cdots
\nonumber \\
&& \times
\left(
\begin{array}{cccc}
 X(m_{P-1}+1)  &  X(m_{P-1}+2)  & \cdots &  X(m_P) \\
 X(m_{P-1}+2)  &  X(m_{P-1}+3)  & \cdots &  X(m_{P-1}+1) 
\end{array}
\right) \ .
\end{eqnarray}
Here the decomposition of $\cal P$ into the product of
the cyclic permutations can be obtained as follows:
\begin{enumerate}
\item[i)\ \ ] Choose an arbitrary integer i $\in$ \{ 1,2, $\cdots$ m,0 \} and 
define $X(1)$ = $i$. 
\item[ii)\ ]
 Put $X(2)$ = $p(X(1))$, $X(3)$ = $p(X(2))$, $\cdots$ .
\item[iii)]
 If $p(X(m_1))$ = $X(1)$, then choose an integer $j$ which is not yet used, 
and put $X(m_1+1)$ = $j$. 
\item[iv)]
Repeat the procedure.
\end{enumerate}
This is a well-known theorem in the permutation group theory. 
One of the cyclic permutations includes 0. 
We put
\begin{equation}
 X(m_{N-1}+1) = 0 \ , 
\end{equation}
for later convenience.
By applying the permutation, the summation of all possible 
full contractions between $\Psi$ and $\Psi^\dagger$ can be written as 
\begin{eqnarray}
&&
\langle \Phi_0^{\rm n} | [
 \Psi_{\mu_1}^\dagger(t_1) 
\hspace{-2em}
      { \rule[-2.5ex]{0.1ex}{1.5ex}
        \hspace{-0.1ex}
        \rule[-2.5ex]{7.5em}{0.1ex}
        \hspace{-0.1ex}
        \rule[-2.5ex]{0.1ex}{1.5ex}
      }
\hspace{-5.5em}
\tau_3
 \Psi_{\nu_1}(t_1)
\hspace{-2em}
      { \rule[-1.7ex]{0.1ex}{1.0ex}
        \hspace{-0.1ex}
        \rule[-1.7ex]{3em}{0.1ex}
        \hspace{-0.1ex}
        \rule[-1.7ex]{0.1ex}{1.0ex}
      }
\hspace{-0.5em}
\cdots
\hspace{1em}
 \Psi_{\mu_m}^\dagger(t_m)
\hspace{-4.5em}
      { \rule[-1.9ex]{0.1ex}{1.0ex}
        \hspace{-0.1ex}
        \rule[-1.9ex]{2em}{0.1ex}
        \hspace{-0.1ex}
        \rule[-1.9ex]{0.1ex}{1.0ex}
      }
\hspace{2.6em}
\tau_3
\Psi_{\nu_m}(t_m)
\hspace{-5.5em}
      { \rule[-2.3ex]{3em}{0.1ex}
        \hspace{-0.1ex}
        \rule[-2.3ex]{0.1ex}{1.2ex}
      }
\hspace{2.6em}
\Psi_{\nu}(t)
\hspace{-5.2em}
      { \rule[-3.0ex]{4em}{0.1ex}
        \hspace{-0.1ex}
        \rule[-3.0ex]{0.1ex}{2.0ex}
      }
\hspace{1.4em}
\Psi_{\mu}^\dagger
\hspace{-5em}
      { \rule[-3.7ex]{4.5em}{0.1ex}
        \hspace{-0.1ex}
        \rule[-3.7ex]{0.1ex}{2.5ex}
      }
\nonumber \\
&& \hspace{1em} + 
\mbox{ all other possible full contractions between $\Psi$ and $\Psi^\dagger$}
 ] | \Phi_0^{\rm n} \rangle_{\rm conn}
\nonumber \\
&& = 
\sum_{ {\cal P} {\rm conn} } A_1 A_2 \cdots A_P \ ,
\end{eqnarray}
where
\begin{eqnarray}
A_{i+1} &=&
-{\rm Tr} \: \left[ \:
\tau_3
\left(
\begin{array}{ll}
 iG_0^{11}{}_{ {\textstyle \nu}  \raisebox{-0.5ex}{$\scriptstyle X(m_i+1)$} \;
               {\textstyle \mu}  \raisebox{-0.5ex}{$\scriptstyle X(m_i+2)$} } &
 iG_0^{12}{}_{ {\textstyle \nu}  \raisebox{-0.5ex}{$\scriptstyle X(m_i+1)$} \;
               {\textstyle \overline{\mu} } \raisebox{-0.5ex}{$\scriptstyle X(m_i+2)$}
              }
 \vspace{1.5ex}\\
 iG_0^{21}{}_{ {\textstyle \overline{\nu}} \raisebox{-0.5ex}{$\scriptstyle X(m_i+1)$} \;
               {\textstyle \mu}            \raisebox{-0.5ex}{$\scriptstyle X(m_i+2)$} } &
 iG_0^{22}{}_{ {\textstyle \overline{\nu}} \raisebox{-0.5ex}{$\scriptstyle X(m_i+1)$} \;
               {\textstyle \overline{\mu}} \raisebox{-0.5ex}{$\scriptstyle X(m_i+2)$} }
 \\
\end{array}
\right)_{\textstyle (t_{X(m_i+1)},t_{X(m_i+2)})}
\right.
\nonumber \\
&& \times
\tau_3
\left(
\begin{array}{ll}
 iG_0^{11}{}_{ {\textstyle \nu}  \raisebox{-0.5ex}{$\scriptstyle X(m_i+2)$} \;
               {\textstyle \mu}  \raisebox{-0.5ex}{$\scriptstyle X(m_i+3)$} } &
 iG_0^{12}{}_{ {\textstyle \nu}  \raisebox{-0.5ex}{$\scriptstyle X(m_i+2)$} \;
               {\textstyle \overline{\mu}} \raisebox{-0.5ex}{$\scriptstyle X(m_i+3)$} }
 \vspace{1.5ex}\\
 iG_0^{21}{}_{ {\textstyle \overline{\nu}} \raisebox{-0.5ex}{$\scriptstyle X(m_i+2)$} \;
               {\textstyle \mu}            \raisebox{-0.5ex}{$\scriptstyle X(m_i+3)$} } &
 iG_0^{22}{}_{ {\textstyle \overline{\nu}} \raisebox{-0.5ex}{$\scriptstyle X(m_i+2)$} \;
               {\textstyle \overline{\mu}} \raisebox{-0.5ex}{$\scriptstyle X(m_i+3)$} }
 \\
\end{array}
\right)_{\textstyle (t_{X(m_i+2)},t_{X(m_i+3)})}
\nonumber \\
&& \times \cdots
\nonumber \\
&& \times
\tau_3
\left(
\begin{array}{ll}
 iG_0^{11}{}_{ {\textstyle \nu} \raisebox{-0.5ex}{$\scriptstyle X(m_{i+1}-1)$} \;
               {\textstyle \mu} \raisebox{-0.5ex}{$\scriptstyle X(m_{i+1})$} } &
 iG_0^{12}{}_{ {\textstyle \nu} \raisebox{-0.5ex}{$\scriptstyle X(m_{i+1}-1)$} \;
               {\textstyle \overline{\mu}} \raisebox{-0.5ex}{$\scriptstyle X(m_{i+1})$} }
 \vspace{1.5ex}\\
 iG_0^{21}{}_{ {\textstyle \overline{\nu}}
                \raisebox{-0.5ex}{$\scriptstyle X(m_{i+1}-1)$} \;
               {\textstyle \mu} 
                \raisebox{-0.5ex}{$\scriptstyle X(m_{i+1})$} } &
 iG_0^{22}{}_{ {\textstyle \overline{\nu}}
                \raisebox{-0.5ex}{$\scriptstyle X(m_{i+1}-1)$} \;
               {\textstyle \overline{\mu}}
                \raisebox{-0.5ex}{$\scriptstyle X(m_{i+1})$} }
 \\
\end{array}
\right)_{\textstyle (t_{X(m_{i+1}-1)},t_{X(m_{i+1})})}
\nonumber \\
&& \times
\left.
\tau_3
\left(
\begin{array}{ll}
 iG_0^{11}{}_{ {\textstyle \nu}
                \raisebox{-0.5ex}{$\scriptstyle X(m_{i+1})$} \;
               {\textstyle \mu}
                \raisebox{-0.5ex}{$\scriptstyle X(m_i+1)$} } &
 iG_0^{12}{}_{ {\textstyle \nu}
                \raisebox{-0.5ex}{$\scriptstyle X(m_{i+1})$} \;
               {\textstyle \overline{\mu}}
                \raisebox{-0.5ex}{$\scriptstyle X(m_i+1)$} }
 \vspace{1.5ex}\\
 iG_0^{21}{}_{ {\textstyle \overline{\nu}}
                \raisebox{-0.5ex}{$\scriptstyle X(m_{i+1})$} \;
               {\textstyle \mu}
                \raisebox{-0.5ex}{$\scriptstyle X(m_i+1)$} } &
 iG_0^{22}{}_{ {\textstyle \overline{\nu}}
                \raisebox{-0.5ex}{$\scriptstyle X(m_{i+1})$} \;
               {\textstyle \overline{\mu}}
                \raisebox{-0.5ex}{$\scriptstyle X(m_i+1)$} } \\
\end{array}
\right)_{\textstyle (t_{X(m_{i+1})},t_{X(m_i+1)})}
\right] \ ,
\nonumber\\
\label{Ai+1}
\end{eqnarray}
where it is assumed that $A_{i+1}$ does not have the indices
$\nu$ $\equiv$ $\nu_0$ and $\mu$ $\equiv$ $\mu_0$. 
In the derivation we have used 
\begin{eqnarray}
\Psi_{\nu_i}(t_i) \Psi_{\mu_j}^\dagger(t_j)
\hspace{-5.0em}
      { \rule[-2.0ex]{0.1ex}{1.0ex}
        \hspace{-0.1ex}
        \rule[-2.0ex]{3em}{0.1ex}
        \hspace{-0.1ex}
        \rule[-2.0ex]{0.1ex}{1.0ex}
      }
\hspace{3em}
&=&
\left(
\begin{array}{ll}
iG_0^{11}{}_{\nu_i \mu_j} (t_i,t_j) &
iG_0^{12}{}_{\nu_i \overline{\mu}_j} (t_i,t_j)
\vspace{1.5ex}
\\
iG_0^{21}{}_{\overline{\nu}_i \mu_j} (t_i,t_j) &
iG_0^{22}{}_{\overline{\nu}_i \overline{\mu}_j} (t_i,t_j)
\end{array}
\right) \nonumber \\
&\equiv&
\left(
\begin{array}{ll}
iG_0^{11}{}_{\nu_i \mu_j}  &
iG_0^{12}{}_{\nu_i \overline{\mu}_j} 
\vspace{1.5ex}
\\
iG_0^{21}{}_{\overline{\nu}_i \mu_j} &
iG_0^{22}{}_{\overline{\nu}_i \overline{\mu}_j} 
\end{array}
\right)_{(t_i,t_j)} \ , 
\end{eqnarray}
and 
\begin{equation}
(a,b) 
\left(
\begin{array}{l}
 c \\
 d
\end{array}
\right)
=
{\rm Tr}
\left(
\begin{array}{l}
 c \\
 d
\end{array}
\right)
(a,b) \ .
\end{equation}
The product of $G_0$ including the suffices $\nu$ and $\mu$ is a 2 by 2 matrix: 
\begin{eqnarray}
A_N &=&
\left(
\begin{array}{ll}
 iG_0^{11}{}_{ {\textstyle \nu}\:
               {\textstyle \mu}
                \raisebox{-0.5ex}{$\scriptstyle X(m_{N-1}+2)$} } &
 iG_0^{12}{}_{ {\textstyle \nu}\:
               {\textstyle \overline{\mu}}
                \raisebox{-0.5ex}{$\scriptstyle X(m_{N-1}+2)$} }
 \vspace{1.5ex}\\
 iG_0^{21}{}_{ {\textstyle \overline{\nu}}\:
               {\textstyle \mu}
               \raisebox{-0.5ex}{$\scriptstyle X(m_{N-1}+2)$} } &
 iG_0^{22}{}_{ {\textstyle \overline{\nu}}\:
               {\textstyle \overline{\mu}}
               \raisebox{-0.5ex}{$\scriptstyle X(m_{N-1}+2)$} } \\
\end{array}
\right)_{\textstyle (t,t_{X(m_{N-1}+2)})}
\nonumber \\
&& \hspace{-1.5cm} \times
\tau_3
\left(
\begin{array}{ll}
 iG_0^{11}{}_{ {\textstyle \nu}
               \raisebox{-0.5ex}{$\scriptstyle X(m_{N-1}+2)$} \;
               {\textstyle \mu}
               \raisebox{-0.5ex}{$\scriptstyle X(m_{N-1}+3)$} } &
 iG_0^{12}{}_{ {\textstyle \nu}
               \raisebox{-0.5ex}{$\scriptstyle X(m_{N-1}+2)$} \;
               {\textstyle \overline{\mu}}
               \raisebox{-0.5ex}{$\scriptstyle X(m_{N-1}+3)$} }
 \vspace{1.5ex}\\
 iG_0^{21}{}_{ {\textstyle \overline{\nu}}
                \raisebox{-0.5ex}{$\scriptstyle X(m_{N-1}+2)$} \;
               {\textstyle \mu}
                \raisebox{-0.5ex}{$\scriptstyle X(m_{N-1}+3)$} } &
 iG_0^{22}{}_{ {\textstyle \overline{\nu}}
                \raisebox{-0.5ex}{$\scriptstyle X(m_{N-1}+2)$} \;
               {\textstyle \overline{\mu}}
                \raisebox{-0.5ex}{$\scriptstyle X(m_{N-1}+3)$} } \\
\end{array}
\right)_{\textstyle (t_{X(m_{N-1}+2)},t_{X(m_{N-1}+3)})}
\nonumber \\
&& \times \cdots
\nonumber \\
&& \times
\tau_3
\left(
\begin{array}{ll}
 iG_0^{11}{}_{ {\textstyle \nu}
                \raisebox{-0.5ex}{$\scriptstyle X(m_N-1)$} \;
               {\textstyle \mu}
                \raisebox{-0.5ex}{$\scriptstyle X(m_N)$} } &
 iG_0^{12}{}_{ {\textstyle \nu}
                \raisebox{-0.5ex}{$\scriptstyle X(m_N-1)$} \;
               {\textstyle \overline{\mu}}
                \raisebox{-0.5ex}{$\scriptstyle X(m_N)$} }
 \vspace{1.5ex}\\
 iG_0^{21}{}_{ {\textstyle \overline{\nu}}
                \raisebox{-0.5ex}{$\scriptstyle X(m_N-1)$} \;
               {\textstyle \mu}
                \raisebox{-0.5ex}{$\scriptstyle X(m_N)$} } &
 iG_0^{22}{}_{ {\textstyle \overline{\nu}}
                \raisebox{-0.5ex}{$\scriptstyle X(m_N-1)$} \;
               {\textstyle \overline{\mu}}
                \raisebox{-0.5ex}{$\scriptstyle X(m_N)$} } \\
\end{array}
\right)_{\textstyle (t_{X(m_N-1)},t_{X(m_N)})}
\nonumber \\
&& \times
\tau_3
\left(
\begin{array}{ll}
 iG_0^{11}{}_{ {\textstyle \nu}
                \raisebox{-0.5ex}{$\scriptstyle X(m_N)$} \;
               {\textstyle \mu}  } &
 iG_0^{12}{}_{ {\textstyle \nu}
                \raisebox{-0.5ex}{$\scriptstyle X(m_N)$} \;
               {\textstyle \overline{\mu}} }
 \vspace{1.5ex}\\
 iG_0^{21}{}_{ {\textstyle \overline{\nu}}
                \raisebox{-0.5ex}{$\scriptstyle X(m_N)$} \;
               {\textstyle \mu} } &
 iG_0^{22}{}_{ {\textstyle \overline{\nu}}
                \raisebox{-0.5ex}{$\scriptstyle X(m_N)$} \;
               {\textstyle \overline{\mu}} } \\
\end{array}
\right)_{\textstyle ( t_{X(m_N)},0 )}
\ , \label{AN}
\end{eqnarray}
with
\begin{eqnarray}
 && \nu = \nu_0 = \nu_{X(m_{N-1}+1)} \ ,  \\
 && \mu = \mu_0 = \mu_{X(m_{N-1}+1)} \ . 
\end{eqnarray}

 For $m$ in Eq.~(\ref{18.1}) there are $m!$ identical terms which are 
obtained to each other by an exchange of the set of the summation 
indices and an integral variable 
($\mu_i$, 
 $\nu_i$, 
 $l_i$, 
 $m_i$, 
 $t_i$)
$\leftrightarrow$
($\mu_j$, 
 $\nu_j$, 
 $l_j$, 
 $m_j$, 
 $t_j$).
They are called topologically equivalent, otherwise the terms are called 
topologically distinct. 
One of the topologically equivalent terms has always
\begin{eqnarray}
\lefteqn{
\hat{\alpha}_{l_1 m_1}(t_1) \,\hat{\alpha}_{l_2 m_2}(t_2)
\hspace{-6.7em}
      { \rule[-2.0ex]{0.1ex}{1.0ex}
        \hspace{-0.1ex}
        \rule[-2.0ex]{4.0em}{0.1ex}
        \hspace{-0.1ex}
        \rule[-2.0ex]{0.1ex}{1.0ex}
      }
\hspace{2.5em}
\;\hat{\alpha}_{l_3 m_3}(t_3) \,\hat{\alpha}_{l_4 m_4}(t_4)
\hspace{-6.7em}
      { \rule[-2.0ex]{0.1ex}{1.0ex}
        \hspace{-0.1ex}
        \rule[-2.0ex]{4.0em}{0.1ex}
        \hspace{-0.1ex}
        \rule[-2.0ex]{0.1ex}{1.0ex}
      }
\hspace{3.3em}
\cdots
\hat{\alpha}_{l_{m-1} m_{m-1}}(t_{m-1}) \,\hat{\alpha}_{l_m m_m}(t_m)
\hspace{-9.5em}
      { \rule[-2.0ex]{0.1ex}{1.0ex}
        \hspace{-0.1ex}
        \rule[-2.0ex]{6.5em}{0.1ex}
        \hspace{-0.1ex}
        \rule[-2.0ex]{0.1ex}{1.0ex}
      }
\hspace{3em}
        }
\nonumber \\
&&= 
(\alpha_{l_1})_0\; (\alpha_{l_2})_0\;
 iD_{l_1 m_1, l_2 m_2}^0(t_1,t_2)
\;(\alpha_{l_3})_0\; (\alpha_{l_4})_0\;
iD_{l_3 m_3, l_4 m_4}^0(t_3,t_4)
\cdots \nonumber \\
&& \mbox{\hspace{1.5ex}}
\times
(\alpha_{l_{m-1}})_0\; (\alpha_{l_m})_0\;
iD_{l_{m-1} m_{m-1}, l_m m_m}^0(t_{m-1},t_m) \ ,
\end{eqnarray}
where
$iD_{l_1 m_1, l_2 m_2}^0(t_1,t_2)$ is the unperturbed phonon Green function. 
Thus Eq.~(\ref{18.1}) becomes
\begin{eqnarray}
\lefteqn{
\left(
\begin{array}{ll}
iG_{\nu\mu}^{11}(t)            & iG_{\nu\overline{\mu}}^{12}(t)\vspace{1ex}\\
iG_{\overline{\nu}\mu}^{21}(t) & iG_{\overline{\nu}\,\overline{\mu}}^{22}(t) \\
\end{array}
\right)
         } \nonumber \\
= &&
\sum_{m=0,2,\cdots}^\infty \left( \frac{-i}{\hbar}\right)^m
\int_{-\infty}^\infty dt_1 \cdots
\int_{-\infty}^\infty dt_m
\nonumber \\
&& \times
\sum_{l_1 m_1} (\alpha_{l_1})_0\; \kappa_{l_1}
\sum_{\nu_1\mu_1}
\langle \mu_1 | F_{l_1 m_1}^\dagger | \nu_1 \rangle
\cdots
\sum_{l_m m_m} (\alpha_{l_m})_0\; \kappa_{l_m}
    \sum_{\nu_m\mu_m} 
\langle \mu_m | F_{l_m m_m}^\dagger | \nu_m \rangle
\nonumber \\
&& \times 
\hspace{-3ex}
\sum_{
\begin{array}{c}
{\cal P} {\scriptstyle\rm  conn} \vspace{-1ex}\\
{\scriptstyle\rm topologically \ distinct}
\end{array}
       }     
\hspace{-3ex}
iD_{l_1m_1,l_2m_2}^0(t_1,t_2) \,
iD_{l_3m_3,l_4m_4}^0(t_3,t_4) 
\cdots
iD_{l_{m-1}m_{m-1},l_mm_m}^0(t_{m-1},t_m) 
\nonumber \\
&& \times
A_1 A_2 \cdots A_P \ .
\end{eqnarray}

 One may put the unperturbed Green functions to be diagonal: 
\begin{eqnarray}
&&
iG_0^{11}{}_{\nu\mu}(t,t^\prime) =
iG_0^{11}{}_\nu(t,t^\prime) \delta_{\mu\nu} \ , \\
&&
iG_0^{22}{}_{\overline{\nu}\,\overline{\mu}}(t,t^\prime) =
iG_0^{22}{}_{\overline{\nu}}(t,t^\prime) \delta_{\mu\nu} \ , \\
&&
iG_0^{12}{}_{\nu\,\overline{\mu}}(t,t^\prime) =
iG_0^{12}{}_{\nu\,\overline{\nu}}(t,t^\prime) \delta_{\mu\nu} \ , \\
&&
iG_0^{21}{}_{\overline{\nu}\,\mu}(t,t^\prime) =
iG_0^{21}{}_{\overline{\nu}\,\nu}(t,t^\prime) \delta_{\mu\nu} \ ,\\
&&
iD_{l_im_i, l_{i+1}m_{i+1}}^0(t_i,t_{i+1}) =
iD_{l_im_i}^0(t_i,t_{i+1}) \delta_{l_il_{i+1}} \delta_{\,m_i\; -m_{i+1}} \ . 
\end{eqnarray}
That is, we can put
\begin{equation}
 \mu_{X(i+1)} = \nu_{X(i)} \ , 
\end{equation}
in Eqs.~(\ref{Ai+1}) and (\ref{AN}). 
Consequently the final form of the perturbed Green function in the 
time representation is given by 
\begin{eqnarray}
\lefteqn{
\left(
\begin{array}{ll}
iG_{\nu\mu}^{11}(t)            & iG_{\nu\overline{\mu}}^{12}(t)\vspace{1.1ex}\\
iG_{\overline{\nu}\mu}^{21}(t) & iG_{\overline{\nu}\,\overline{\mu}}^{22}(t) \\
\end{array}
\right)
         } \nonumber \\
= &&
\sum_{m=0,2,\cdots}^\infty \left( \frac{-i}{\hbar}\right)^m
\int_{-\infty}^\infty dt_1 \cdots
\int_{-\infty}^\infty dt_m
\nonumber \\
&& \times
\sum_{l_1 m_1}
\sum_{l_2 m_2}
\cdots
\sum_{l_{m-1} m_{m-1}}
\sum_{\nu_1\nu_2\cdots\nu_m}\hspace{-1.5ex}\raisebox{0.7ex}{$^\prime$}
\nonumber \\
&& \times
\;(\alpha_{l_1})_0\; \kappa_{l_1}
\langle \nu_{\overline{p}(1)} | F_{l_1 m_1}^\dagger | \nu_1 \rangle
\;(\alpha_{l_1})_0\; \kappa_{l_1}
\langle \nu_{\overline{p}(2)} | F_{l_1\, -m_1}^\dagger | \nu_2 \rangle
\nonumber \\
&& \times
\;(\alpha_{l_3})_0\; \kappa_{l_3}
\langle \nu_{\overline{p}(3)} | F_{l_3 m_3}^\dagger | \nu_3 \rangle
\;(\alpha_{l_3})_0\; \kappa_{l_3}
\langle \nu_{\overline{p}(4)} | F_{l_3\, -m_3}^\dagger | \nu_4 \rangle
\nonumber \\
&& \times \cdots
\nonumber \\
&& \times
\;(\alpha_{l_{m-1}})_0\; \kappa_{l_{m-1}}
\langle \nu_{\overline{p}(m-1)} | F_{l_{m-1} m_{m-1}}^\dagger | \nu_{m-1} \rangle
\;(\alpha_{l_{m-1}})_0\; \kappa_{l_{m-1}}
\langle \nu_{\overline{p}(m)} | F_{l_{m-1}\, -m_{m-1}}^\dagger | \nu_m \rangle
\nonumber \\
&& \times
\hspace{-3ex}
\sum_{
\begin{array}{c}
{\cal P} {\scriptstyle\rm  conn} \vspace{-5pt}\\
{\scriptstyle\rm topologically \ distinct}
\end{array}
       }     
\hspace{-4.0ex}
iD_{l_1m_1}^0(t_1,t_2) \,
iD_{l_3m_3}^0(t_3,t_4) 
\cdots
iD_{l_{m-1}m_{m-1}}^0(t_{m-1},t_m) 
\nonumber \\
&& \times
A_1 A_2 \cdots A_P \ ,
\label{30}
\end{eqnarray}
where $\langle \nu_{\overline{p}(X(m_{N-1}+1))}|$ = $ \langle \nu_{X(m_N)} | $
is replaced by $\langle \mu|$, and 
$| \nu_{X(m_{N-1}+1)}\rangle$ = $| \nu_0 \rangle $ 
is replaced by $|\nu\rangle$ in the 
matrix elements of $F_{lm}^\dagger$. 
$\sum_{\nu_1\nu_2\cdots\nu_m}^{\prime}$ does not contain 
$\nu_{X(m_N)}$. 
$\overline{p}(i)$ denotes $p^{-1}(i)$. 
$A_{i}$ takes the form 
\begin{eqnarray}
\lefteqn{ A_{i\neq N} } \nonumber \\
&=&
-{\rm Tr} \: \left[ \:
\tau_3
\left(
\begin{array}{ll}
 iG_0^{11}{}_{ {\textstyle \nu}
                \raisebox{-0.5ex}{$\scriptstyle X(m_{i-1}+1)$} } &
 iG_0^{12}{}_{ {\textstyle \nu}
                \raisebox{-0.5ex}{$\scriptstyle X(m_{i-1}+1)$} \;
               {\textstyle \overline{\nu}}
                \raisebox{-0.5ex}{$\scriptstyle X(m_{i-1}+1)$} }
 \vspace{1.5ex}\\
 iG_0^{21}{}_{ {\textstyle \overline{\nu}}
                \raisebox{-0.5ex}{$\scriptstyle X(m_{i-1}+1)$} \;
               {\textstyle \nu}
                \raisebox{-0.5ex}{$\scriptstyle X(m_{i-1}+1)$} } &
 iG_0^{22}{}_{ {\textstyle \overline{\nu}}
                \raisebox{-0.5ex}{$\scriptstyle X(m_{i-1}+1)$} } \\
\end{array}
\right)_{\textstyle (t_{X(m_{i-1}+1)},t_{X(m_{i-1}+2)})}
\right.
\nonumber \\
&& \times
\tau_3
\left(
\begin{array}{ll}
 iG_0^{11}{}_{ {\textstyle \nu}
                \raisebox{-0.5ex}{$\scriptstyle X(m_{i-1}+2)$} } &
 iG_0^{12}{}_{ {\textstyle \nu}
                \raisebox{-0.5ex}{$\scriptstyle X(m_{i-1}+2)$} \;
               {\textstyle \overline{\nu}}
                \raisebox{-0.5ex}{$\scriptstyle X(m_{i-1}+2)$} }
 \vspace{1.5ex}\\
 iG_0^{21}{}_{ {\textstyle \overline{\nu}}
                \raisebox{-0.5ex}{$\scriptstyle X(m_{i-1}+2)$} \;
               {\textstyle \nu}
                \raisebox{-0.5ex}{$\scriptstyle X(m_{i-1}+2)$} } &
 iG_0^{22}{}_{ {\textstyle \overline{\nu}}
                \raisebox{-0.5ex}{$\scriptstyle X(m_{i-1}+2)$} } \\
\end{array}
\right)_{\textstyle (t_{X(m_{i-1}+2)},t_{X(m_{i-1}+3)})}
\nonumber \\
&& \times \cdots
\nonumber \\
&& \times
\left.
\tau_3
\left(
\begin{array}{ll}
 iG_0^{11}{}_{ {\textstyle \nu}
                \raisebox{-0.5ex}{$\scriptstyle X(m_i)$} } &
 iG_0^{12}{}_{ {\textstyle \nu}
                \raisebox{-0.5ex}{$\scriptstyle X(m_i)$} \;
               {\textstyle \overline{\nu}}
                \raisebox{-0.5ex}{$\scriptstyle X(m_i)$} }
 \vspace{1.5ex}\\
 iG_0^{21}{}_{ {\textstyle \overline{\nu}}
                \raisebox{-0.5ex}{$\scriptstyle X(m_i)$} \;
               {\textstyle \nu}
                \raisebox{-0.5ex}{$\scriptstyle X(m_i)$} } &
 iG_0^{22}{}_{ {\textstyle \overline{\nu}}
                \raisebox{-0.5ex}{$\scriptstyle X(m_i)$} } \\
\end{array}
\right)_{\textstyle (t_{X(m_i)},t_{X(m_{i-1}+1)})}
\right] \ ,
\nonumber \\
\end{eqnarray}
\begin{eqnarray}
A_N &=&
\left(
\begin{array}{ll}
 iG_0^{11}{}_{\textstyle \nu} &
 iG_0^{12}{}_{ {\textstyle \nu} \: {\textstyle \overline{\nu}} }
 \vspace{1.5ex}\\
 iG_0^{21}{}_{ {\textstyle \overline{\nu}} \: {\textstyle \nu} } &
 iG_0^{22}{}_{ {\textstyle \overline{\nu}} } \\
\end{array}
\right)_{\textstyle (t,t_{X(m_{N-1}+2)})}
\nonumber \\
&& \hspace{-1.5cm} \times
\tau_3
\left(
\begin{array}{ll}
 iG_0^{11}{}_{ {\textstyle \nu}
                \raisebox{-0.5ex}{$\scriptstyle X(m_{N-1}+2)$} } &
 iG_0^{12}{}_{ {\textstyle \nu}
                \raisebox{-0.5ex}{$\scriptstyle X(m_{N-1}+2)$} \;
               {\textstyle \overline{\nu}}
                \raisebox{-0.5ex}{$\scriptstyle X(m_{N-1}+2)$} }
 \vspace{1.5ex}\\
 iG_0^{21}{}_{ {\textstyle \overline{\nu}}
                \raisebox{-0.5ex}{$\scriptstyle X(m_{N-1}+2)$} \;
               {\textstyle \nu}
                \raisebox{-0.5ex}{$\scriptstyle X(m_{N-1}+2)$} } &
 iG_0^{22}{}_{ {\textstyle \overline{\nu}}
                \raisebox{-0.5ex}{$\scriptstyle X(m_{N-1}+2)$} } \\
\end{array}
\right)_{\textstyle (t_{X(m_{N-1}+2)},t_{X(m_{N-1}+3)})}
\nonumber \\
&& \times \cdots
\nonumber \\
&& \times
\tau_3
\left(
\begin{array}{ll}
 iG_0^{11}{}_{ {\textstyle \nu}
                \raisebox{-0.5ex}{$\scriptstyle X(m_N-1)$} } &
 iG_0^{12}{}_{ {\textstyle \nu}
                \raisebox{-0.5ex}{$\scriptstyle X(m_N-1)$} \;
               {\textstyle \overline{\nu}}
                \raisebox{-0.5ex}{$\scriptstyle X(m_N-1)$} }
 \vspace{1.5ex}\\
 iG_0^{21}{}_{ {\textstyle \overline{\nu}}
                \raisebox{-0.5ex}{$\scriptstyle X(m_N-1)$} \;
               {\textstyle \nu}
                \raisebox{-0.5ex}{$\scriptstyle X(m_N-1)$} } &
 iG_0^{22}{}_{ {\textstyle \overline{\nu}}
                \raisebox{-0.5ex}{$\scriptstyle X(m_N-1)$} } \\
\end{array}
\right)_{\textstyle (t_{X(m_N-1)},t_{X(m_N)})}
\nonumber \\
&& \times
\tau_3
\left(
\begin{array}{ll}
 iG_0^{11}{}_{\textstyle \mu} &
 iG_0^{12}{}_{ {\textstyle \mu} \; {\textstyle \overline{\mu}} }
 \vspace{1.5ex}\\
 iG_0^{21}{}_{ {\textstyle \overline{\mu}} \; {\textstyle \mu} } &
 iG_0^{22}{}_{ {\textstyle \overline{\mu}} } \\
\end{array}
\right)_{\textstyle ( t_{X(m_N)},0 )}
\ .  \nonumber \\
\end{eqnarray}
The perturbed Green function in Lehmann representation is obtained 
\begin{eqnarray}
\lefteqn{
\frac{1}{\hbar}
\left(
\begin{array}{ll}
  iG_{\nu\mu}^{11}(\omega)
& iG_{\nu\overline{\mu}}^{12}(\omega)\vspace{1.5ex}\\
  iG_{\overline{\nu}\mu}^{21}(\omega)
& iG_{\overline{\nu}\,\overline{\mu}}^{22}(\omega) \\
\end{array}
\right)
         } \nonumber \\
&=&
\lim_{\delta \rightarrow +0}
\frac{1}{\hbar}
\int_{-\infty}^\infty dt \: e^{i(\omega + i(2\theta(t)-1)\delta) t/\hbar}
\left(
\begin{array}{ll}
  iG_{\nu\mu}^{11}(t)
& iG_{\nu\overline{\mu}}^{12}(t)\vspace{1.5ex}\\
  iG_{\overline{\nu}\mu}^{21}(t)
& iG_{\overline{\nu}\,\overline{\mu}}^{22}(t) \\
\end{array}
\right)
          \nonumber \\
&=&
\sum_{m=0,2,\cdots}^\infty
\left( \frac{-i}{\hbar} \right)^m
\hspace{-3ex}
\sum_{
\begin{array}{c}
{\cal P} {\scriptstyle\rm  conn} \vspace{-1ex}\\
{\scriptstyle\rm topologically \ distinct}
\end{array}
       }     
\hspace{-3.5ex}
\int_{-\infty}^\infty \frac{dw_1}{2\pi}
\int_{-\infty}^\infty \frac{dw_2}{2\pi}
\cdots
\int_{-\infty}^\infty \frac{dw_m}{2\pi}
\nonumber \\
&& \times
\sum_{l_1m_1}
\sum_{l_3m_3}
\cdots
\sum_{l_{m-1}m_{m-1}}
\sum_{1,2,\cdots,m}{}\hspace{-1.7ex}\raisebox{0.7ex}{$^\prime$}
\nonumber \\
&& \times
\;(\alpha_{l_1})_0\; \kappa_{l_1}
\langle \overline{p}(1) | F_{l_1m_1}^\dagger | 1 \rangle
\;(\alpha_{l_1})_0\; \kappa_{l_1}
\langle \overline{p}(2) | F_{l_1\,-m_1}^\dagger | 2 \rangle
\nonumber \\
&& \times
\;(\alpha_{l_3})_0\; \kappa_{l_3}
\langle \overline{p}(3) | F_{l_3m_3}^\dagger | 3 \rangle
\;(\alpha_{l_3})_0\; \kappa_{l_3}
\langle \overline{p}(4) | F_{l_3\,-m_3}^\dagger | 4 \rangle
\;\cdots
\nonumber \\
&& \times
\;(\alpha_{l_{m-1}})_0\; \kappa_{l_{m-1}}
\langle \overline{p}(m-1) | F_{l_{m-1}m_{m-1}}^\dagger | m-1 \rangle
\;(\alpha_{l_{m-1}})_0\; \kappa_{l_{m-1}}
\langle \overline{p}(m)   | F_{l_{m-1}\,-m_{m-1}}^\dagger | m   \rangle
\nonumber \\ 
&& \times
\frac{i}{\hbar} D_{l_1m_1}^0(-w_1+w_{\overline{p}(1)})
\frac{i}{\hbar} D_{l_3m_3}^0(-w_3+w_{\overline{p}(3)})
\cdots
\frac{i}{\hbar} D_{l_{m-1}m_{m-1}}^0(-w_{m-1}+w_{\overline{p}(m-1)})
\nonumber \\
&& \times
{\cal A}_1 {\cal A}_2 \cdots {\cal A}_P
\nonumber \\
&& \times
  2\pi\hbar^2\, \delta(w_1+w_2-w_{\overline{p}(1)}-w_{\overline{p}(2)})
\,2\pi\hbar^2\, \delta(w_3+w_4-w_{\overline{p}(3)}-w_{\overline{p}(4)})
\cdots
\nonumber \\
&& \times
2\pi\hbar^2\, \delta(w_{m-1}+w_m-w_{\overline{p}(m-1)}-w_{\overline{p}(m)}) \ .
\label{LehmannG}
\end{eqnarray}
We have used an abbreviation $|i\rangle$ = $|\nu_i\rangle$. 
As in the time-representation (\ref{30}), 
$\langle \overline{p}(X(m_{N-1}+1))|$ = $\langle X_{m_N} | $
is replaced by $\langle \mu|$,
and 
$|X(m_{N-1}+1)\rangle$ = $| 0 \rangle$ 
is replaced by $|\nu\rangle$ in 
the matrix elements of $F_{lm}^\dagger$. 
$\sum_{1,2,\cdots,m}^{\prime}$ does not contain $X(m_N)$. 
Note $w_0$ $=$ $\omega$. 
$\cal A$ can be written 
\begin{eqnarray}
{\cal A}_{i\neq N} &=& 
-{\rm Tr}\: \mbox{\Large[}\:
\tau_3
\frac{i}{\hbar} G_{X(m_{i-1}+1)}^0(w_{X(m_{i-1}+1)})
\tau_3
\frac{i}{\hbar} G_{X(m_{i-1}+2)}^0(w_{X(m_{i-1}+2)})
\nonumber \\
&& \times
\cdots
\tau_3
\frac{i}{\hbar} G_{X(m_i)}^0(w_{X(m_i)})
\:\mbox{\Large ]}
\ , 
\label{calAi}
\end{eqnarray}
where
\begin{equation}
G_i^0(w)
\equiv
\left(
\begin{array}{ll}
G_0^{11}{}_{\nu_i}(w)                 &
G_0^{12}{}_{\nu_i\overline{\nu}_i}(w) \vspace{1.5ex}\\
G_0^{21}{}_{\overline{\nu}_i\nu_i}(w) &
G_0^{22}{}_{\overline{\nu}_i}(w) \\
\end{array}
\right)
\ .
\end{equation}
For $i$ = $N$, 
\begin{eqnarray}
{\cal A}_{N}
&=& 
\frac{i}{\hbar} G_0^0(w_0)
\tau_3
\frac{i}{\hbar} G_{X(m_{N-1}+2)}^0(w_{X(m_{N-1}+2)})
\nonumber \\
&& \times
\cdots
\tau_3
\frac{i}{\hbar} G_{X(m_N)}^0(w_{X(m_N)})
\ , 
\label{calAN}
\end{eqnarray}
where 
\begin{eqnarray}
&& G_{X(m_{N-1}+1)}^0(w_{X(m_{N-1}+1)}) = G_0^0(w_0)  \equiv G_\nu^0(\omega) \ , \\ 
&& G_{X(m_N)}^0(w_{X(m_N)}) \equiv G_\mu^0(\omega) \ . 
\label{44}
\end{eqnarray}
If $N$ = 1, one may put $m_{N-1}$ = 0. 
Now the diagram rule for $(1/\hbar)G_\mu(\omega)$ is clear
from Eqs.~(\ref{LehmannG}) -- (\ref{44}): 
\begin{enumerate}
\item[i)\ \ ]
Draw connected topologically-distinct diagrams with 
the unperturbed nucleon Green functions, phonon Green functions and 
the interaction points, and
assign 
$(1/\hbar)G_\mu^0(w)$,
$(i/\hbar)D_{lm}^0(w)$, 
$ (\alpha_l)_0\,\kappa_l \langle\nu|F_{lm}^\dagger|\mu\rangle$ 
for emitting a phonon with ($lm$) and
$ (\alpha_l)_0\,\kappa_l \langle\nu|F_{l\,-m}^\dagger|\mu\rangle$ 
for absorbing a phonon ($lm$), respectively, 
with adequate suffices $\mu_1$, $l_1m_1$ etc. 
The definition of ``connected'' is that any part of the diagram is connected to 
the other part by either nucleon or phonon Green function. 
Assign intermediate energies in such a way that 
the interaction points conserve the energy. 
\item[ii)\ ]
Make products of the nucleon Green functions in the order indicated by 
the diagram 
putting $\tau_3$ at the connecting points of the nucleon Green functions.
\item[iii)]
If the product is a closed loop, take $-$Tr of the product. 
\item[iv)]
Take summations with respect to the intermediate single-particle states, 
$lm$ and the integral 
$\int_{-\infty}^\infty d\omega_i/2\pi$
with respect to the intermediate energies $\omega_i$'s. 
\end{enumerate}
ii) and iii) are the points different from the usual rule without 
the anomalous Green function. 

 A simple example of diagram is shown in Fig.~A-1.

\begin{figure}
\begin{center}

\epsfig{file=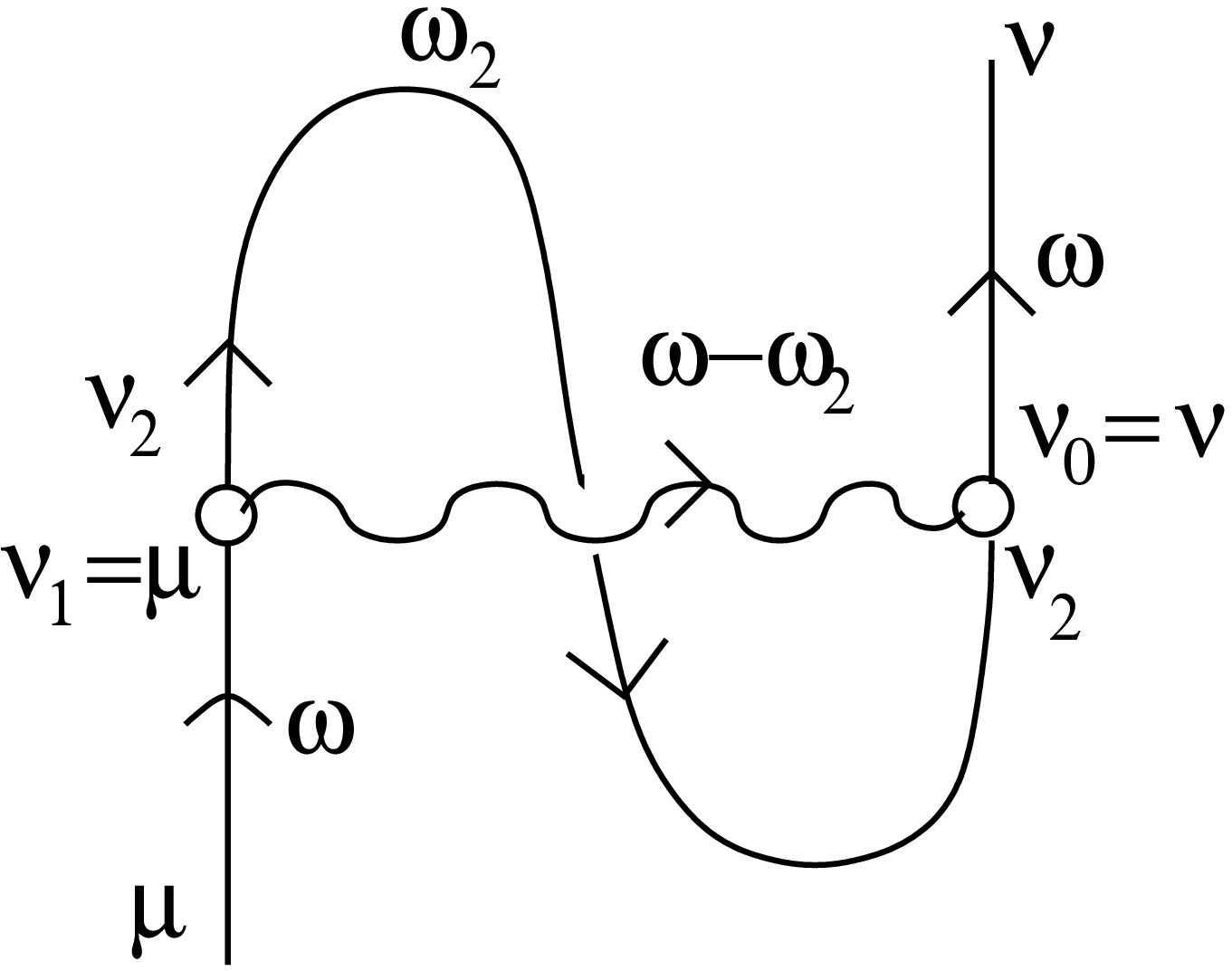,height=4.0cm} 

\vspace{2ex}

\parbox{13cm}{
      \baselineskip=1.1ex
      \small

Figure A-1.
An example of diagram. 
 }

\end{center}
\end{figure}

Using the diagram rule i) -- iv), 
one can obtain the equation
\begin{eqnarray}
\frac{1}{\hbar} G_{\nu\mu}(\omega)
&=& \int \frac{d\omega_2}{2\pi} \sum_{lm} \sum_{\nu_2}
\frac{1}{\hbar} G_\nu    ^0(\omega  ) \tau_3 
\frac{1}{\hbar} G_{\nu_2}^0(\omega_2) \tau_3 
\frac{1}{\hbar} G_{\mu  }^0(\omega  )
\nonumber\\
&& \times
(\alpha_l)_0\, \kappa_l \langle \nu_2 | F_{lm}^\dagger | \mu   \rangle 
(\alpha_l)_0\, \kappa_l \langle \nu   | F_{l\,-m}^\dagger | \nu_2 \rangle 
\nonumber\\
&& \times
\frac{i}{\hbar} D_{lm}^0 (\omega-\omega_2) \ , 
\end{eqnarray}
The self-energy is defined by removing the entering and exiting nucleon 
Green function from the diagram. 

 We have two comments on the above derivation. 
The final result of the diagram rule is identical to that 
of the original formulation \cite{Sc64}. 
The difference is that we have 
$\Psi_\mu (\cdots) \Psi_\nu
     { 
\hspace{-8.6ex}
        \rule[-1.5ex]{0.1ex}{0.7ex}
        \hspace{-0.1ex}
        \rule[-1.5ex]{3.5em}{0.1ex}
        \hspace{-0.1ex}
        \rule[-1.5ex]{0.1ex}{0.7ex}
\hspace{0.5ex}
      }
$
$\neq$ 0. 
If an extension of the Nambu-Gor'kov formulation is made for a two-body 
force dependent on the spin,
derivation of the diagram rule would be more complicated. 
Second, the diagram rule derived here is mathematically equivalent to 
the ways proposed in \cite{Mi67,Be65}. 
Migdal et al.~do not use the 2$\times$2 matrix form. 
\par

%
%

\vspace{2em}

\noindent
{\large \bf Appendix B -- spherically symmetric case --}

 Let us consider the self-energy given by Fig.~B-1 under the spherical 
symmetry of the system. 

\begin{figure}
\begin{center}

\epsfig{file=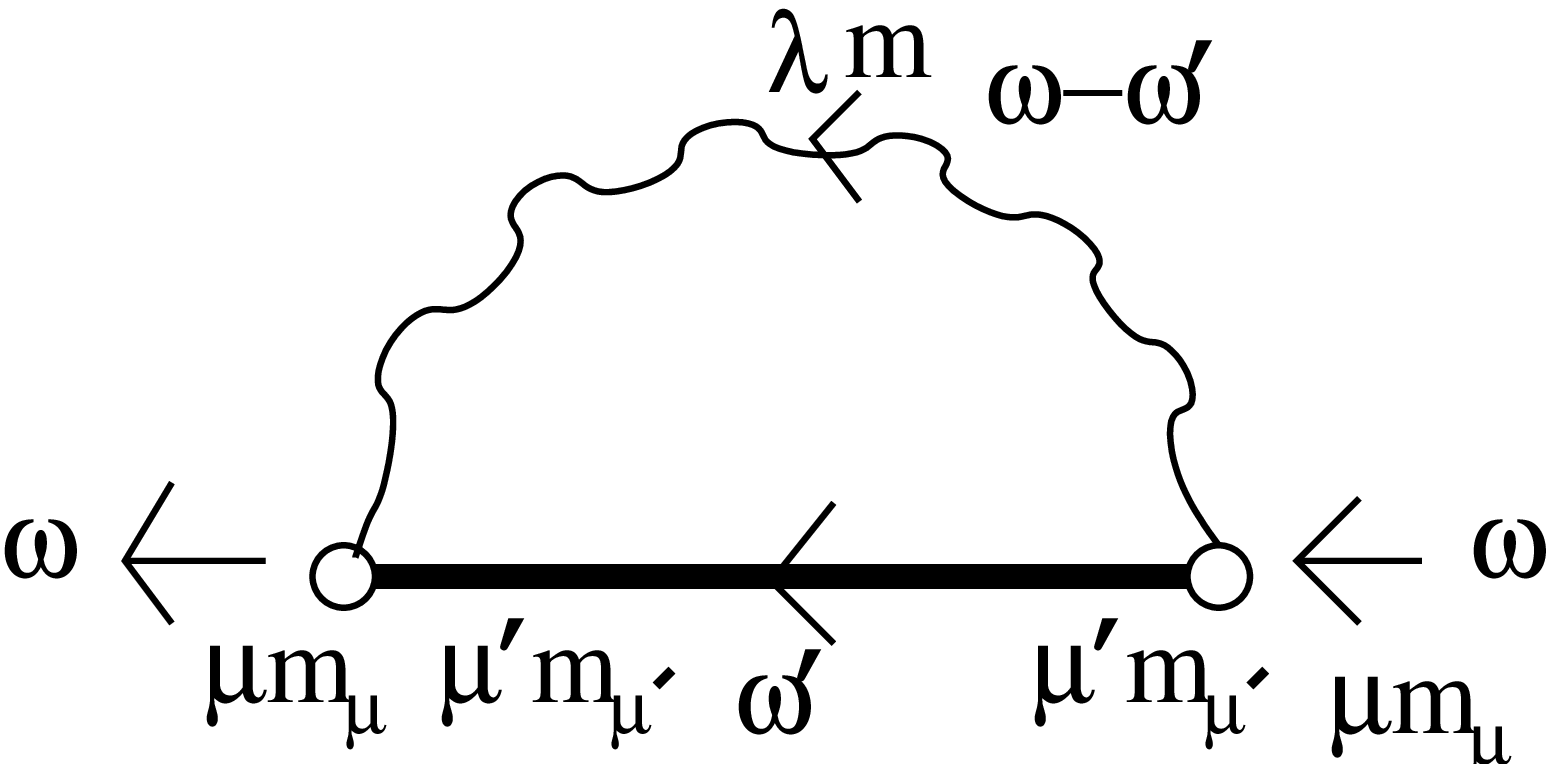,height=3.0cm} 

\vspace{2ex}

\parbox{13cm}{
      \baselineskip=1.1ex
      \small

Figure B-1.
A self-energy.
 }

\end{center}
\end{figure}

Using the diagram rule derived in Appendix A, 
we can write 
\begin{eqnarray}
\hbar \Sigma_{\mu}(\omega)  
&=& 
\int_{-\infty}^\infty \frac{d\omega^\prime}{2\pi}
\sum_{\lambda m} \sum_{ \mu^\prime m_{\mu^\prime} }
\tau_3
\frac{1}{\hbar}
G_{\mu^\prime}(\omega^\prime)
\tau_3
\nonumber \\
&&\times
(\alpha_\lambda)_0\, \kappa_\lambda
\langle \mu^\prime m_{\mu^\prime} |
 F_{\lambda m}^\dagger
| \mu m_\mu \rangle 
(\alpha_\lambda)_0\, \kappa_\lambda
\langle \mu m_\mu |
 F_{\lambda -m}^\dagger
| \mu^\prime m_{\mu^\prime} \rangle 
\frac{i}{\hbar} D_{\lambda m}^0 (\omega - \omega^\prime) \ .
\nonumber \\
\label{slf-e.1}
\end{eqnarray}
For the notations used, see Appendix A. 
Note that $\mu$ in Appendix A corresponds to $(\mu m_\mu)$ = $(nljm)_\mu$ here. 
$m$ is the $z$-component of the multipolarity $\lambda$, 
which is assumed to have only one phonon mode in this appendix
for simplicity. 
The vertex matrix element can be written 
%
%
\begin{equation}
(\alpha_\lambda)_0\, \kappa_\lambda
\langle \mu m_\mu| F_{\lambda m} | \mu^\prime m_{\mu^\prime}\rangle
=
-\sqrt{\frac{\hbar}{2\Omega_\lambda B_\lambda}}
(-)^{j_\mu - m_\mu}
\left(
\begin{array}{ccc}
j_\mu & \lambda & j_{\mu^\prime} \\
-m_\mu & m      & m_{\mu^\prime}
\end{array}
\right)
\langle \mu || R_0 \frac{dU(r)}{dr} Y_{\lambda} || \mu^\prime \rangle 
\ , \label{vertex-aI}
\end{equation}
It is possible to make the vertex matrix elements real. 

For the unperturbed phonon Green function, we have 
%
%
\begin{eqnarray}
i D_{\lambda m}^0(t)
&\stackrel{d}{=}& 
\langle \Phi_{\rm ph} | {\rm T}[ 
 ( c_{\lambda m}^\dagger(t) + c_{\overline{\lambda m}}(t) )
 ( c_{\lambda\,-m}^\dagger + c_{\overline{\lambda\,-m}} )
 ] | \Phi_{\rm ph} \rangle  \\
&=&
\left\{
\begin{array}{cc}
(-)^m\, e^{-i\Omega_\lambda t}, & t>0 \\
(-)^m\, e^{ i\Omega_\lambda t}, & t<0  
\end{array}
\right.
\ , 
\end{eqnarray}
where the phonon creation and annihilation operators have relations 
\begin{eqnarray}
&& c_{\lambda m}^\dagger(t) = c_{\lambda m}^\dagger e^{ i\Omega_\lambda t} \ ,\\ 
&& c_{\overline{\lambda m}} = (-)^m\,c_{\lambda\,-m} \ . 
\end{eqnarray}
$|\Phi_{\rm ph}\rangle$ is a phonon vacuum state. 
The Green function in Lehmann representation is given by 
\begin{eqnarray}
iD_{\lambda m}^0 (\omega)
&\stackrel{d}{=}& 
\lim_{\delta \rightarrow +0}
\int_{-\infty}^\infty dt \,
e^{i\omega t/\hbar}\, i D_{\lambda m}^0(t)
e^{-\delta(2\theta(t)-1)t/\hbar} \nonumber\\
&=&
(-)^m\, iD_\lambda^0(\omega) \ , \label{Dlm} 
\end{eqnarray}
\begin{equation}
iD_\lambda^0(\omega) = 
\frac{i\hbar}{\omega - \hbar\Omega_\lambda + i\delta\hbar} - 
\frac{i\hbar}{\omega + \hbar\Omega_\lambda - i\delta\hbar}        \ . \label{Dl}
\end{equation}
Inserting Eqs.~(\ref{vertex-aI}) and (\ref{Dlm}) to Eq.~(\ref{slf-e.1}), 
we obtain 
\begin{eqnarray}
\hbar \Sigma_{\mu}(\omega)  
&=& 
\int_{-\infty}^\infty \frac{d\omega^\prime}{2\pi}
\sum_{\lambda} \sum_{\mu^\prime}
\tau_3
\frac{1}{\hbar}
G_{\mu^\prime}(\omega^\prime)
\tau_3
\nonumber \\
&&\times
\frac{\hbar}{2\Omega_\lambda B_\lambda}
\frac{1}{2j_\mu + 1}
\left|
\langle \mu || R_0 \frac{dU(r)}{dr} Y_{\lambda} || \mu^\prime \rangle 
\right|^2
\frac{i}{\hbar} D_{\lambda}^0 (\omega - \omega^\prime) \ . \label{slf-e.3}
\end{eqnarray}

\vspace{2em}

\noindent
{\large \bf Appendix C -- general form of particle Green function  --}

\nopagebreak
%
 The particle Green functions in the time representation are defined 
\begin{eqnarray}
G_\mu^{11}(t)
= -i \langle \Psi_0 |
 {\rm T} [ c_{\mu m_\mu}(t)\, c_{\mu m_\mu}^\dagger ]
| \Psi_0 \rangle ,
\label{G11t}\\
G_\mu^{22}(t)
= -i \langle \Psi_0 | {\rm T}
 [ c_{ \overline{\mu m_\mu} }^\dagger(t)\, c_{ \overline{\mu m_\mu} } ]
| \Psi_0 \rangle ,
\label{G22t}
\end{eqnarray}
where 
$|\Psi_0\rangle$ is an $N$-particle ground state
with $N$ being an even number.
The time-dependent creation operator of the single-particle is defined by 
$$ c_{\mu m_\mu}^\dagger(t) =
e^{iH^\prime t/\hbar}
c_{\mu m_\mu}^\dagger
e^{-iH^\prime t/\hbar} \ ,$$
$$H^\prime = H - \varepsilon_F\hat{N}\ .$$ 
$H$ is the many-body Hamiltonian. 
$\varepsilon_F$ is the Fermi level, and 
$\hat{N}$ denotes the particle-number operator. 
$\overline{(\mu m_\mu)}$ indicates the time-reversed state of $(\mu m_\mu)$. 
Since $| \Psi_0 \rangle$ is spherically symmetric, 
the Green function is a scalar. 
Eqs.~(\ref{G11t}) and (\ref{G22t}) yield 

\noindent
i) t $>$ 0
\begin{eqnarray}
G_\mu^{11}(t)
&=& 
-i \langle \Psi_0 |
 c_{\mu m_\mu}(t)\, c_{\mu m_\mu}^\dagger
| \Psi_0 \rangle \nonumber \\
&=&
-i \sum_i e^{ i(E_0 - E_i)t/\hbar }
\langle \Psi_0 | c_{\mu m_\mu}          | \Psi_i \rangle
\langle \Psi_i | c_{\mu m_\mu}^\dagger  | \Psi_0 \rangle \ , \\
G_\mu^{22}(t)
&=&
-i \langle \Psi_0 |
c_{ \overline{\mu m_\mu} }^\dagger(t)\, c_{ \overline{\mu m_\mu} }
| \Psi_0 \rangle \nonumber \\
&=&
-i \sum_i e^{ i(E_0 - E_i)t/\hbar }
\langle \Psi_0 | c_{ \overline{\mu m_\mu} }^\dagger | \Psi_i \rangle
\langle \Psi_i | c_{ \overline{\mu m_\mu} }         | \Psi_0 \rangle \ , 
\end{eqnarray}
ii) t $<$ 0
\begin{eqnarray}
G_\mu^{11}(t)
&=& 
-i \langle \Psi_0 |
(-)c_{\mu m_\mu}^\dagger\, c_{\mu m_\mu}(t)
| \Psi_0 \rangle \nonumber \\
&=&
 i \sum_i e^{ i(E_i - E_0)t/\hbar }
\langle \Psi_0 | c_{\mu m_\mu}^\dagger  | \Psi_i \rangle
\langle \Psi_i | c_{\mu m_\mu}          | \Psi_0 \rangle \ , \\
G_\mu^{22}(t)
&=&
-i \langle \Psi_0 |
(-)c_{ \overline{\mu m_\mu} }\, c_{ \overline{\mu m_\mu} }^\dagger(t)
| \Psi_0 \rangle \nonumber \\
&=&
 i \sum_i e^{ i(E_i - E_0)t/\hbar }
\langle \Psi_0 | c_{ \overline{\mu m_\mu} }         | \Psi_i \rangle
\langle \Psi_i | c_{ \overline{\mu m_\mu} }^\dagger | \Psi_0 \rangle \ .
\end{eqnarray}
$E_0$ is the ground-state eigenenergy of $H^\prime$. 
$\{\Psi_i\}$ indicates a complete set of a many-body space which 
consists of $N\pm 1$,$N\pm 3$ $\cdots$-particle states. 
$E_i$ is the eigenenergy of $|\Psi_i\rangle$ for $H^\prime$. 

 The Green functions in Lehmann representation read 
\begin{eqnarray}
G_\mu^{11}(\omega)
&\stackrel{d}{=}&
\lim_{\eta\rightarrow +0} \int_{-\infty}^\infty  dt\, e^{i\omega t/\hbar}\,
G_\mu^{11}(t) \,
e^{-(2\theta(t)-1)\eta t}
\nonumber \\
&=&
\sum_i
\frac{\left| \langle \Psi_0 | c_{\mu m_\mu} | \Psi_i \rangle \right|^2 }
 { \omega/\hbar + (E_0 - E_i)/\hbar + i\eta }
+
\sum_i
\frac{\left| \langle \Psi_0 | c_{\mu m_\mu}^\dagger | \Psi_i \rangle \right|^2 }
 { \omega/\hbar + (E_i - E_0)/\hbar - i\eta } \ ,\label{G11w} \\
G_\mu^{22}(\omega)
 &\stackrel{d}{=}&
\lim_{\eta\rightarrow +0} \int_{-\infty}^\infty dt\, e^{i\omega t/\hbar}\,
G_\mu^{22}(t)\, e^{ -( 2\theta(t)-1 )\eta t }
\nonumber\\
&=&
\sum_i
\frac{\left| \langle \Psi_0 |
c_{ \overline{\mu m_\mu} }^\dagger
| \Psi_i \rangle \right|^2 }
{ \omega/\hbar + (E_0-E_i)/\hbar + i\eta }
+
\sum_i
\frac{\left| \langle \Psi_0 |
 c_{ \overline{\mu m_\mu} }
| \Psi_i \rangle \right|^2 }
{ \omega/\hbar + (E_i-E_0)/\hbar - i\eta }
\ . 
\label{G22w}
\end{eqnarray}
Thus it is seen that we can put 
\begin{eqnarray}
\frac{1}{\hbar} G_\mu^{11}(\omega) &=& 
\sum_a
\left(
\frac{R_{\mu a}^{11}( \omega_{G+}^{\mu a})}{ \omega - \omega_{G+}^{\mu a} }
+
\frac{R_{\mu a}^{11}(-\omega_{G+}^{\mu a})}{ \omega + \omega_{G+}^{\mu a} }
\right)
e^{ i\omega\eta} \ , \\
\frac{1}{\hbar} G_\mu^{22}(\omega) &=& 
\sum_a
\left(
\frac{R_{\mu a}^{22}( \omega_{G+}^{\mu a})}{ \omega - \omega_{G+}^{\mu a} }
+
\frac{R_{\mu a}^{22}(-\omega_{G+}^{\mu a})}{ \omega + \omega_{G+}^{\mu a} }
\right)
e^{-i\omega\eta} \ .
\end{eqnarray}
The factor
$e^{i\omega\eta}$ ( $e^{-i\omega\eta}$ )
in
$G_\mu^{11}(\omega)$ ( $G_\mu^{22}(\omega)$ )
comes from the definition of the Green functions at $t$ = 0:
\begin{eqnarray}
&&G_\mu^{11}(t=0) = \lim_{t\rightarrow -0} G_\mu^{11}(t) \ ,\\
&&G_\mu^{22}(t=0) = \lim_{t\rightarrow +0} G_\mu^{22}(t) \ .
\end{eqnarray}
See section 7-2 in \cite{Sc64}. 
From the time-reversal invariance it follows that
\begin{eqnarray}
&&E_i = E_{\overline i} \ , \label{EiEibar} \\
&&\hat{T} |\Psi_0 \rangle = \hat{T}^{-1} |\Psi_0 \rangle = |\Psi_0\rangle\ ,
\label{TPsi0} \\
&&\left\{ |\Psi_i\rangle \right\}_{i=\cdots}
=
\left\{ |\Psi_{\overline i}\rangle \right\}_{{\overline i}=\cdots} \ ,
\label{iibar}
\end{eqnarray}
where $\hat{T}$ is the time-reversal operator. 
In addition we have
\begin{equation}
\hat{T}^{-1} | \Psi_i \rangle
= -\hat{T} |\Psi_i \rangle 
= - |\Psi_{\overline i} \rangle \ .
\label{TPsii}
\end{equation}
The negative sign is due to that $|\Psi_i\rangle$ is an odd particle-number 
state. 
By using Eqs.~(\ref{EiEibar}) -- (\ref{TPsii}), Eq.~(\ref{G22w})
becomes 
\begin{eqnarray}
G_\mu^{22}(\omega)
&=&
\sum_i
\frac{\left| \langle \hat{T}^{-1}\Psi_0 |
c_{\mu m_\mu}^\dagger
\hat{T}^{-1} | \Psi_i \rangle \right|^2 }
{ \omega/\hbar + (E_0-E_i)/\hbar + i\eta }
+
\sum_i
\frac{\left| \langle \hat{T}^{-1}\Psi_0 |
c_{\mu m_\mu} 
\hat{T}^{-1} | \Psi_i \rangle \right|^2 }
{ \omega/\hbar + (E_i-E_0)/\hbar - i\eta }
\nonumber\\ 
&=&
\sum_i
\frac{\left| \langle \Psi_0 | c_{\mu m_\mu}^\dagger | \Psi_i \rangle \right|^2 }
{ \omega/\hbar + (E_0-E_i)/\hbar + i\eta }
+
\sum_i
\frac{\left| \langle \Psi_0 | c_{\mu m_\mu}         | \Psi_i \rangle \right|^2 }
{ \omega/\hbar + (E_i-E_0)/\hbar - i\eta }
\ .
\end{eqnarray}
This equation implies
\begin{equation}
G_\mu^{22}(-\omega) = -G_\mu^{11}(\omega)\ .
\end{equation}
For the residues we have the relations
\begin{eqnarray}
&&
R_{\mu a}^{22} ( \omega_{G+}^{\mu a})
= R_{\mu a}^{11} (-\omega_{G+}^{\mu a}) \ , \\
&&
R_{\mu a}^{22} (-\omega_{G+}^{\mu a}) = R_{\mu a}^{11} ( \omega_{G+}^{\mu a}) \ . 
\end{eqnarray}

 The anomalous Green functions are defined as 
\begin{eqnarray}
G_\mu^{12} (t) &=&
-i \langle \Psi_0 | 
 {\rm T}[ c_{\mu m_\mu} (t) c_{ \overline{\mu m_\mu} } ]
 | \Psi_0 \rangle \ ,\\ 
G_\mu^{21} (t) &=& 
-i \langle \Psi_0 | 
 {\rm T} [ c_{ \overline{\mu m_\mu} }^\dagger (t) c_{\mu m_\mu}^\dagger ]
 | \Psi_0 \rangle \ .
\end{eqnarray}
In the same manner as discussed for $G_\mu^{11}(\omega)$ and 
$G_\mu^{22}(\omega)$, it turns out that
\begin{eqnarray}
G_\mu^{12}(\omega)
&=&
\sum_i
\frac{
 \langle \Psi_0 | c_{\mu m_\mu} | \Psi_i \rangle
 \langle \Psi_i | c_{ \overline{\mu m_\mu} } | \Psi_0 \rangle
 }
{ \omega/\hbar + ( E_0 - E_i )/\hbar + i\eta }
-
\sum_i
\frac{
 \langle \Psi_0 | c_{\mu m_\mu}              | \Psi_i \rangle^\ast
 \langle \Psi_i | c_{ \overline{\mu m_\mu} } | \Psi_0 \rangle^\ast
 }
{ \omega/\hbar + ( E_i - E_0 )/\hbar - i\eta }
\ ,\nonumber \\
             \\
G_\mu^{21}(\omega)
&=&
\sum_i
\frac{
 \langle \Psi_0 | c_{ \overline{\mu m_\mu} }^\dagger | \Psi_i \rangle
 \langle \Psi_i | c_{\mu m_\mu}^\dagger              | \Psi_0 \rangle
 }
{ \omega/\hbar + ( E_0 - E_i )/\hbar + i\eta }
-
\sum_i
\frac{
 \langle \Psi_0 | c_{ \overline{\mu m_\mu} }^\dagger | \Psi_i \rangle^\ast
 \langle \Psi_i | c_{\mu m_\mu}^\dagger              | \Psi_0 \rangle^\ast
 }
{ \omega/\hbar + ( E_i - E_0 )/\hbar - i\eta }
\ .\nonumber\\
\end{eqnarray}
Thus one can put 
\begin{eqnarray}
\frac{1}{\hbar}G_\mu^{12}(\omega) &=& 
\sum_a
\left(
\frac{R_{\mu a}^{12}( \omega_{G+}^{\mu a})}{ \omega - \omega_{G+}^{\mu a} }
+
\frac{R_{\mu a}^{12}(-\omega_{G+}^{\mu a})}{ \omega + \omega_{G+}^{\mu a} }
\right)
\ , \\
\frac{1}{\hbar}G_\mu^{21}(\omega) &=& 
\sum_a
\left(
\frac{R_{\mu a}^{21}( \omega_{G+}^{\mu a})}{ \omega - \omega_{G+}^{\mu a} }
+
\frac{R_{\mu a}^{21}(-\omega_{G+}^{\mu a})}{ \omega + \omega_{G+}^{\mu a} }
\right)
\ ,
\end{eqnarray}
with the relations
\begin{eqnarray}
&&
R_{\mu a}^{12} (-\omega_{G+}^{\mu a})
= -{R_{\mu a}^{12}}^\ast (\omega_{G+}^{\mu a}) \ , \\
&&
R_{\mu a}^{21} (-\omega_{G+}^{\mu a})
= -{R_{\mu a}^{21}}^\ast ( \omega_{G+}^{\mu a}) \ ,\\
&&
R_{\mu a}^{21} ( \omega_{G+}^{\mu a})
= -{R_{\mu a}^{12}} (-\omega_{G+}^{\mu a})
= {R_{\mu a}^{12}}^\ast ( \omega_{G+}^{\mu a}) \ , \\
&&
R_{\mu a}^{21} (-\omega_{G+}^{\mu a})
= -R_{\mu a}^{12} ( \omega_{G+}^{\mu a})
= {R_{\mu a}^{12}}^\ast (-\omega_{G+}^{\mu a}) \ .
\end{eqnarray}
It is worthy to note that a relation holds regardless of the 
time-reversal invariance of $|\Psi_0\rangle$: 
\begin{equation}
R_{\mu a}^{11}(\omega_{G+}^{\mu a}) R_{\mu a}^{22}(\omega_{G+}^{\mu a}) 
= \left| R_{\mu a}^{12}(\omega_{G+}^{\mu a}) \right|^2 \ .
\label{Z11Z22Z12}
\end{equation}

 Let us assume that the energy of $|\Psi_i\rangle$ is equal to that 
of $|\Psi_{i^\prime}\rangle$ with a different particle number incidentally. 
Then we can introduce the basis state
$|\Psi_i\rangle$ which has a mixed particle number. 
In this way it is possible that 
$R_{\mu a}^{11}( \omega_{G+}^{\mu a})$ and 
$R_{\mu a}^{11}(-\omega_{G+}^{\mu a})$ are non-zero simultaneously 
without causing a problem, 
if the system is ideally large. 

\vspace{2em}

\noindent
{\large \bf Appendix D -- formula of pole and residue --}

\nopagebreak
%
As is mentioned in the text, we determine the real part of the pole 
of the perturbed nucleon Green function approximately by the condition
\begin{equation}
\det {\overline G}_\mu^{-1} (\pm{\rm Re}\,\omega_{G+}^{\mu a}) = 0 \ ,
\label{det=0}
\end{equation}
where ${\overline G}_\mu^{-1}(\omega)$
is a matrix in which the non-hermitian components of $G_\mu^{-1}(\omega)$
were eliminated. 
The advantage of this method is that the pole search is a one-dimensional 
problem. 

 Using Eq.~(\ref{G11}) in the text, we obtain
\begin{equation}
\left(
\frac{d}{d\omega}
\frac{\hbar}{G_\mu^{11}(\omega)}
\right)^{-1}
=
-
\frac{
 \left\{ 
\displaystyle
 \sum_a
 \left(
 \frac{R_{\mu a}^{11}( \omega_{G+}^{\mu a})}{\omega-\omega_{G+}^{\mu a}}
 +
 \frac{R_{\mu a}^{11}(-\omega_{G+}^{\mu a})}{\omega+\omega_{G+}^{\mu a}}
 \right)
\right\}^2
}
{
\displaystyle
 \sum_a
 \left(
 -\frac{R_{\mu a}^{11}( \omega_{G+}^{\mu a})}{ (\omega-\omega_{G+}^{\mu a})^2 }
 -\frac{R_{\mu a}^{11}(-\omega_{G+}^{\mu a})}{ (\omega+\omega_{G+}^{\mu a})^2 }
 \right)
}
\ .
\end{equation}
Therefore it is found that 
\begin{equation}
\lim_{\omega \rightarrow \pm\omega_{G+}^{\mu a}}
\left(
\frac{d}{d\omega}
\frac{\hbar}{G_\mu^{11}(\omega)}
\right)^{-1}
=
R_{\mu a}^{11}(\pm\omega_{G+}^{\mu a}) \ ,
\end{equation}
and in accordance with Eq.~(\ref{det=0}) we approximate 
\begin{equation}
R_{\mu a}^{11}(\pm\omega_{G+}^{\mu a}) 
\simeq
\lim_{\omega \rightarrow \pm{\rm Re}\,\omega_{G+}^{\mu a}}
\left(
\frac{d}{d\omega}
\frac{\hbar}{{\overline G}_\mu^{11}(\omega)}
\right)^{-1} \ .
\end{equation}
The following equation may be more convenient in the numerical calculation: 
\begin{equation}
R_{\mu a}^{11}(\pm\omega_{G+}^{\mu a}) 
\simeq
\left.
\frac
 { \omega + \varepsilon_\mu^0 - \varepsilon_F
  - \hbar {\overline \Sigma}_\mu^{22}(\omega) }
 { \frac{d}{d\omega}\,\hbar^2\det {\overline G}_\mu^{-1}(\omega) }
\right|_{\omega=\pm{\rm Re}\,\omega_{G+}^{\mu a}}
\ ,
\label{old143}
\end{equation}
where ${\overline\Sigma}_\mu(\omega)$ is defined in the same way as 
${ {\overline G}_\mu }^{-1}$.

 For the pairing part we have 
\begin{eqnarray}
R_{\mu a}^{12}(\pm\omega_{G+}^{\mu a}) 
&\simeq&
\lim_{\omega \rightarrow \pm{\rm Re}\,\omega_{G+}^{\mu a}}
\left(
\frac{d}{d\omega}
\frac{\hbar}{{\overline G}_\mu^{12}(\omega)}
\right)^{-1}
\nonumber\\
&=&
\left.
\frac
 { {\overline \Sigma}_\mu^{12}(\omega) }
 { \frac{d}{d\omega}\,\hbar\det {\overline G}_\mu^{-1}(\omega) }
\right|_{\omega=\pm{\rm Re}\,\omega_{G+}^{\mu a}}
\ .
\end{eqnarray}
For the imaginary part of the pole it is derived that 
\begin{eqnarray}
{\rm Im} 
\frac{\hbar}{G_\mu^{11}(\pm{\rm Re}\,\omega_{G+}^{\mu a})}
R_{\mu a}^{11}(\pm\omega_{G+}^{\mu a})
&=&
{\rm Im}
\frac
{
R_{\mu a}^{11}(\pm\omega_{G+}^{\mu a})
}
{\displaystyle \sum_{a^\prime}
\left( 
\frac
 {R_{\mu a}^{11}(\omega_{G+}^{\mu a^\prime})}
 {\pm{\rm Re}\,\omega_{G+}^{\mu a}-\omega_{G+}^{\mu a^\prime}}
+
\frac
 {R_{\mu a}^{11}(-\omega_{G+}^{\mu a^\prime})}
 {\pm{\rm Re}\,\omega_{G+}^{\mu a}+\omega_{G+}^{\mu a^\prime}}
\right)
}
\nonumber \\
&\simeq&
\mp{\rm Im}\, \omega_{G+}^{\mu a} \ .
\end{eqnarray}
This approximation is based on a condition that 
the imaginary part of the pole is very small. 

\vspace{2em}

\noindent
{\large \bf Appendix E -- total energy --}

%
 Derivation of the equation of the total energy is possible 
in the same way as that for the case without the pairing. 
(See sections 7 (Chap.3) and 46 (Chap.12) in \cite{FW71}). 
We put the Hamiltonian 
\begin{eqnarray}
H &=&
\sum_{\mu m_\mu} {\tilde\varepsilon}_\mu^0 c_{\mu m_\mu}^\dagger c_{\mu m_\mu}
+
\varepsilon_F\langle {\hat N} \rangle
\nonumber \\
&&+
\sum_{\lambda m} \hat{\alpha}_{\lambda m}
\sum_{\mu m_\mu \nu m_\nu}
\langle \mu m_\mu | \kappa_\lambda F_{\lambda m}^\dagger | \nu m_\nu \rangle
c_{\mu m_\mu}^\dagger c_{\nu m_\nu}
\nonumber\\
&&+
\sum_{\lambda m}
\hbar\Omega_{\lambda}(c_{\lambda m}^\dagger c_{\lambda m} + \frac{1}{2})
\ .\label{H}
\end{eqnarray}
The expectation value of the particle number 
$\langle \hat{N} \rangle$  was taken in advance. 
For the particle-phonon coupling, see Appendix A. 
$\hbar\Omega_\lambda$ is the phonon energy. 
We consider 
\begin{eqnarray}
&&c_{\mu m_\mu}(t) = e^{iHt/\hbar} c_{\mu m_\mu} e^{-iHt/\hbar}\ , \\
&& {\hat\alpha}_{\lambda m}(t) =
e^{iHt/\hbar} \hat{\alpha}_{\lambda m} e^{-iHt/\hbar}\ . 
\end{eqnarray}
Then it follows that
\begin{equation}
i\hbar \frac{\partial}{\partial t} c_{\mu m_\mu}(t)
=
{\tilde\varepsilon}_\mu^0 c_{\mu m_\mu} (t)
+
\sum_{\lambda m} \hat{\alpha}_{\lambda m}(t)
\sum_{\nu m_\nu}
\langle \mu m_\mu | \kappa_\lambda F_{\lambda m}^\dagger | \nu m_\nu \rangle 
c_{\nu m_\nu}(t) \ .
\end{equation}
Multiplying $c_{\mu m_\mu}^\dagger(t^\prime)$, $t^\prime$ $>$ $t$,
 from the left, using the definition of the particle 
Green function in the time representation and putting $t$ = 0, 
one finds 
\begin{eqnarray}
\lefteqn{
 \lim_{t^\prime \rightarrow +0}
 \left(
 \left.
 i\hbar\frac{\partial}{\partial t}
 \sum_{\mu m_\mu} (-i)G_\mu^{11}(t,t^\prime)\right|_{t=0}
 \right)
 } \nonumber \\
&&=
\langle \sum_{\mu m_\mu}
{\tilde \varepsilon}_\mu^0 c_{\mu m_\mu}^\dagger c_{\mu m_\mu} \rangle
+
\left\langle
\sum_{\lambda m} \hat{\alpha}_{\lambda m}
\sum_{\mu m_\mu \nu m_\nu}
\langle \mu m_\mu | \kappa_\mu F_{\lambda m}^\dagger | \nu m_\nu \rangle 
c_{\mu m_\mu}^\dagger c_{\nu m_\nu}
\right\rangle
\ .
\end{eqnarray}
By using the Green function in Lehmann representation, 
it turns out that
\begin{eqnarray}
\langle H \rangle 
&=&
-i \lim_{\eta \rightarrow +0 }
\sum_{\mu m_\mu}
\int_{-\infty}^\infty \frac{d\omega}{2\pi}
\frac{1}{\hbar} \omega G_\mu^{11}(\omega) e^{i\omega\eta}
+\varepsilon_F \langle \hat{N} \rangle
+
\langle \sum_{\lambda m}
\hbar \Omega_{\lambda}
( c_{\lambda m}^\dagger c_{\lambda m} + 1/2 ) 
\rangle \nonumber\\
&=&
\sum_{\mu m_\mu} \sum_{a}
(-\omega_{G+}^{\mu a}) R_{\mu a}^{11}(-\omega_{G+}^{\mu a})
+\varepsilon_F \langle \hat{N} \rangle
+
\langle H_0^{\rm ph} \rangle \ ,
\end{eqnarray}
where $ H_0^{\rm ph} $ denotes the fourth term in Eq.~(\ref{H}).  
$\omega$ in our notation always indicates energy. 
The last equation is derived for the ground state, for which
the poles with positive infinitesimal imaginary part have negative real part.
( ${\rm Re}\,\omega_{G+}^{\mu a}$ $>$ 0 )

 It is possible to rewrite the total energy by using Dyson equation 
as follows:
\begin{eqnarray}
\lefteqn{
 \langle H \rangle 
 -\langle H_0^{\rm ph} \rangle
}\nonumber\\
&=&
\varepsilon_F \langle \hat{N} \rangle
-i\sum_{\mu m_\mu}
\frac{1}{2}
\lim_{\eta\rightarrow +0}
\int_{-\infty}^\infty \frac{d\omega}{2\pi}
{\rm Tr}\left[
\left(
\begin{array}{cc}
\omega & 0 \\
0      & \omega
\end{array}
\right)
\frac{1}{\hbar} G_\mu(\omega)
\left(
\begin{array}{cc}
e^{i\eta\omega} & 0 \\
0               & e^{-i\eta\omega}
\end{array}
\right)
\right]
\nonumber\\
&=&
\varepsilon_F \langle \hat{N} \rangle 
-i\sum_{\mu m_\mu} \frac{1}{2} \lim_{\eta\rightarrow +0}\int_{-\infty}^\infty
\frac{d\omega}{2\pi}
{\rm Tr}
\left[
\left\{\rule{0ex}{3.5ex}
\hbar G_\mu^{-1}(\omega) 
\right.
\right.
\nonumber\\
&&+
\left.
\left.
\left(
\begin{array}{cc}
\tilde{\varepsilon}_\mu^0  & 0 \\
0                          & -\tilde{\varepsilon}_\mu^0
\end{array}
\right)
+
\hbar\Sigma_\mu(\omega)
\right\}
\frac{1}{\hbar} G_\mu(\omega)
\left(
\begin{array}{cc}
e^{i\eta\omega} & 0 \\
0               & e^{-i\eta\omega} 
\end{array}
\right)
\right]
\nonumber\\
&=&
\sum_{\mu m_\mu} \sum_a
\varepsilon_{\mu}^0 R_{\mu a}^{11}(-\omega_{G+}^{\mu a})
-i\sum_{\mu m_\mu} \frac{1}{2} \lim_{\eta\rightarrow +0}\int_{-\infty}^\infty
\frac{d\omega}{2\pi}
\left(
\hbar \Sigma_\mu^{11}(\omega) \frac{1}{\hbar} G_\mu^{11}(\omega)e^{i\eta\omega}
\right.\nonumber\\
&&
+ \hbar\Sigma_\mu^{12}(\omega) \frac{1}{\hbar} G_\mu^{21}(\omega)e^{i\eta\omega}
+ \hbar \Sigma_\mu^{21}(\omega) \frac{1}{\hbar} G_\mu^{12}(\omega)e^{-i\eta\omega}
\nonumber\\
&&
\left.
+ \hbar\Sigma_\mu^{22}(\omega) \frac{1}{\hbar} G_\mu^{22}(\omega)e^{-i\eta\omega}
\right) \ ,
\end{eqnarray}
We have used the relation between $G_\mu^{11}(\omega)$ and $G_\mu^{22}(\omega)$   
(Appendix C). 
A reasonable definition of the pairing energy may be 
\begin{equation}
E_{\rm pair}
=
-i\sum_{\mu m_\mu} \frac{1}{2} \lim_{\eta\rightarrow +0}\int_{-\infty}^\infty
\frac{d\omega}{2\pi}
\left(
 \hbar\Sigma_\mu^{12}(\omega) \frac{1}{\hbar} G_\mu^{21}(\omega)e^{i\eta\omega}
+\hbar \Sigma_\mu^{21}(\omega) \frac{1}{\hbar} G_\mu^{12}(\omega)e^{-i\eta\omega}
\right) \ .
\end{equation}
A factor $i/2$ is necessary in the diagramatic derivation. 
If one puts 
$\hbar\Sigma_\mu^{12}(\omega) \simeq \hbar\Sigma_\mu^{12}(\omega_{G+}^{\mu a_0})$ 
in addition to the conditions which are used for deriving Eq.~(\ref{gapDycmp}), 
then we have 
\begin{equation}
E_{\rm pair} \simeq \sum_\mu(2j_\mu + 1) u_\mu v_\mu \Delta_\mu \ . 
\label{epairbcs}
\end{equation}
It is noted that Eq.~(\ref{epairbcs}) differs from the known pairing energy 
in the BCS approximation by a factor 2. 
This is because the particle-phonon interaction introduced in our calculation 
is not of the two-body type (see Eq.~(\ref{h_int}), \cite{FW71} and its chap.~12).

\vspace{2em}

\noindent
{\large \bf Appendix F -- pairing gap --}


 The equation to determine the pole is given by 
\begin{eqnarray}
0 &=& \hbar^2{\rm Re\,} \det G_\mu^{-1}({\rm Re\,}\omega_{G+}^{\mu a}) \nonumber\\
&=&
\left|
\begin{array}{cc}
{\rm Re\,}\omega_{G+}^{\mu a} - \tilde{\varepsilon}_\mu^0
- {\rm Re\,}\hbar\Sigma_\mu^{11}({\rm Re\,}\omega_{G+}^{\mu a})
&
-\hbar\Sigma_\mu^{12}({\rm Re\,}\omega_{G+}^{\mu a})
\vspace{1.5ex}
\\
-\hbar\Sigma_\mu^{21}({\rm Re\,}\omega_{G+}^{\mu a})
&
{\rm Re\,}\omega_{G+}^{\mu a} + \tilde{\varepsilon}_\mu^0
- {\rm Re\,}\hbar\Sigma_\mu^{22}({\rm Re\,}\omega_{G+}^{\mu a})
\end{array}
\right|\ .\nonumber\\
\label{det}
\end{eqnarray}
We define functions
\begin{eqnarray}
&& Z_\mu(\omega_{G+}^{\mu a}) \stackrel{d}{=}
1 - \frac{1}{\omega_{G+}^{\mu a}}
{\rm Re\,}\hbar{\Sigma_\mu^{11}}^{\rm odd}(\omega_{G+}^{\mu a}) \ ,\label{calC} \\
&&{\Sigma_{\mu}^{11}}^{\rm odd}(\omega_{G+}^{\mu a})
\stackrel{d}{=}
\frac{1}{2}
\left(
\Sigma_{\mu}^{11}( \omega_{G+}^{\mu a})
-
\Sigma_{\mu}^{11}(-\omega_{G+}^{\mu a})
\right)
\ ,\\
&&{\Sigma_{\mu}^{11}}^{\rm even}(\omega_{G+}^{\mu a})
\stackrel{d}{=}
\frac{1}{2}
\left(
\Sigma_{\mu}^{11}( \omega_{G+}^{\mu a})
+
\Sigma_{\mu}^{11}(-\omega_{G+}^{\mu a})
\right) \ .
\end{eqnarray}
By dividing by $(\, Z_\mu({\rm Re\,}\omega_{G+}^{\mu a})\, )^2$, 
Eq.~(\ref{det}) can be written 
\begin{equation}
0 = 
\left|
\begin{array}{cc}
{\rm Re}\,\omega_{G+}^{\mu a}
- \frac{
 \tilde{\varepsilon}_\mu^0
  + {\rm Re\,}\hbar{\Sigma_\mu^{11}}^{\rm even}({\rm Re\,}\omega_{G+}^{\mu a}) }
 {Z_\mu({\rm Re\,}\omega_{G+}^{\mu a})}
&
-\frac{ \hbar\Sigma_\mu^{12}({\rm Re\,}\omega_{G+}^{\mu a}) }
      { Z_\mu({\rm Re\,}\omega_{G+}^{\mu a}) }
\vspace{1.5ex}
\\
-\frac{ \hbar\Sigma_\mu^{21}({\rm Re\,}\omega_{G+}^{\mu a}) }
      { Z_\mu({\rm Re\,}\omega_{G+}^{\mu a}) }
&
{\rm Re\,}\omega_{G+}^{\mu a}
+\frac{
 \tilde{\varepsilon}_\mu^0
 +{\rm Re\,}\hbar{\Sigma_\mu^{11}}^{\rm even}({\rm Re\,}\omega_{G+}^{\mu a}) }
 { Z_\mu({\rm Re\,}\omega_{G+}^{\mu a}) }
\end{array}
\right|\ . \label{det2}
\end{equation}
We have used 
$\Sigma_\mu^{22}(\omega)$ = $-\Sigma_\mu^{11}(-\omega)$.
The diagonal elements of Eq.~(\ref{det2}) has the structure
of the BCS equation 
for a time-reversal-invariant state. 
Therefore it is usual to define the pairing gap by
\begin{equation}
\Delta_\mu = 
\frac{ \hbar \Sigma_\mu^{12}({\rm Re\,}\omega_{G+}^{\mu a_0}) }
     { Z_\mu({\rm Re\,}\omega_{G+}^{\mu a_0}) }
\ . \label{gap}
\end{equation}
( See sections 7-2 in \cite{Sc64}, 10 in \cite{Al82} and \cite{Av99}. )
$a_0$ in Eq.~(\ref{gap}) indicates the quasiparticle pole. 
From Eq.~(\ref{det2}) it is also seen that 
\begin{equation}
 Z_\mu({\rm Re\,}\omega_{G+}^{\mu a_0}) = 
\frac{ 1 }{ {\rm Re\,}\omega_{G+}^{\mu a_0} }
\sqrt{
 (\, \tilde{\varepsilon}_\mu^0
  + {\rm Re\,}\hbar {\Sigma_\mu^{11}}^{\rm even}({\rm Re\,}\omega_{G+}^{\mu a_0})\, )^2
 + \left| \hbar\Sigma_\mu^{12}({\rm Re\,}\omega_{G+}^{\mu a_0})\right|^2
}
\ .
\label{gap2}
\end{equation}

 In the same way the perturbed single-particle energy is given by 
\begin{equation}
\widetilde{\varepsilon}_\mu^1 = 
 \frac{ 1 }{ Z_\mu({\rm Re\,}\omega_{G+}^{\mu a_0}) }
(\, \widetilde{\varepsilon}_\mu^0
+ {\rm Re\,}\hbar {\Sigma_\mu^{11}}^{\rm even}({\rm Re\,}\omega_{G+}^{\mu a_0})\,) \ .
\label{per-spe}
\end{equation}

\end{document}